\documentclass[aps,prc,showpacs,amssymb,amsmath,amsfonts,nofootinbib,floatfix]{revtex4-2}
\usepackage{graphicx}% Include figure files

\makeatletter
\def\p@subsection{}

\def\p@subsubsection{}
\makeatother
\usepackage{natbib}
\usepackage[usenames]{color}
\usepackage{epsfig,graphicx}
\usepackage{epstopdf}
\usepackage{amssymb,amsmath,eqparbox}
\usepackage{subfig}
\usepackage{epstopdf}
\usepackage{enumitem}
\usepackage{comment}
\usepackage[bookmarks={false}]{hyperref}
\usepackage{enumitem}
\usepackage{bbm}

\definecolor{grey}{rgb}{0.9,0.9,0.9}
\definecolor{black}{rgb}{0,0,0}

\hypersetup{pdfstartview={XYZ null null 1.}}

\newcommand{\be}{\begin{eqnarray}}
\newcommand{\ee}{\end{eqnarray}}
\newcommand{\bc}{\begin{center}}
\newcommand{\ec}{\end{center}}

\newcommand{\Ocal}{\mathcal{O}}

\newcommand{\Real}{\text{Re}}
\newcommand{\Imag}{\text{Im}}

\newcommand{\HISKP}{Helmholtz-Institut f\"{u}r Strahlen- und Kernphysik, Universit\"{a}t Bonn, D-53115 Bonn, Germany, Germany}
\newcommand{\RudjerBoskovic}{Rudjer Bo\v{s}kovi\'{c} Institute, Bijeni\v{c}ka cesta 54, P.O. Box 180, 10002 Zagreb, Croatia}
\newcommand{\Mainz}{Institut f\"{u}r Kernphysik, Universit\"{a}t Mainz, D-55099 Mainz, Germany}
\newcommand{\Tesla}{Tesla Biotech d.o.o., Mandlova 7, 10000 Zagreb, Croatia}

\begin{document}

\allowdisplaybreaks

%\title{Single-energy partial-wave analysis of $K^{+} \Lambda$ photoproduction using the AA/PWA-method}
\title{Application of the single-channel, single-energy AA/PWA method to $K^{+} \Lambda$ photoproduction}
\author{A.~{\v{S}}varc}\email[Corresponding author: ]{svarc@irb.hr}
\affiliation{\RudjerBoskovic,\Tesla}
\author{Y.~Wunderlich}
\affiliation{\HISKP}
\author{L.~Tiator}
\affiliation{\Mainz}

\date{\today}
\begin{abstract}
\vspace*{0.5cm} The new single-channel, single-energy partial wave analysis method based on a simultaneous use of amplitude and
partial wave analysis called AA/PWA, developed and tested on $\eta$ photoproduction in ref.~\cite{Svarc2020}, is applied to the
$K^{+} \Lambda$ photoproduction for the center-of-mass energy range of 1625 MeV $< W <$ 2296 MeV. A complete set of multipoles
has been created. The advantages of the method have been confirmed, and a comparison with the only existing single-energy partial
wave analysis of  $K^{+} \Lambda$ photoproduction given in refs.~\cite{Anisovich2017,Anisovich2017a} is presented. We confirm the
size and shape of Bonn-Gatchina multipoles, but we do not confirm the unambiguous interpretation of the structure in the $M_{1-}$
multipole as a $N(1880) \frac{1}{2}^{+}$ resonance. The decisive role of the self-consistency of the world database is
emphasized.

\end{abstract}

\maketitle

\newpage

\section{Introduction} \label{sec:Intro}
Single-channel, singe-energy partial wave analysis (SC-SE-PWA) has always been of particular interest primarily for
experimentalists, but also for theorists. For experimentalists it seemed to be the most direct way to convert measured data to
physically interpretable partial waves, and for theorists it seemed to be the most direct way to test the validity of their
theoretical approach. Many attempts have been made to prove the uniqueness of SC-SE-PWA~\cite{Martin} even in the case of a
single elastic channel, and they culminated with research by Ion Sabba Stefanescu, who has formulated necessary conditions for
the uniqueness of SC-SE-PWA in the elastic domain\footnote{The proof of uniqueness requires the multivariate analyticity of the
amplitude as a function of two Mandelstam variables.}~\cite{Stefanescu}. However, as the search for nucleon resonances basically
takes place in the inelastic region, the continuum-ambiguity problem became of utter importance~\cite{Continuum-ambiguity}. This
discussion has started a long time ago, but has never been completely finished. Recently, this problem was reopened by our group,
and culminated with the conclusion that each single-channel PWA in the inelastic region is inherently model dependent, as it
depends on the free energy- and angle-dependent continuum ambiguity phase which leaves all observables invariant, but the
angular-dependent part of the ambiguity mixes partial waves. Hence this makes the quantum numbers of resonances unidentifiable
without additional information~\cite{Svarc2021}. Since the full information on the phase can be obtained only from all possible
inelastic channels, it remains undetermined in single-channel models, and the only way to make a single-channel PWA unique is to
fix this phase to some known value. Here we have to distinguish the following two cases: single-channel energy-dependent PWA
(SC-ED-PWA) and single-channel energy-independent PWA (SC-SE-PWA). In SC-ED-PWA the phase is automatically determined by the
analyticity of the continuous ED model, and in SC-SE-PWA it is absolutely free, so we have to take it over from some theoretical
calculation. Observe that the common denominator of both cases is that the analysis has to be extended to multiple channels, as
fixing the continuum-ambiguity phase is only possible by restoring multi-channel unitarity, and that can only be done in
analyzing all available channels for this reaction. So, in both cases, we have to use coupled-channels models, but even there the
missing phase can only be at least approximately determined as all possible, open two-body channels are never known. The problem
totally collapses when three-body channels are involved, so we have to face and live with the fact that each single-channel PWA
is inherently model dependent. This automatically means that SC-SE-PWA is also inherently model dependent. So, when we compare
different PWAs, we have to match the reaction-amplitude phases first.

Constraining SC-SE-PWA has for decades been done by either fixing some partial waves to values from some theoretical model, or
penalizing some or all partial waves to the particular constraining theoretical model. This is the traditional way. A strong step
forward was done by the Karlsruhe-Helsinki group in the 1980s~\cite{Hoehler84} when the problem was raised to the level of
reaction amplitudes where phase ambiguities appear. Following the work of Ion Stefanescu~\cite{Stefanescu} on the importance of
analyticity in two Mandelstam variables, fixed-$t$ analyticity was introduced in $\pi N$ elastic scattering. Analyticity in $t$
was achieved using the manifestly analytic Pietarinen decomposition of invariant amplitudes and fitting the free parameters of
the decomposition to the data base transformed from [W$_{\rm fixed}$, $\theta$] coordinates into [$t_{\rm fixed}$, W]
coordinates\footnote{Observables are traditionally analyzed in [W$_{\rm fixed}$, $\theta$] space.}, while analyticity in
Mandelstam $s$ is enforced by using the traditional partial wave decomposition. The free continuum-ambiguity phase was
predetermined by the choice of starting values in the $t$-variable minimization. As the reaction-amplitude phase is fairly well
known for the elastic scattering, this method resulted in the KH80- and KH84 solutions for elastic $\pi N$ partial
waves~\cite{Hoehler84}, which have been accepted and used for decades. The same method was recently revived and applied with
great success to pion photoproduction~\cite{Osmanovic2019,Osmanovic2021}. Luckily, in pion photoproduction this phase is for
sufficiently low energies linked to the well-determined phase of elastic $\pi N$ scattering due to Watson's
theorem~\cite{Watson}. Therefore, in refs.~\cite{Osmanovic2019,Osmanovic2021} the authors also do not face the problem of the
unknown phase. However, for all other reactions where Watson's theorem breaks down (like $\eta$ or $K \Lambda$ photoproduction)
this is not true, so the phase stays poorly determined, and SC-SE-PWA stays model dependent. Triggered by the fact that the
continuum-ambiguity phase in SC method in the inelastic domain has to be constrained in particular for reactions where Watson's
theorem breaks down, a new method based on only one, i.e. the Mandelstam $s$ variable has been developed in
ref.~\cite{Svarc2020}. However, in this model we openly acknowledge the problem that the continuum-ambiguity phase is unknown,
and from the very start constrain it to a phase of some chosen theoretical coupled-channels model. The rest is very similar to
the fixed-$t$ analyticity procedure, but restricted to the $s$ channel only. The procedure is a two-step process applied to the
same database: the first step consists of an amplitude analysis of the database where the moduli of the reaction amplitudes are
fitted while the reaction-amplitude phases are fixed to the values of a particular ED coupled-channels model (this eliminates the
continuum ambiguity); the second step is a standard truncated partial wave analysis (TPWA), where the reaction amplitudes are
forced to be close to the reaction amplitudes of the first step using penalization techniques. In this way the continuity in
energy is ensured through the continuous phase, and continuity in angle is ensured by the TPWA. Let us observe that the proposed
method relies on using minimal theory dependence, which is given by fixing the phase only. Unfortunately, this works perfectly
well only for the ideal case when all observables are self-consistent. This has been shown in ref.~\cite{Svarc2018} for a
complete set of pseudo-observables (numeric data) for $\eta$ photoproduction generated by the ETA-MAID model~\cite{ETA-MAID15a}
which are by definition self-consistent. The free phase in unconstrained SC-SE-PWA is replaced by the original ETA-MAID phase,
and the continuous generating multipoles were exactly reproduced. Unfortunately, real data are never self-consistent, so when the
method is applied to real data some discontinuities in partial waves might appear. Therefore, all scatter of the result is the
consequence of experimental errors and data inconsistencies. The method has been developed and tested on a world database of
$\eta$ photoproduction, and presented in ref.~\cite{Svarc2020}. As the results for $\eta$ photoproduction were
stable~\cite{Svarc2020}, and the pole content of the obtained solution looked very reasonable~\cite{Svarc2021}, in this paper we
used this method for the $K \Lambda$ photoproduction reaction where the isospin structure is identical. However, the main
advantage of  $K \Lambda$ photoproduction is that we have access to results from complementary analyses, to which our results can
be compared. First of all, there exists a very confident theoretical coupled-channels ED model by the Bonn-Gatchina group to
start with~\cite{BG-web}. Second, results of SC-SE-PWA made by the same group are published~\cite{Anisovich2017,Anisovich2017a},
so we have direct numbers to compare our results with. The situation is even more favorable. The Bonn-Gatchina SC-SE-PWA was done
in the standard way: only lower partial waves were left free, while all higher partial waves were fixed to the Bonn-Gatchina ED
model, while we offer the simultaneous variation of all multipoles within the framework of AA/PWA method. So, results and
advantages/disadvantages of both approaches can be directly compared. Third, but not the least important is that we had access to
a full $K \Lambda$ photoproduction database in numeric form from the Bonn-Gatchina web page~\cite{BG-web}. So, as all input can
be made identical, the benefits of the new approach could be clearly detected. We discuss similarities and differences, and point
out to the reasons why this is so.

\section{The AA/PWA method} \label{sec:SummaryPWAAA}
In the AA/PWA method, from the very start we openly accept the fact that the overall continuum-ambiguity phase of the reaction
amplitudes in any inelastic SC analysis is by default undetermined as it depends on other channels~\cite{Svarc2021},  and  we fix
it to a phase of some chosen theoretical coupled-channels model. However, let us stress that the way it is done in our paper is
only an approximation. The amplitude phases contain two parts: the first part are relative phases which are determined by single-
and double-polarization observables, and can be uniquely determined in the SC model\footnote{Let us remember that in pseudoscalar
meson photoproduction, the four observables $d\sigma/d\Omega$, $\Sigma$, $T$ and $P$ determine the absolute values of the four
transversity amplitudes(see Table~\ref{tab:PhotoproductionObservables} in appendix~\ref{sec:PhotoproductionFormalism}), and the
remaining four observables in a complete set determine the three remaining relative phases.}, and the second part is the
continuum-ambiguity phase which is unknown in any SC approach~\cite{Svarc2020}.  In principle, we should only fix the unknown
continuum ambiguity phase to a theoretical model. However, as the separation of each reaction-amplitudes phase in relative- and
continuum ambiguity part is unknown, we opted to fully fix the phases of all 4 reaction amplitudes. So, we expect that the model
we use will be fairly good, and fit also the non-measured spin observables. The rest is very similar to procedures implementing
fixed-$t$ analyticity~\cite{Hoehler84,Osmanovic2018,Osmanovic2019,Osmanovic2021}, but restricted to the $s$ channel only. The
procedure is a two-step process applied to the same data base: the first step is an amplitude analysis of the database, where the
moduli of reaction amplitudes are fitted while the reaction-amplitude phases are fixed to the values of a particular ED
coupled-channels model (this eliminates the continuum ambiguity); and the second step is a standard truncated partial wave
analysis (TPWA) where the reaction amplitudes are forced to be close to the reaction amplitudes resulting from the first step,
using a penalization technique. In this way the continuity in energy is ensured through the continuous phase, and continuity in
angle is ensured via the TPWA. All scatter of the result is, hence, the consequence of experimental errors and data
inconsistencies, as in ref.~\cite{Svarc2018} it has been shown that fixing only the phase results in a smooth and unique solution
in the case where a self-consistent database has been generated in the form of pseudo observables from a known model. The AA/PWA
method has been developed and tested on a world database for $\eta$ photoproduction, and in details presented in
ref.~\cite{Svarc2020}.
 \\ \\ \noindent
For the convenience of the reader, we summarize the essence of the method mostly relying on the text in ref.~\cite{Svarc2021}.
  \\ \\ \indent
In ref.~\cite{Svarc2020} we have formulated a single-channel, single-energy partial wave analysis (SC-SE-PWA) procedure of
determining reaction amplitude via fitting scattering data when the number of equations may be less than number of unknown
quantities, which combines amplitude- and partial wave analyses into one logical sequence, and directly from the data generates a
set of continuous partial waves using a minimally model-dependent input~(AA/PWA). We have demonstrated that by controlling the
reaction-amplitude phase, and freely varying the reaction-amplitude partial waves, we obtain a continuous solution with far
better agreement with the used data base than the original energy dependent (ED) model.
\\ \\ \indent
The most standard, classic approach is the one where one penalizes partial waves by requiring that fitted partial waves reproduce
the observable $\cal O$ and are at the same time close to some partial waves taken from a theoretical model:
\begin{eqnarray} \label{Eq1} \chi^2(W) & = & \sum_{i=1}^{N_{data}}w^i \left[ {\cal O}^{exp}_i (W,\Theta_i) - {\cal O}^{th}_i ({\cal
M}^{fit}(W),\Theta_i) \right]^2 + \lambda_{pen} \sum_{j=1}^{N_{mult}}\left| {\cal M}^{fit}_j(W)- {\cal M}^{th}_j(W) \right|^2 \ee
where \be {\cal M} & \stackrel{def}{=} & \left\{ {\cal M}_0, {\cal M}_1, {\cal M}_2, ..., {\cal M}_{N_{mult}} \right\} \nonumber
\end{eqnarray}
$w_i$  is the statistical weight and $N_{mult}$ is the number of partial waves (multipoles),  ${\cal M}^{fit}$  are fitting
parameters and ${\cal M}^{th}$ are continuous functions taken from a particular theoretical model (For a detailed outline of how
the observables~${\cal O}_{i}$ are composed in terms of multipoles for pseudoscalar meson photoproduction, see
appendix~\ref{sec:PhotoproductionFormalism}.). Instead, we use the possibility to make the penalization function independent of a
particular model as first formulated in the Karlsruhe-Helsinki $\pi N$ elastic PWA by G. H\"{o}hler in the
mid-80es~\cite{Hoehler84}. Partial waves which are inherently model dependent are replaced with a penalization function which was
constructed from reaction amplitudes which can be in principle directly linked to experimental data without any model in the
amplitude-reconstruction procedure. So, the equation~\eqref{Eq1} was changed to:
\begin{eqnarray} \label{Eq2}
\chi^2(W) & = & \sum_{i=1}^{N_{data}}w^i \left[ {\cal O}^{exp}_i (W,\Theta_i) - {\cal O}^{th}_i ({\cal M}^{fit}(W),\Theta_i)
\right]^2 + {\cal P}(W)  \\ \nonumber {\cal P}(W)    &=& \lambda_{pen} \sum_{i=1}^{N_{data}} \sum_{k=1}^{N_{amp}}\left| {\cal
A}_k({\cal M}^{fit}(W),\Theta_i)- {\cal A}_k^{pen}(W,\Theta_i) \right|^2\,,
\end{eqnarray}
where ${\cal A}_k$ is the generic name for any of reaction amplitudes (invariant, helicity, transversity $\ldots$). ${\cal
A}_k({\cal M}^{fit}(W),\Theta_i)$ is the reaction-amplitude value generated by the fitted multipoles, and ${\cal
A}_k^{pen}(W,\Theta_i) $ is the penalizing function coming from the amplitude analysis. However, one is now facing two
challenges: to get reaction amplitudes which fit the data, and also to make them continuous. In the Karlsruhe-Helsinki
case~\cite{Hoehler84,Osmanovic2018,Osmanovic2019,Osmanovic2021}, this was accomplished by implementing fixed-$t$ analyticity and
fitting the data base for fixed $t$. So, the first step of the KH fixed-$t$ approach was to create the data base ${\cal
O}(W)|_{t=fixed}$  using the measured base $ {\cal O}(\cos \, \theta)|_{W=fixed}$, and then to fit them with a manifestly
analytic representation of the reaction amplitudes for a fixed $t$. Manifest analyticity was implemented by using the Pietarinen
decomposition of reaction amplitudes.  Then the second step was to perform a penalized PWA defined by Eq.~\eqref{Eq2} in the
fixed-$W$ channel where the penalizing factor ${\cal A}_k({\cal M}^{pen}(W,\Theta_i)) $ was obtained in the first step in a
fixed-$t$ channel. In that way a stabilized SE PWA was performed. This approach was revived recently for SE PWA of $\eta p$ and
$\pi^0 p$ photoproduction and very recently also for pion photoproduction in full isospin by the Mainz-Tuzla-Zagreb
collaboration, and analyzed in details in refs.~\cite{Osmanovic2018,Osmanovic2019,Osmanovic2021}.
\\ \\
We propose an alternative.
\\ \\
We also use Eq.~\eqref{Eq2}, but the penalizing factor ${\cal P}(W)$ is generated by an amplitude analysis performed in the same
fixed-$W$ representation, and not in the fixed-$t$ one. The phase is in our approach openly recognized as undeterminable, and
taken over from the chosen coupled-channels ED model. This simplifies the procedure significantly, and avoids quite some
theoretical assumptions on the behavior in the fixed-$t$ representation.
\\ \\
We also propose a 2-step process as in ref.~\cite{Hoehler84,Osmanovic2018,Svarc2021}:
\begin{itemize}[leftmargin=2.cm]
  \item[\emph{Step 1:}] \hspace*{1.cm} \\
   Amplitude analysis of experimental data in fixed-$W$ system to generate penalizing factor  ${\cal P}(W)$
  \item[\emph{Step 2:}] \hspace*{1.cm} \\
   Penalized PWA using Eq.~\eqref{Eq2} with the penalization factor from
         \emph{Step 1}.
\end{itemize}

And now we are bound to say something about the importance of the reaction-amplitudes phase. The continuum ambiguity forbids to
conclude onto the correct phase in any single-channel analysis because the loss of probability flux to other channels starts
after the first inelastic threshold opens. The only way to solve the continuum ambiguity problem is to reintroduce the unitarity
constraint in the context of a coupled-channels formalism. If we pick the phase in a single-channel analysis arbitrarily by hand,
we are departing from the genuine 'true' phase, the phase in which partial waves do not mix, and we introduce a pole-shift from
one partial wave into another via the angular dependent part of the continuum ambiguity (see refs.~\cite{Svarc2018,Svarc2018a}).
However, each coupled-channels model by construction results in the non-pole-mixing solution. Namely, some form of interaction
introducing poles is formulated, and the background contribution is included. Then, the data in all channels are simultaneously
fitted forcing the phase to be the correct one, and the non-pole-mixing situation is established. Background contributions
automatically enforce the phase to be a non-mixing one. It is needless to say that all coupled-channels models should end up with
the same phase in the ideal case, but incompleteness of the data forbids that to happen. Therefore, phases of different models
\cite{BoGa,Juelich,MAID,GWU/SAID} are somewhat different, and we cannot avoid this. However, fixing the phase to the phase of a
particular model ensures to obtain the non-mixing pole solution; departure from it automatically enforce pole mixing, so the
analytic structure of such a solution is spoiled. So, we can chose a different phase, a phase coming from any model, but it has
to be the proper phase originating from that model. A free, uncontrolled departure from ED model phase is not allowed.
\\ \\ \indent
Of course the question is purely quantitative: How much can we depart from the 'true' phase in an uncontrolled way to maintain
the correct analytic properties. In other words, the question is how much we are allowed to reduce the importance of the penalty
function and maintain the correct analyticity.
\\ \\ \indent
All relevant formulae and more details on the photoproduction formalism are given in appendix~\ref{sec:PhotoproductionFormalism}
of this paper.

\section{Application of AA/PWA to $K^{+} \Lambda$ photoproduction data} \label{sec:KLambdaDataAnalysis}
As the AA/PWA method worked so well on $\eta$ photoproduction, we have decided to test it on the next natural candidate reaction,
and that is $K^+ \Lambda$ photoproduction. This reaction has the same isospin structure as $\eta$ photoproduction, it has a rich
database, so the technical effort involved in adapting the AA/PWA scheme was minimal. However, there is one big advantage: we
have results from other complementary analyses to compare with. Namely, four years ago the Bonn-Gatchina group made a classic
SC-SE-PWA analysis of $K^+ \Lambda$ photoproduction. In ref.~\cite{Anisovich2017}, the first four multipoles ($E_{0^+}$,
$M_{1^-}$, $E_{1^+}$, and $M_{1^+}$) were let free, while all higher multipoles were forced by a penalty function to stay close
to the Bonn-Gatchina ED theoretical coupled-channels model, while in the forthcoming reference~\cite{Anisovich2017a} the next
three multipoles ($E_{2^-}$, $M_{2^-}$, and $M_{2^+}$) were released in addition. In this paper we focus on comparing our results
with the results of ref.~\cite{Anisovich2017}.
\\ \\ \indent
However, to do so we have to use the identical data base, and identical Bonn-Gatchina (BG) ED model constraining partial waves.
This turned out not to be a problem, as the data base is in numerical form given on two very nice web
pages~\cite{BG-web,GWU-web}. Unfortunately, choosing BG ED multipoles turned out to be much more difficult. For some reason, the
particular BG ED solution used for both BG publications~\cite{Anisovich2017,Anisovich2017a} is not given on the BG web page, so
we obtained these numbers via private communication~\cite{Anisovich-priv-com}. As the reader will see later, this turned out to
be extremely important as this solution was especially tuned to fit $K^+ \Lambda$ data, and the absolute normalization of all
multipoles is somewhat different. This is trivially visible for the dominant $E_{0^+}$ multipole, where the value of used BG ED
solution was notably larger than either of the solutions BG2014-2 or BG2019 given on their web page.

\subsection{Description of the database} \label{sec:DescriptionDatabase}
In Table~\ref{tab:expdata} we give our data base which is in numeric form taken over from the Bonn-Gatchina and
George-Washington-University web pages~\cite{BG-web,GWU-web}:
\begin{table*}[htb]
\begin{center}
\caption{\label{tab:expdata} Experimental data from CLAS, and GRAAL used in our PWA. Note that the observables $C_x$ and $C_z$
are measured in a rotated coordinate frame~\cite{Bradford}. They are related to the standard observables $C_{x'}$ and $C_{z'}$ in
the $c.m.$ frame by an angular rotation: $C_x= C_{z'} \sin(\theta)+ C_{x'} \cos(\theta)$, and $C_z= C_{z'} \cos(\theta)- C_{x'}
\sin(\theta)$, see ref.~\cite{Anisovich2017a}. }
\bigskip
\begin{ruledtabular}
\begin{tabular}{ccccccc}
 Obs.        & $N$ & $E_{c.m.}$~[MeV] & $N_E$  & $\theta_{cm}$~[deg] & $N_\theta$ & Reference    \\
\hline
 $d\sigma/d\Omega \equiv \sigma_0$ & $3615$ & $1625-2295$ & $268$  & $28 - 152$ & $5-19$ & CLAS(2007)~\cite{Bradford}, CLAS(2010)~\cite{McCracken} \\
 $\Sigma$   & $ 400$ & $1649 - 2179$ & $ 34$  & $35 - 143$ & $6-16$ & GRAAL(2007)~\cite{Lleres}, CLAS(2016)~\cite{Paterson} \\
 $T$        & $ 408$ & $1645 - 2179$ & $ 34$  & $31 - 142$ & $6-16$ &  GRAAL(2007)~\cite{Lleres},CLAS(2016)~\cite{Paterson}  \\
 $P$        & $ 1597$ & $1625 - 2295$ & $ 78$  & $28 - 143$ & $6-18$ & CLAS(2010)~\cite{McCracken},  GRAAL(2007)~\cite{Lleres} \\
 $O_{x'}$   & $ 415$   & $1645 - 2179$ & $ 34 $  & $31 - 143$ & $6-16$ & GRAAL(2007)~\cite{Lleres}, CLAS(2016)~\cite{Paterson} \\
 $O_{z'}$   & $ 415$  & $1645 - 2179 $ & $ 34 $  & $31 - 143$ & $6-16$ &  GRAAL(2007)~\cite{Lleres}, CLAS(2016)~\cite{Paterson} \\
 $C_x$     & $ 138$  & $1678 - 2296$ & $ 14 $  & $31 - 139$ & $9 $ &  CLAS(2007)~\cite{Bradford} \\
 $C_z$      & $ 138 $ & $1678 - 2296 $ & $ 14 $  & $31 - 139$ & $9$ &  CLAS(2007)~\cite{Bradford}
\end{tabular}
\end{ruledtabular}
\end{center}
\end{table*}
\clearpage As we see from the Table~\ref{tab:expdata}, we have a situation at hand which is very similar to $\eta$
photoproduction:
\begin{enumerate}
\item We have eight measured observables at our disposal, and unfortunately, identically as in $\eta$ photoproduction,
this is still not a complete set of observables (some observables from either the beam-target or the target-recoil categories are
missing). For details, see Appendix~{\ref{sec:SolutionTheory}.\ref{sec:LowerEnergiesSolutionTheory}}.
\item We have a strong dominance of $d\sigma/d\Omega$ data over all other observables.
\item Only four observables out of eight are given in the full, analyzed energy range of 1625 MeV $< W_{c.m.} <$ 2296 MeV,
and these are $d\sigma/d\Omega$, $P$, $C_x$ and $C_z$. The remaining four observables  $\Sigma$, $T$,  $O_{x'}$, and $O_{z'}$ are
measured only up to \mbox{$\approx 2180$ MeV}. This might create unwanted discontinuities at this energy\footnote{Eight
observables at lower energies might create slightly different multipoles than only four at energies above  $\approx 2180$ MeV.
So, the transition may not be smooth for all multipoles at this energy.}. For details see
Appendix~{\ref{sec:SolutionTheory}.\ref{sec:HigherEnergiesSolutionTheory}}.
\end{enumerate}

Even a superficial glimpse at Table~\ref{tab:expdata} tells us that the measured data are given at different energies and
different angles, so some data re-binning is in order. Standardly, data binning consists of using the data not at the exact
energy where they were taken, but in the energy interval $ W_{exact}-\Delta/2 < W_{exact} <  W_{exact} + \Delta/2$ where $\Delta$
is the energy bin-width. However, in the case of scarce data this procedure might introduce an unwanted dissipation of data,
resulting in possible discontinuities between energy bins. Therefore, we have adopted and used an altogether different method.
Instead, we rely on a two-dimensional (2D) data interpolation. We simultaneously interpolate experimental data and their
corresponding experimental error in energy $W$ and angle $\Theta$ using a standard Mathematica routine~\cite{Mathematica}, and
use interpolated values instead of binned ones.
\\  \noindent
Our interpolation strategy is as follows:
 \begin{itemize}
      \item Energy grid: \\
             The analysis is performed on a collection of energies where at least one polarization observable apart from the
             cross section~$\sigma_{0}$ was measured (141 energy points).
      \item Angular grid: \\
       The analysis is done on the following pre-chosen fixed values of 16 points: \\
       \mbox{cos($\Theta$) = \{-0.7, -0.6, -0.5, -0.4, -0.3, -0.2, -0.1, 0., 0.1, 0.2, 0.3, 0.4, 0.5, 0.6, 0.7, 0.80 \} }.
 \end{itemize}

\begin{figure}[h!]
\bc
\includegraphics[width=0.4\textwidth]{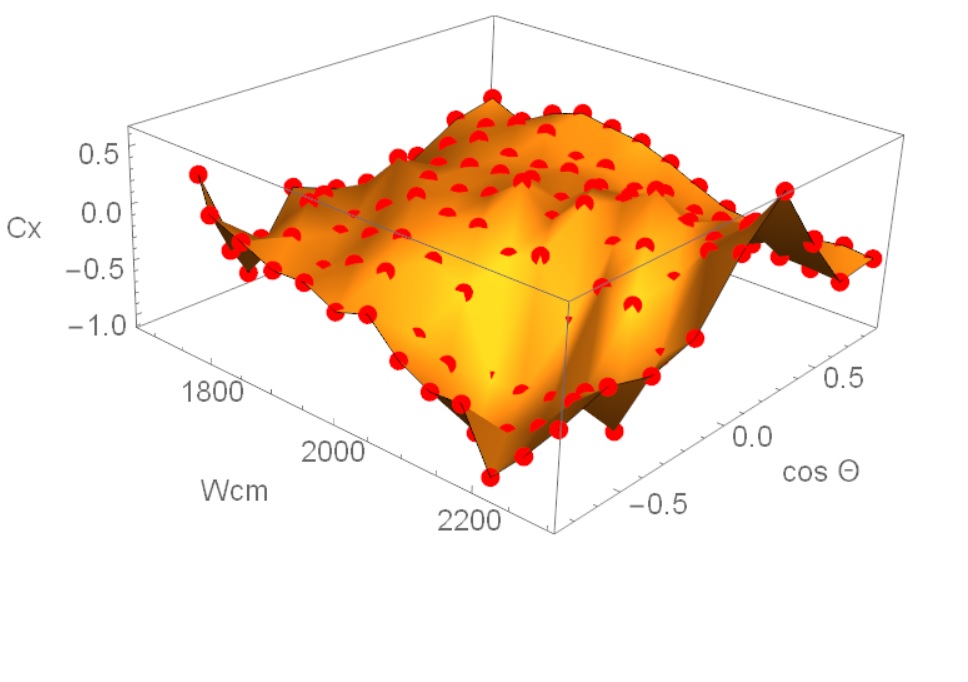} \hspace{0.5cm}
\includegraphics[width=0.4\textwidth]{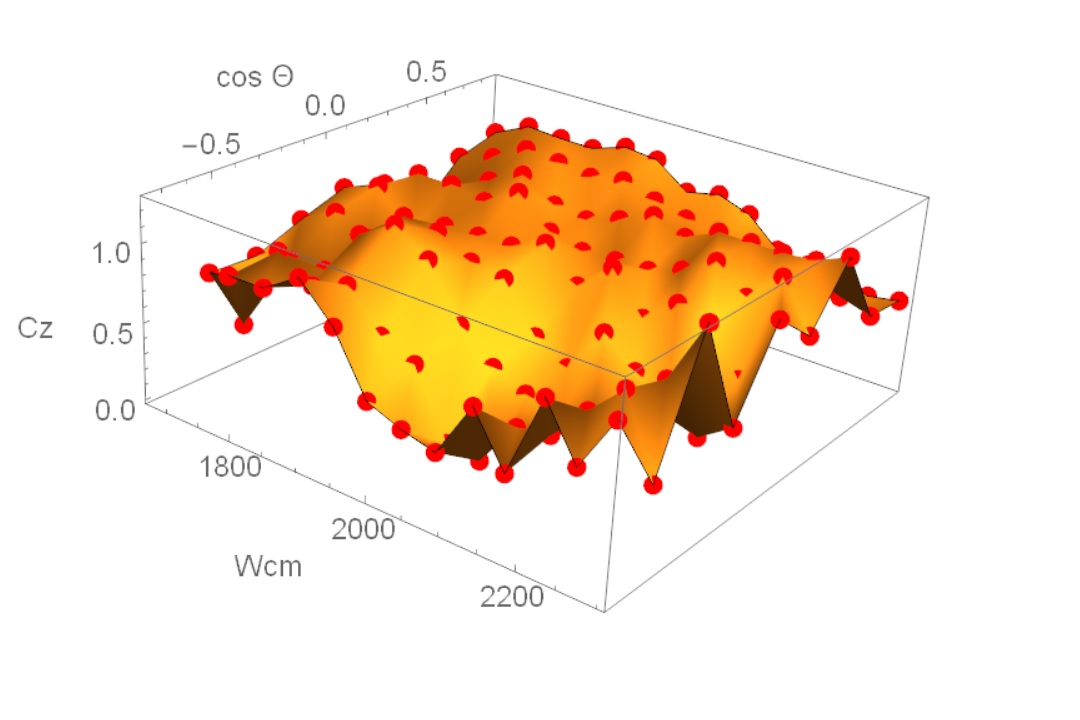}
\caption{\label{CxCz} Quality of the 2D interpolation for $C_x$ and $C_z$ observables.
The red symbols are measured values, and the orange surface shows the interpolated values. (Color online)}
\ec
\end{figure}

So, let us summarize. The AA/PWA method is not done on the realistic, measured energy and angular values, but at interpolated
values of all observables instead. Instead, some interpolation has to be done since we are doing a single-energy analysis, i.e.
at least energies between analyzed observables have to match (which they don't do for the measured data sets). This has some
advantages and possibly some drawbacks. The main advantage is that the number of analyzed points is increased. In the standard
binning method, the number of analyzed energy and angular points is directly limited by the number of measured points for the
least known observable. So, when we perform the analysis with the energy-binning technique, we can make an analysis using a
maximum number of observables only on a small number of points, on points where the least known observable is measured. We can
never benefit from the vast amount of energy- and angular points where all other energies are measured. On the rest of energies
and angles the number of used observables is smaller, and the uncertainty introduced into the analysis hence grows. However, when
we use interpolation technique, we have all measured observables at all analyzing points as interpolated values, and the
confidence into our analysis depends on the quality of the interpolation. So, it is of utter importance to have a confident
interpolation for the "worst" observable where the separation between measured points is the farthest. In our case the "worst"
observables are $C_x$ and $C_z$ (see Table~\ref{tab:expdata}), and in Fig.~\ref{CxCz} we show the quality of the corresponding
interpolations. We are of the opinion that the interpolation of these two observables is fairly good.

Of course, the natural drawback is that we introduce additional unmeasured points, so the contribution of poorly measured
observables to the overall $\chi^2$ grows. However, this effect of overstressing the statistical importance of poorly measured
observables is present in the binning technique, but is introduced differently (by increasing the weighting factor for these
observables). So, the idea is similar, but for the interpolation method implemented in a more confident way.

Our fitting strategy reads as follows:
\begin{itemize}
\item The AA step is done on interpolated energies (141 points) and interpolated angles (16 points).
\item The TPWA minimization step for obtaining multipoles is also done on interpolated energies and pre-chosen angular values.
However, the statistical data analysis to obtain $\chi^2$ is done at exact energies and exact number of angles (energy dependent)
for each observable (cf. Table~\ref{tab:expdata}).
\end{itemize}

\subsection{Choosing the energy-dependent constraining model} \label{sec:EDmodel}
\begin{figure}[h!]
\bc
\includegraphics[width=0.08\textwidth]{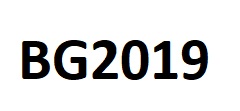} \\
\includegraphics[width=0.23\textwidth]{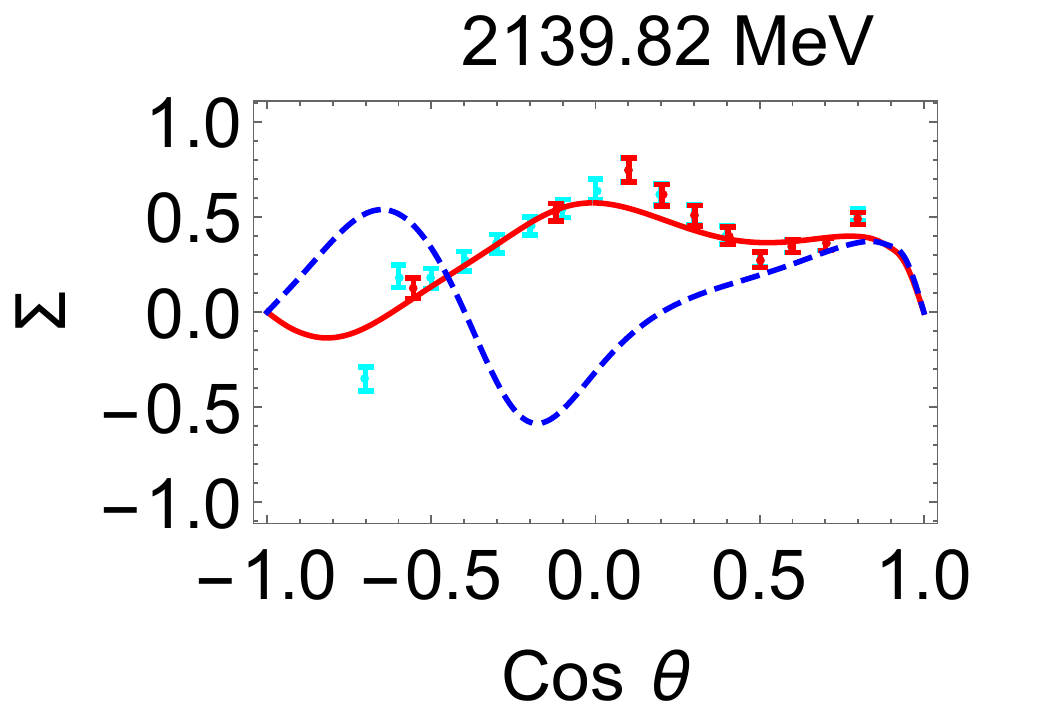} \hspace{0.2cm}
\includegraphics[width=0.23\textwidth]{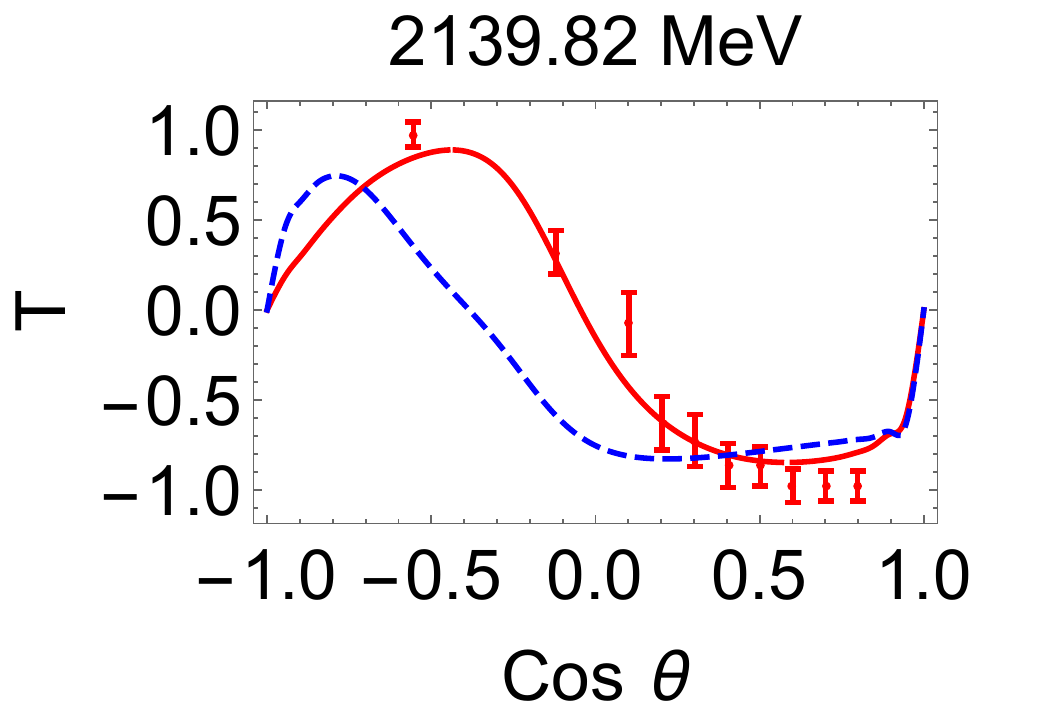} \hspace{0.2cm}
\includegraphics[width=0.23\textwidth]{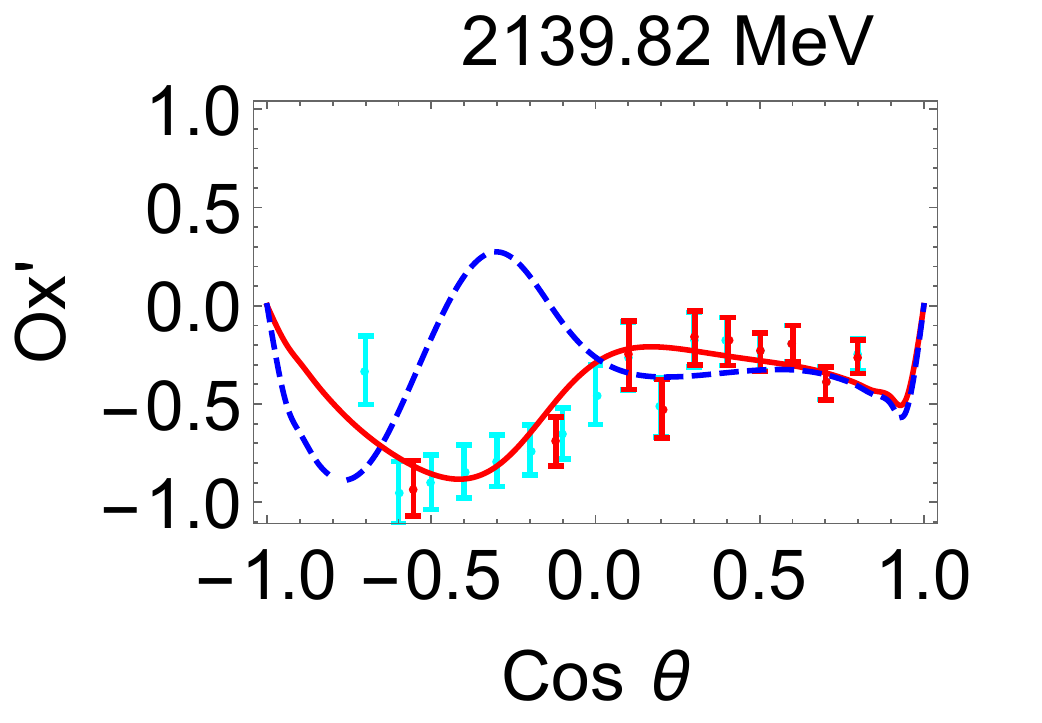} \hspace{0.2cm}
\includegraphics[width=0.23\textwidth]{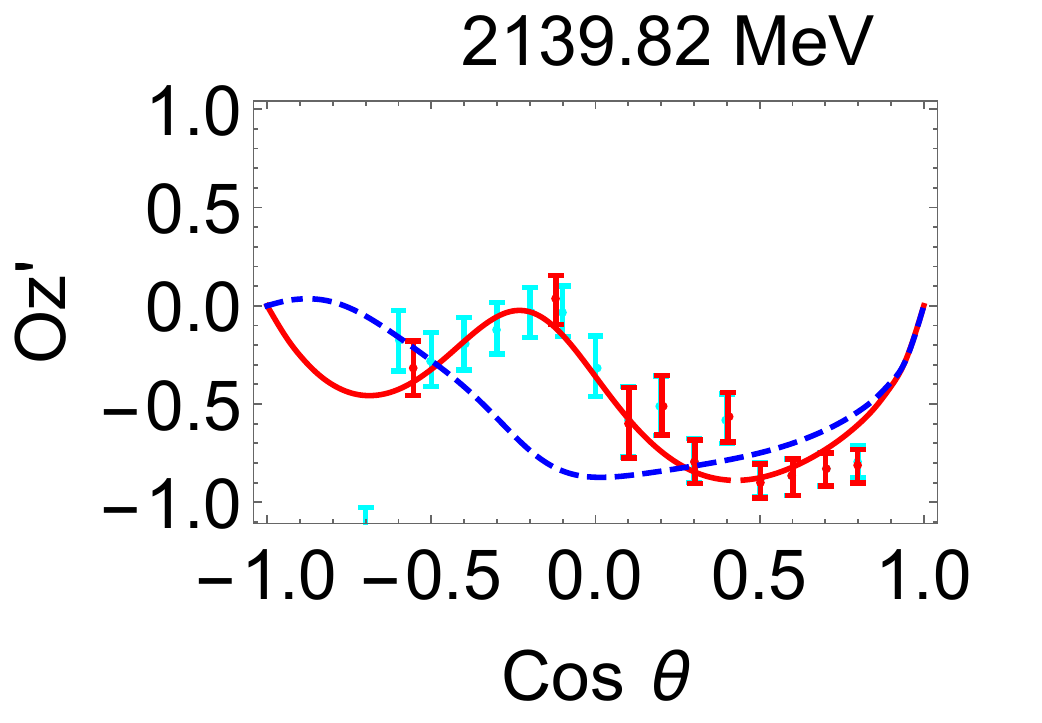} \\
\includegraphics[width=0.08\textwidth]{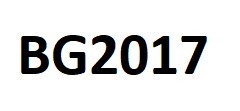} \\
\includegraphics[width=0.23\textwidth]{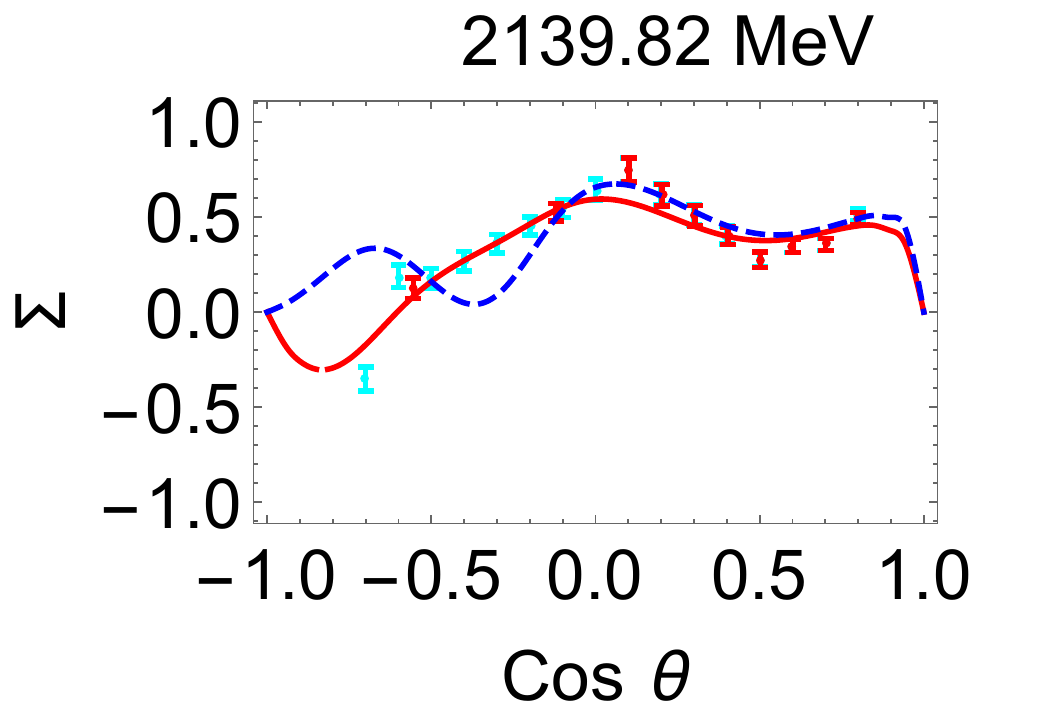} \hspace{0.2cm}
\includegraphics[width=0.23\textwidth]{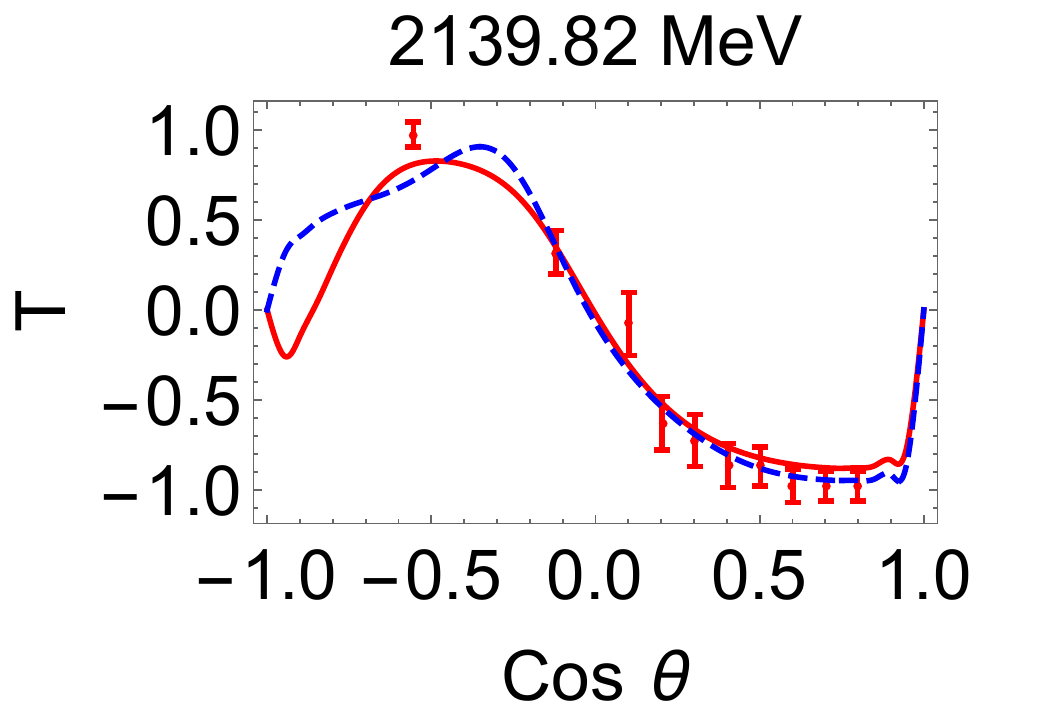} \hspace{0.2cm}
\includegraphics[width=0.23\textwidth]{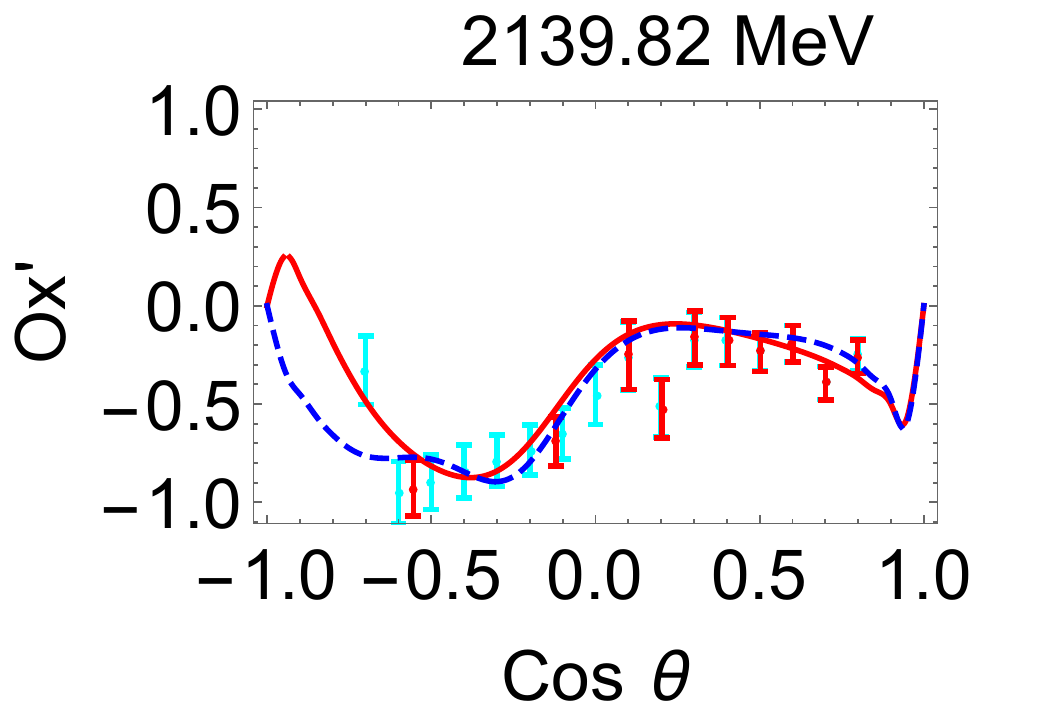} \hspace{0.2cm}
\includegraphics[width=0.23\textwidth]{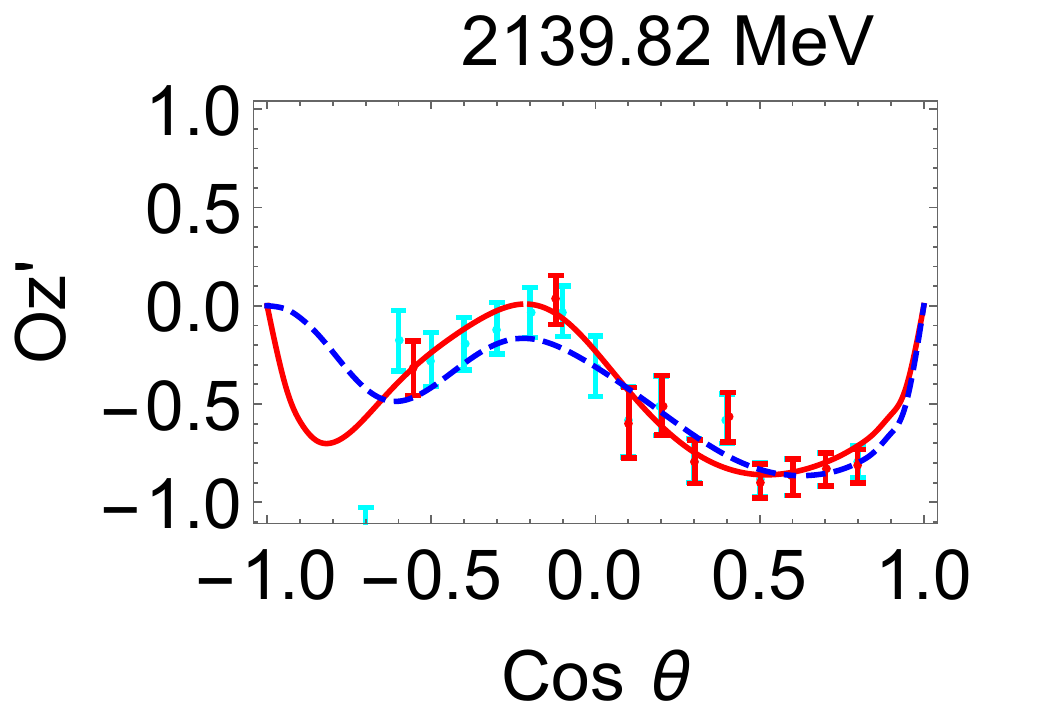} \\
\caption{\label{BG2019spinobs} The quality of the BG2017 and BG2019 solutions for the polarization observables $\Sigma$, $T$,
$O_{x'}$ and $O_{z'}$ is illustrated at one randomly chosen energy. Red symbols are measured data, cyan symbols are interpolated
data, the dashed blue line is the result of the BG2019 and BG2017 model respectively, and the red line is the result of our fit.}
\ec
\end{figure}

Our first intention was to take the solution BG2019 from the Bonn-Gatchina web page~\cite{BG-web} as a constraining solution. To
our surprise, this solution fits the polarization observables $\Sigma$, $T$, $O_{x'}$ and $O_{z'}$ rather poorly at higher
energies. We show the discrepancy of that solution with the above-mentioned four observables at one randomly chosen higher energy
in Fig.~\ref{BG2019spinobs}. However, when the SE-PWA has been performed in refs.~\cite{Anisovich2017,Anisovich2017a}, another
Bonn-Gatchina ED model was used to constrain the higher partial waves, and this solution was different from BG2019. We call this
solution the BG2017 model. As one of the goals of the present paper is to compare the results of our AA/PWA method with results
from Bonn-Gatchina publications, it is natural to take the same constraining input, but to be used on the level of
reaction-amplitude phases. As seen in refs.~\cite{Anisovich2017,Anisovich2017a} the agreement of the BG2017 model with
polarization observables is very good, and this is very important for choosing the phase as we fix relative phases in addition to
the continuum-ambiguity phase. Unfortunately, we realized that this solution is never given anywhere in numbers explicitly, and
we got it only via private communication~\cite{Anisovich-priv-com}. As our model requires fairly good agreement with all data,
the natural model of choice for the constraining phase had to be the officially unpublished BG2017 model, in spite of the fact
that it is older, and actually never published.

\vspace*{-1.cm}
\subsection{Results} \label{sec:Results}
\begin{figure}[h!]
\bc
\includegraphics[width=0.37\textwidth]{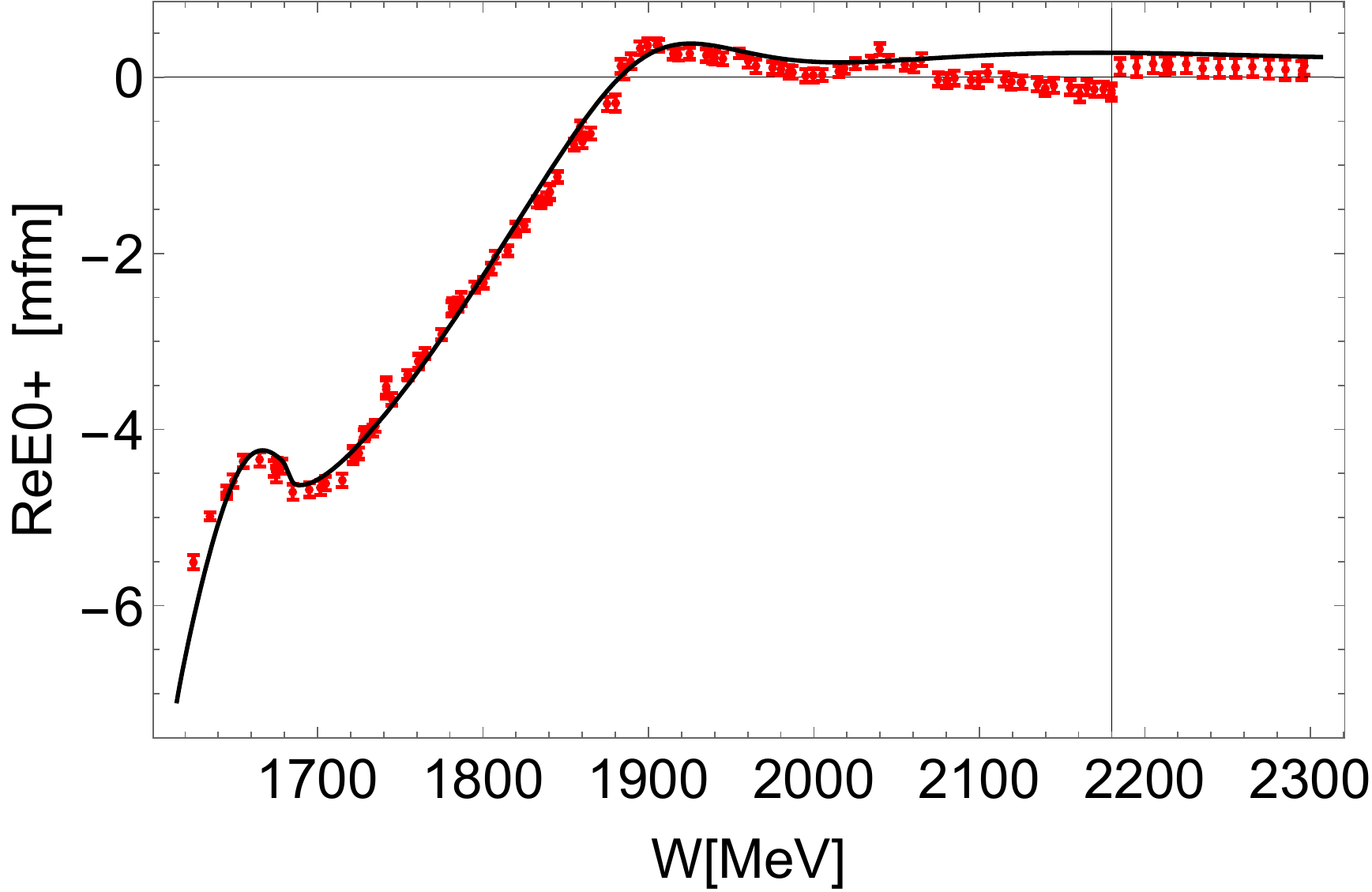} \hspace{0.5cm}
\includegraphics[width=0.37\textwidth]{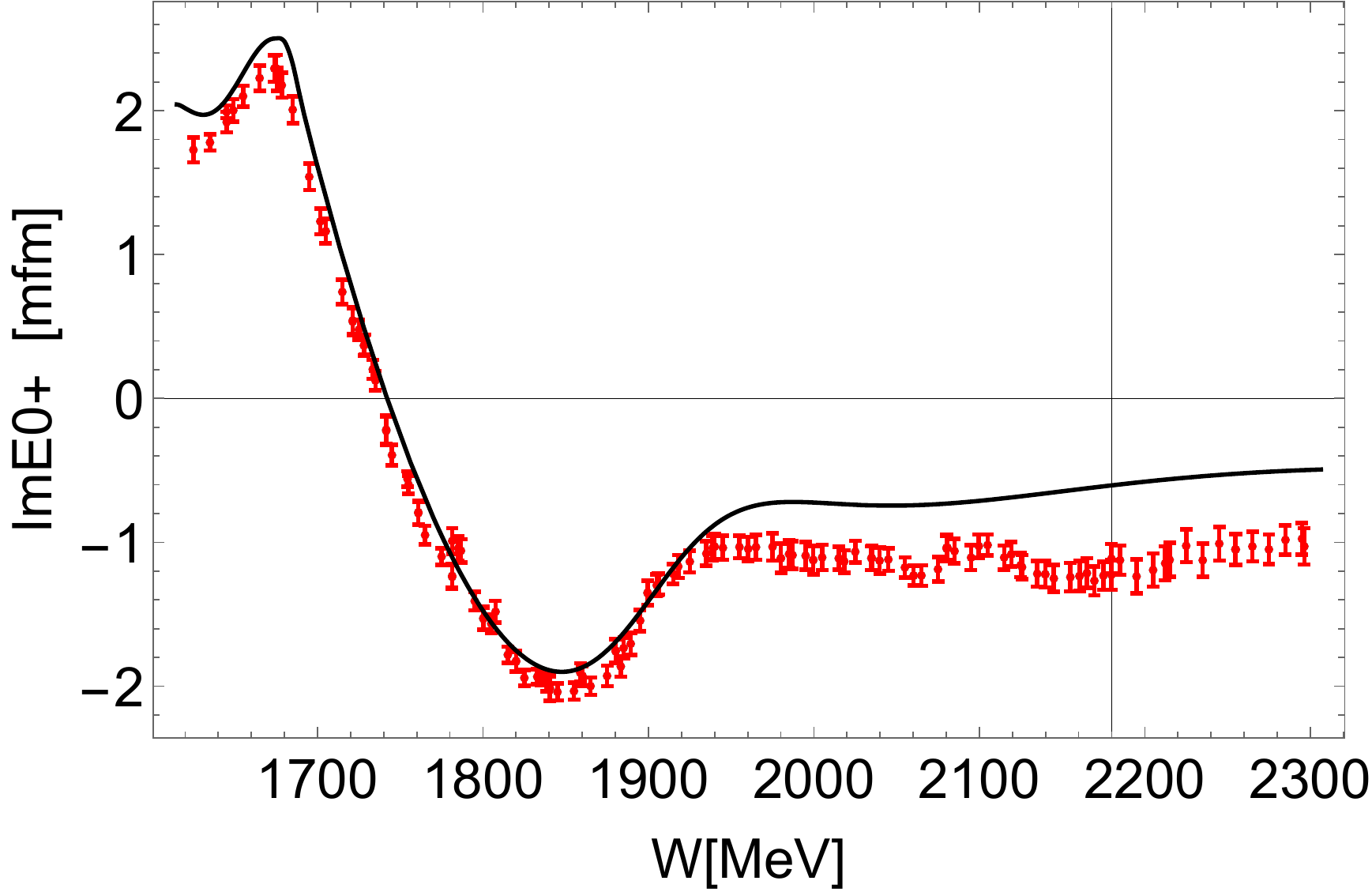}  \\
\includegraphics[width=0.37\textwidth]{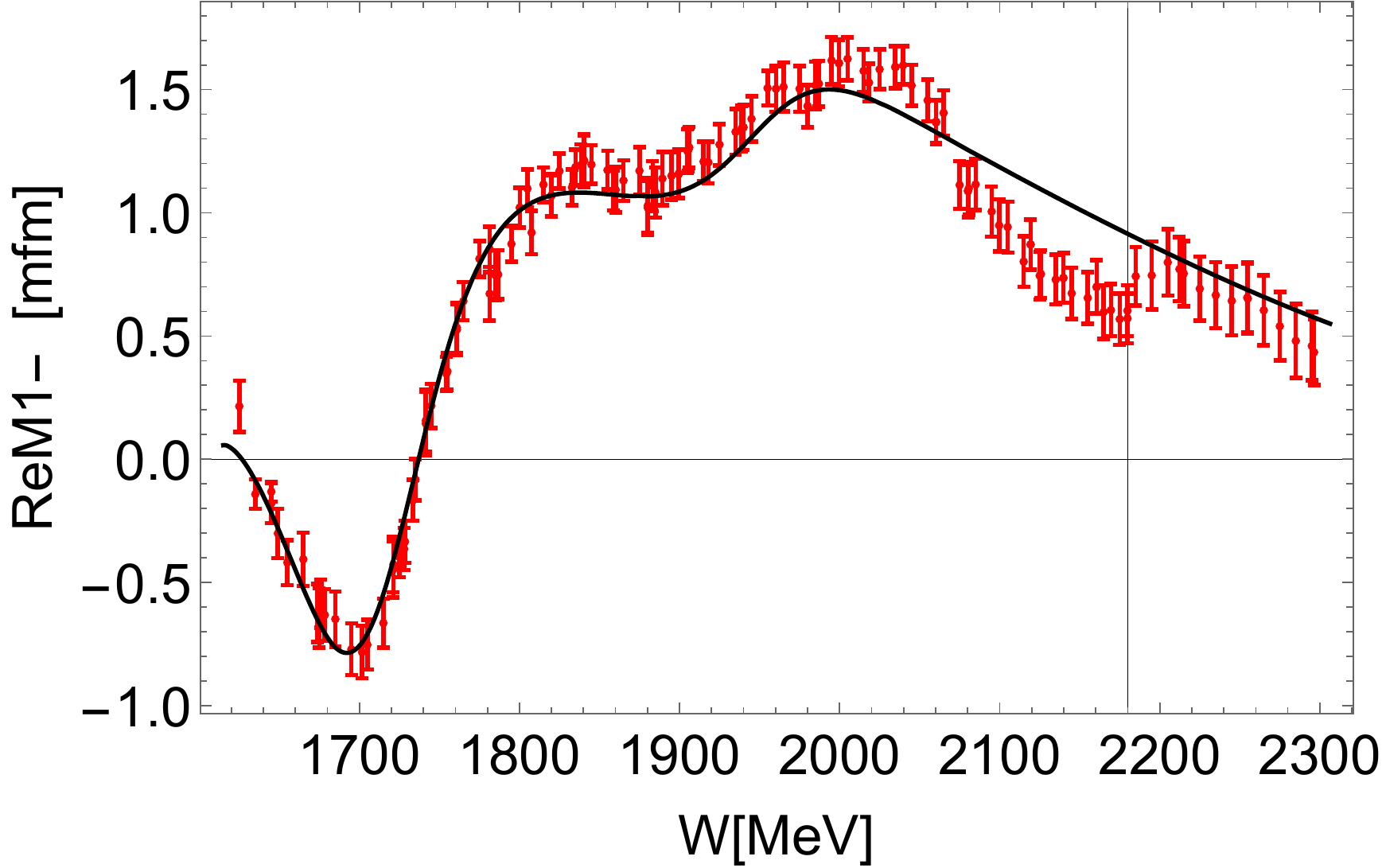} \hspace{0.5cm}
\includegraphics[width=0.37\textwidth]{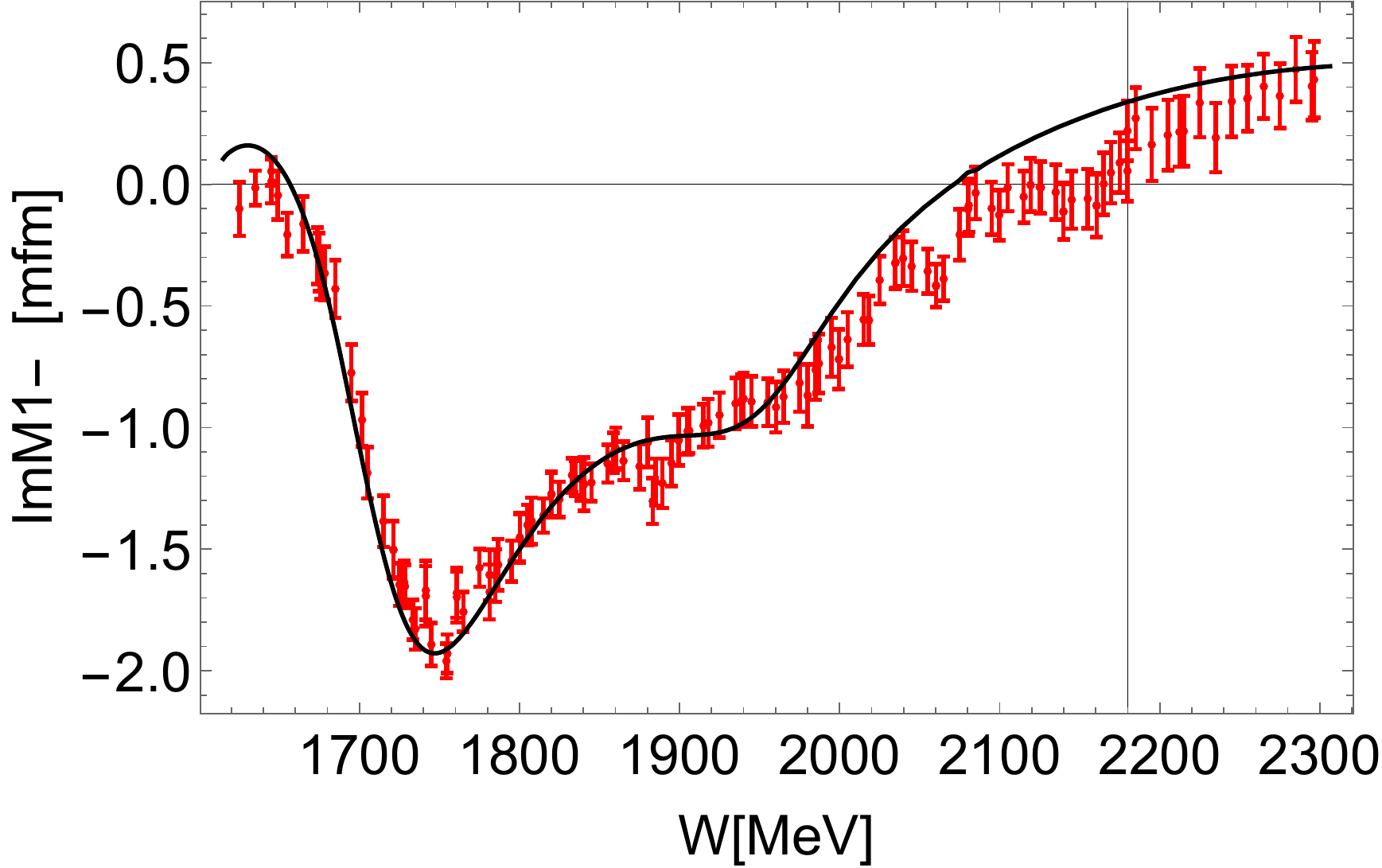}  \\
\includegraphics[width=0.37\textwidth]{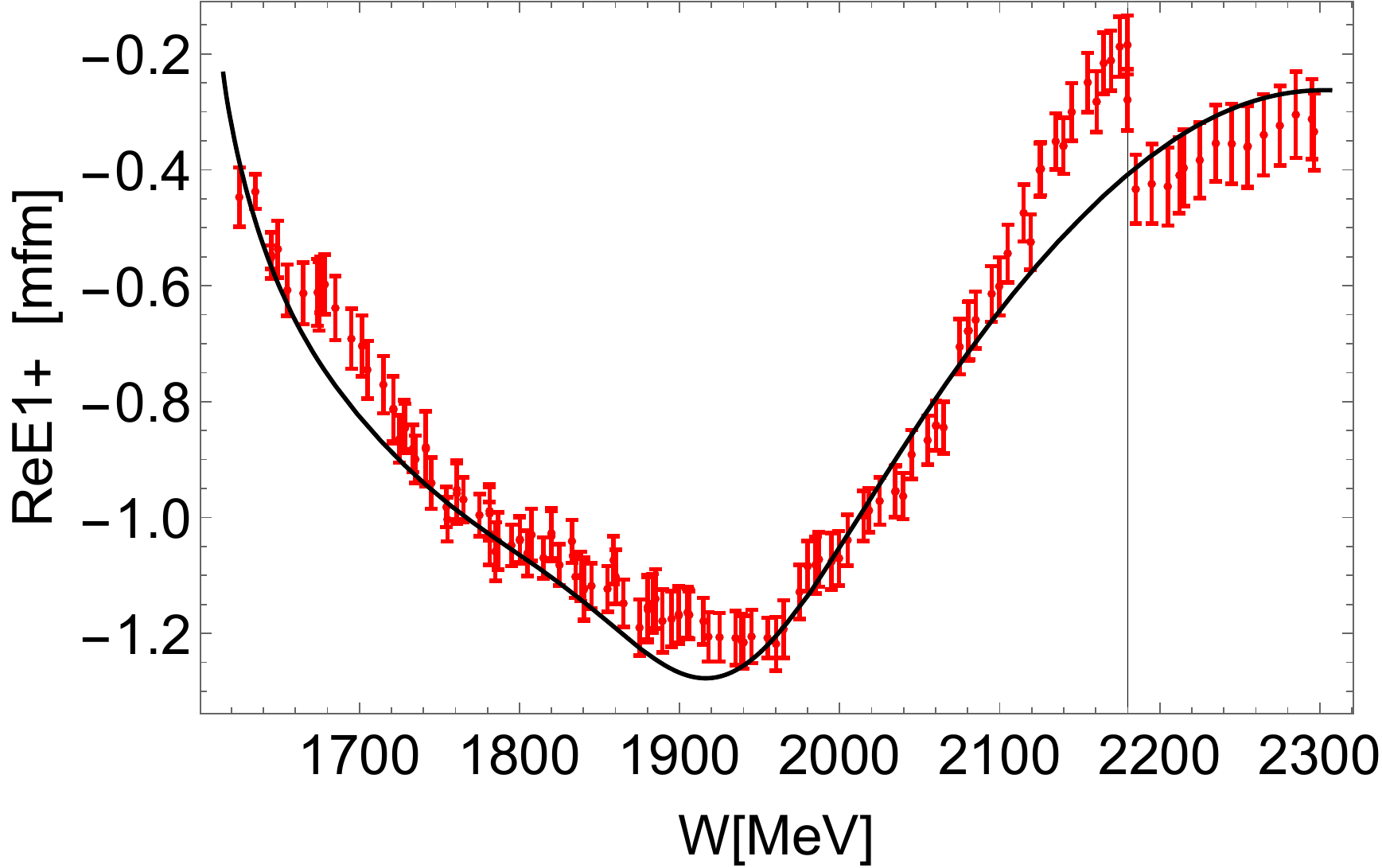} \hspace{0.5cm}
\includegraphics[width=0.37\textwidth]{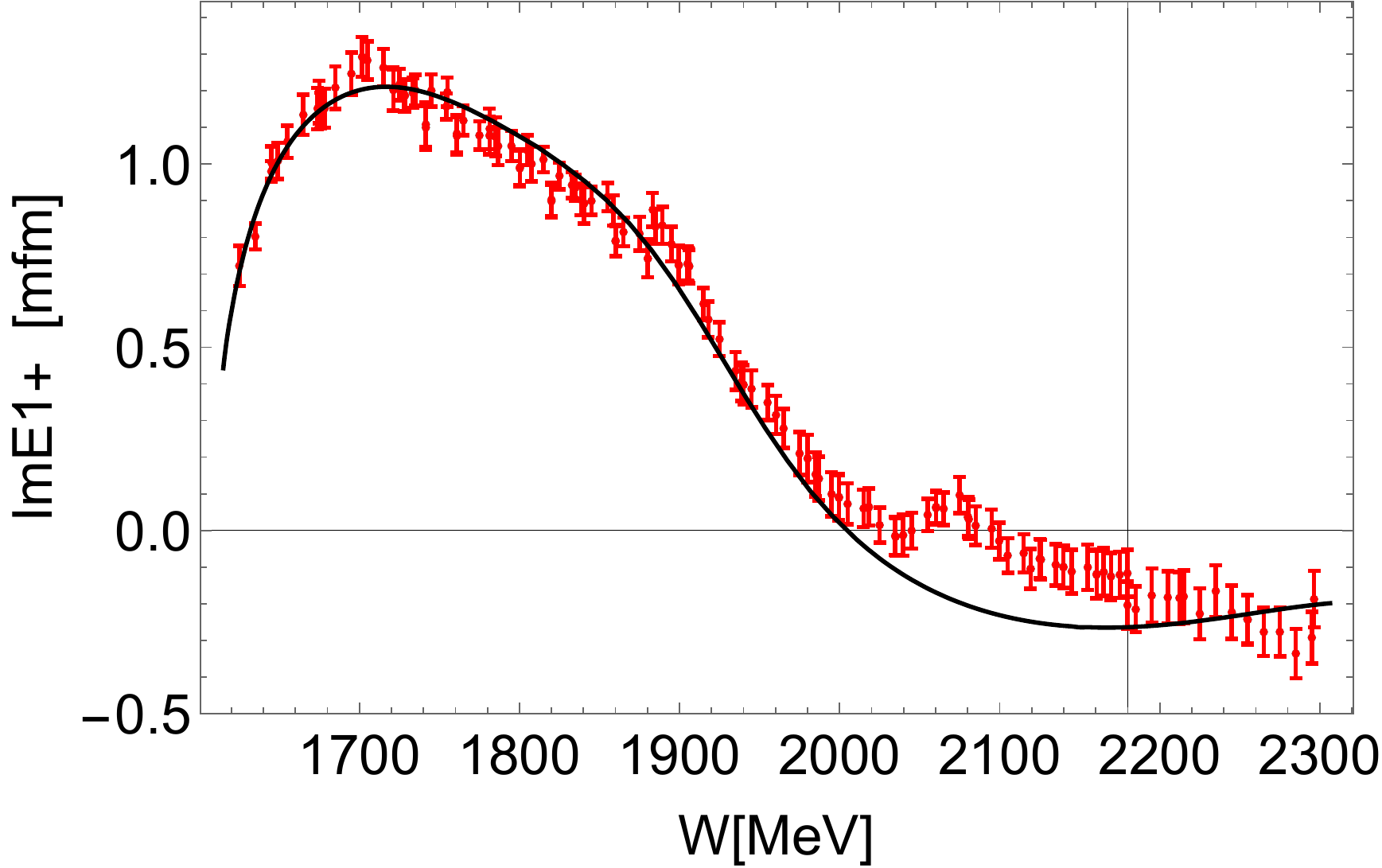}  \\
\includegraphics[width=0.37\textwidth]{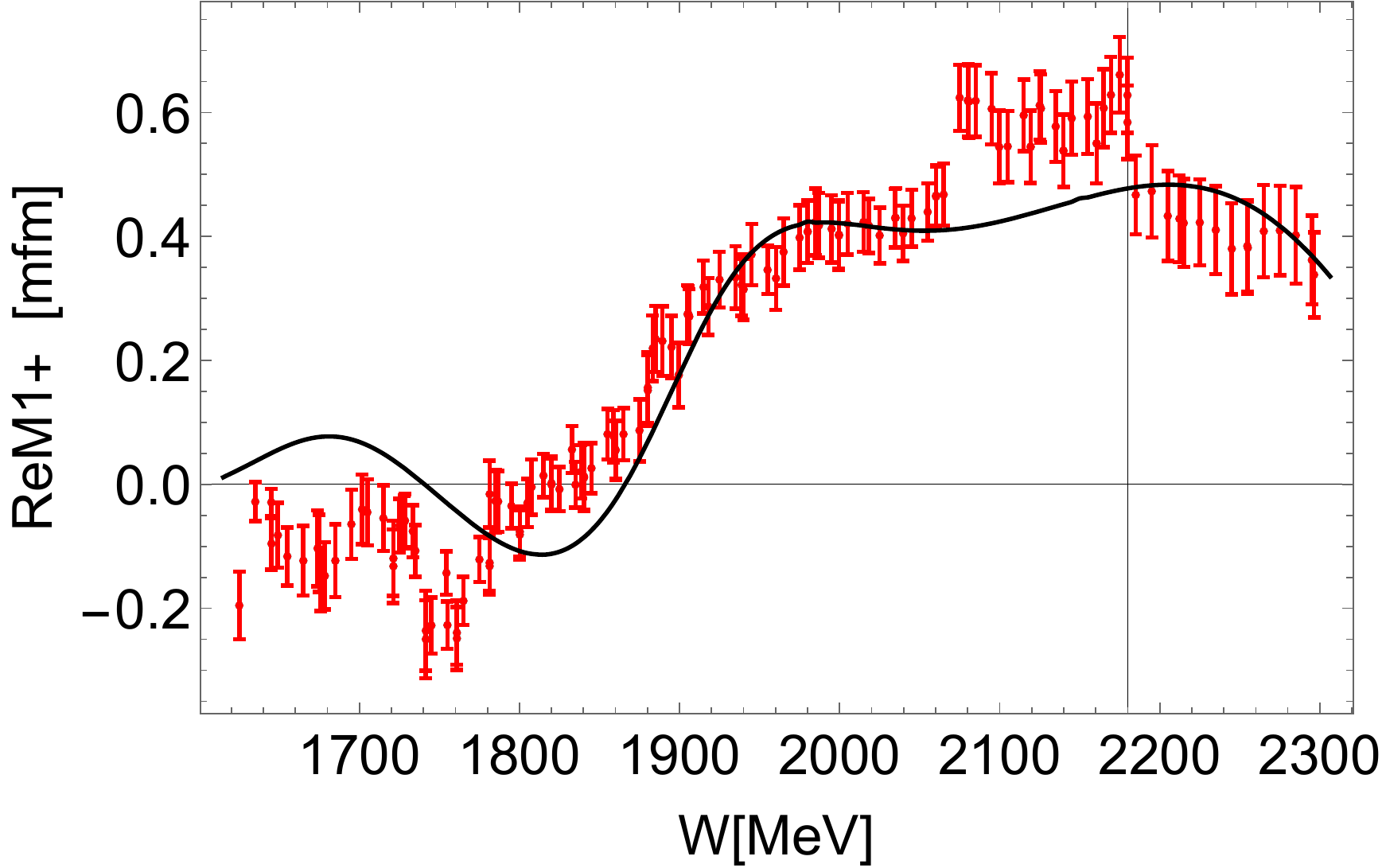} \hspace{0.5cm}
\includegraphics[width=0.37\textwidth]{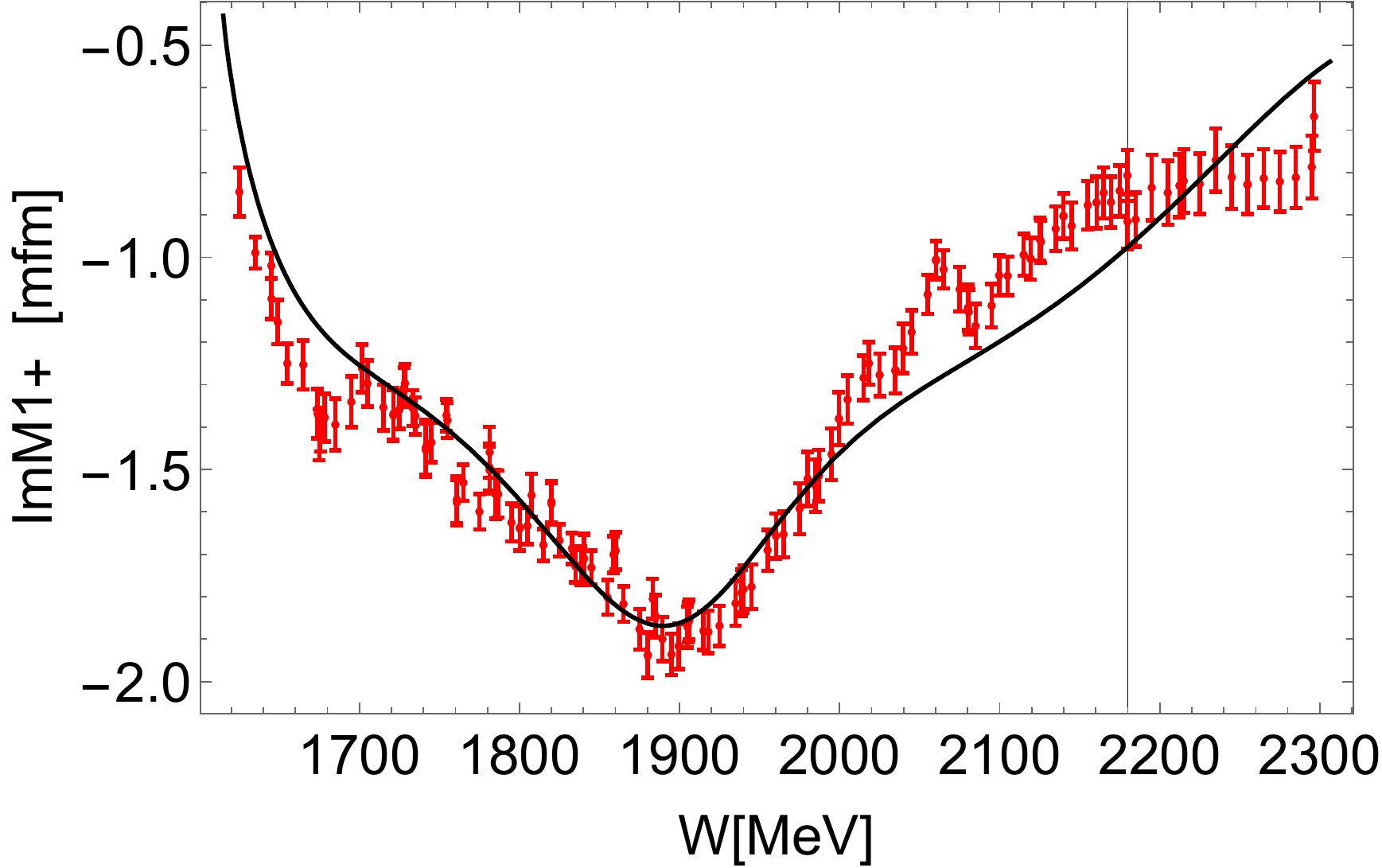}  \\
\includegraphics[width=0.37\textwidth]{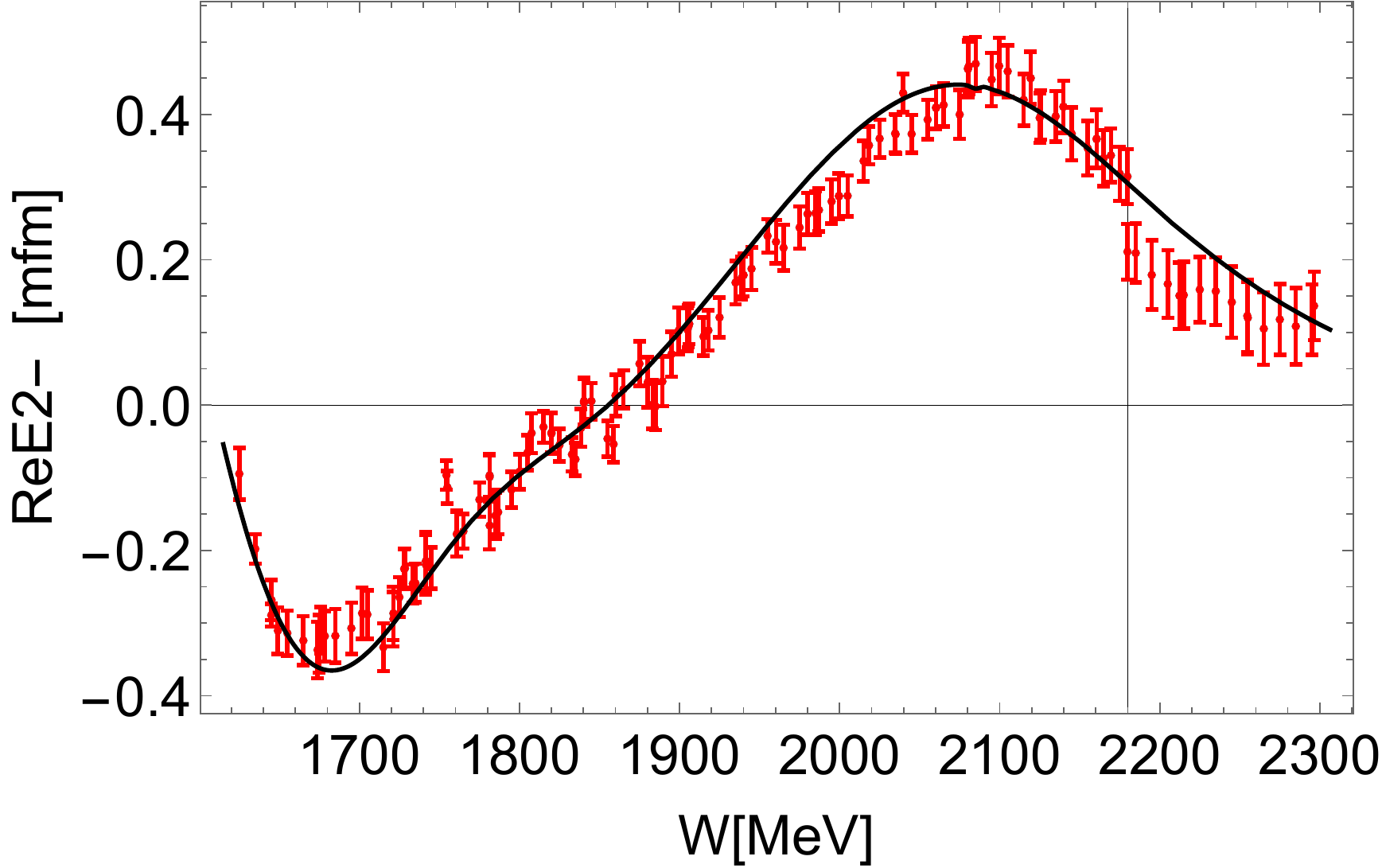} \hspace{0.5cm}
\includegraphics[width=0.37\textwidth]{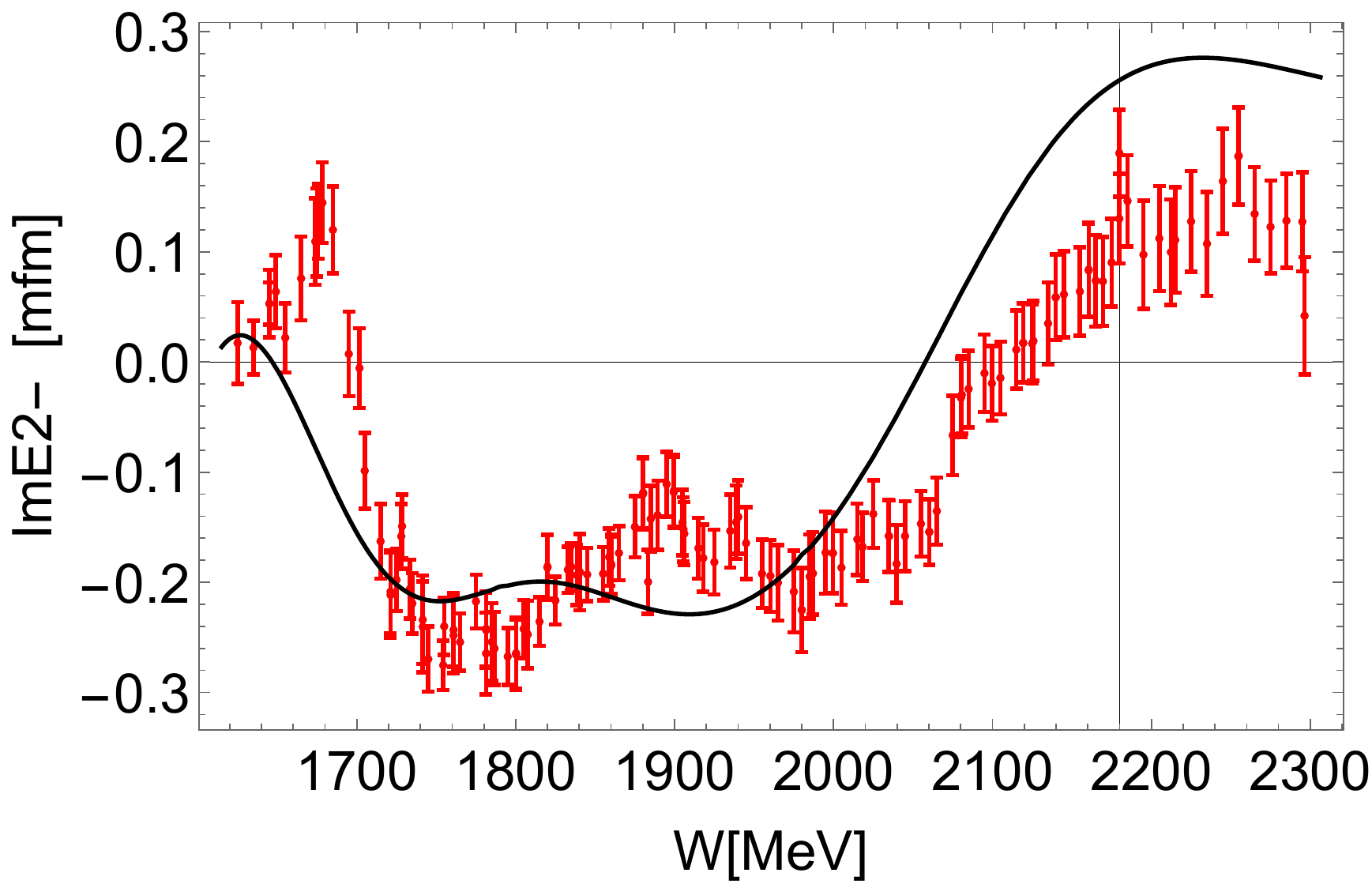}  \\
\caption{\label{Multipoles:a}(Color online) The multipoles for the $L=0$, $1$ and $2$ partial waves of our AA/PWA solution are
shown. Red discrete symbols correspond to our solution, and the black full line gives the BG2017 ED solution for comparison. The
thin vertical black line marks the energy where only 4 observables are measured instead of 8 (cf. Table~\ref{tab:expdata}). } \ec
\end{figure}

\begin{figure}[h!]
\bc
\includegraphics[width=0.37\textwidth]{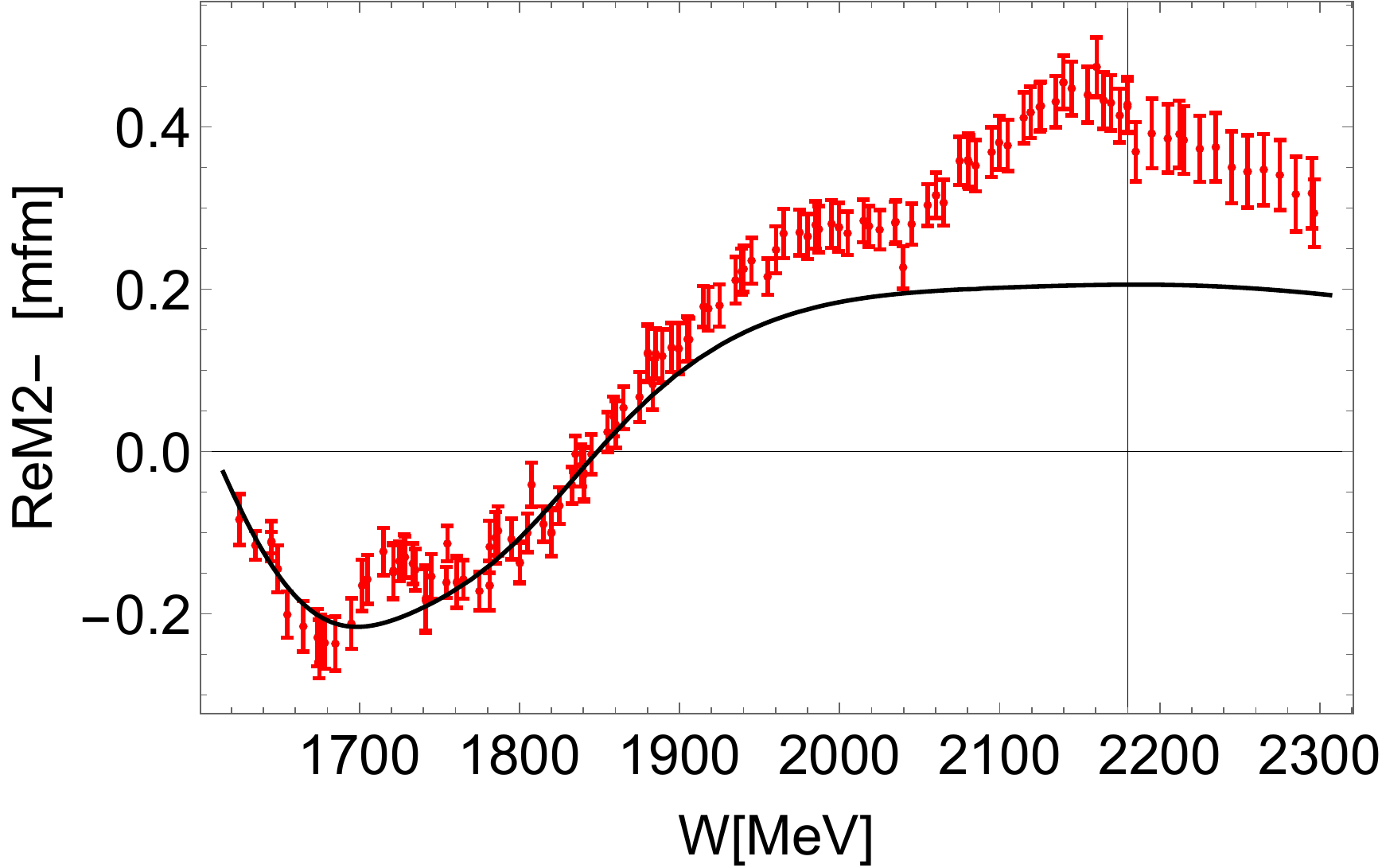} \hspace{0.5cm}
\includegraphics[width=0.37\textwidth]{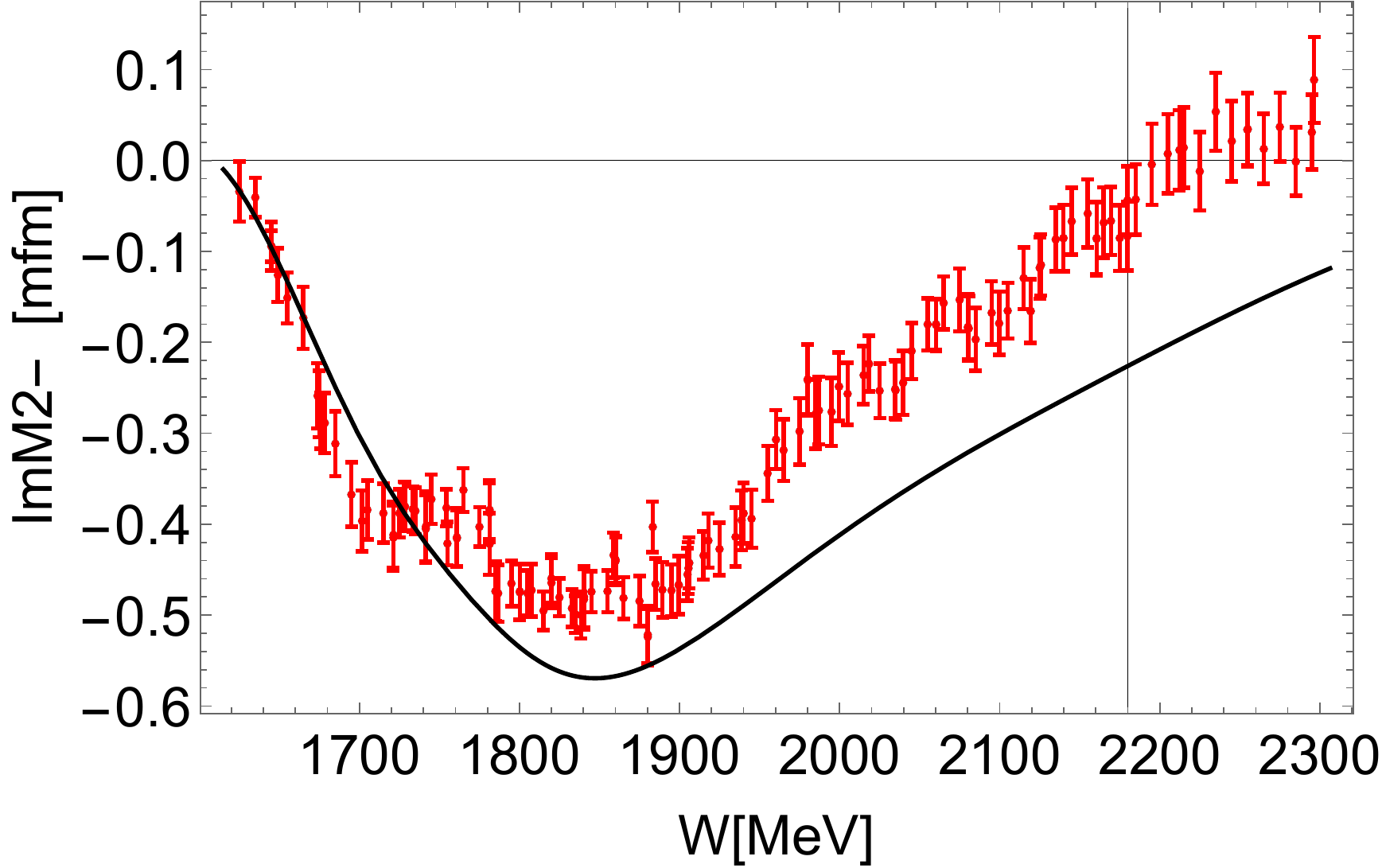}  \\
\includegraphics[width=0.37\textwidth]{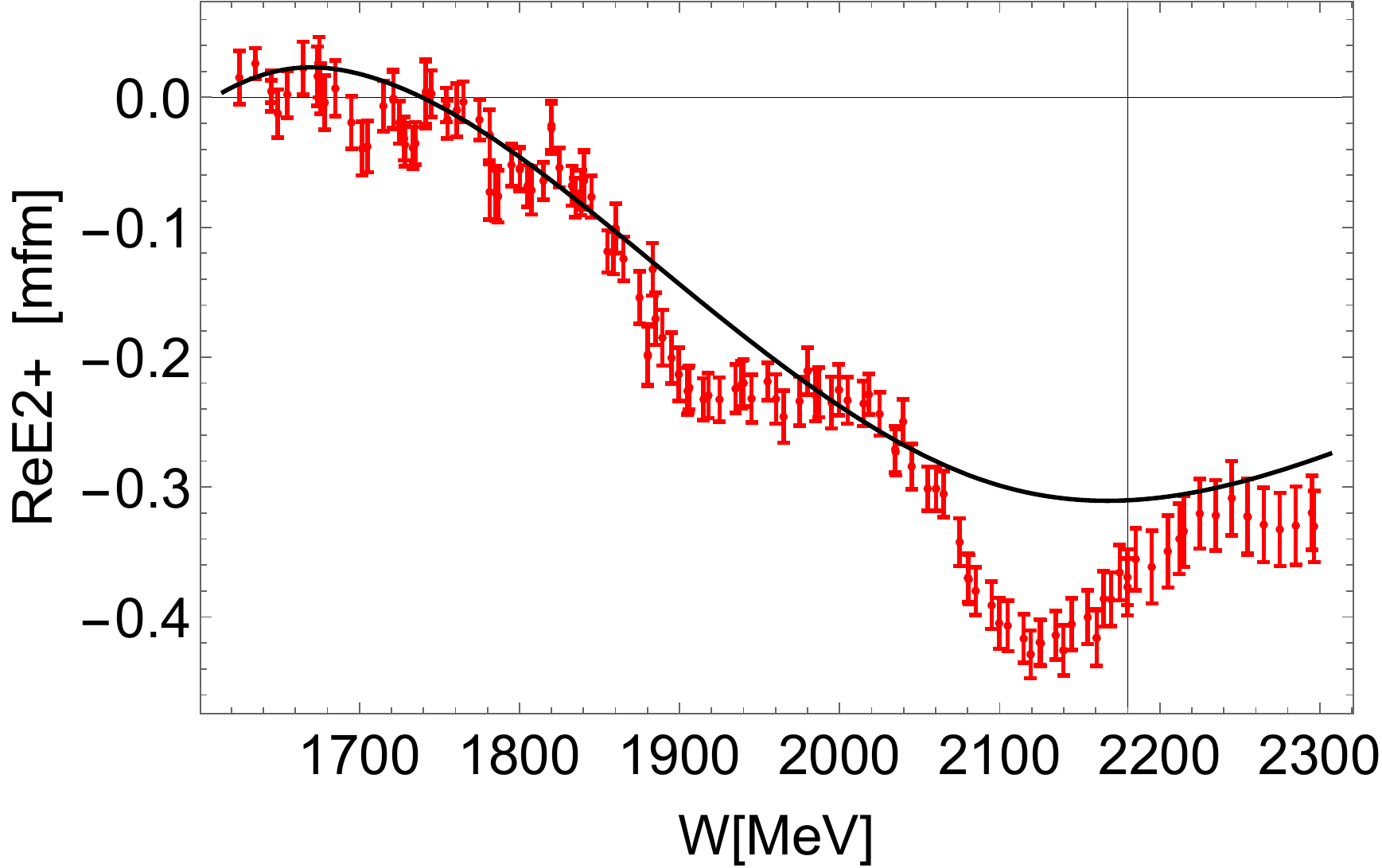} \hspace{0.5cm}
\includegraphics[width=0.37\textwidth]{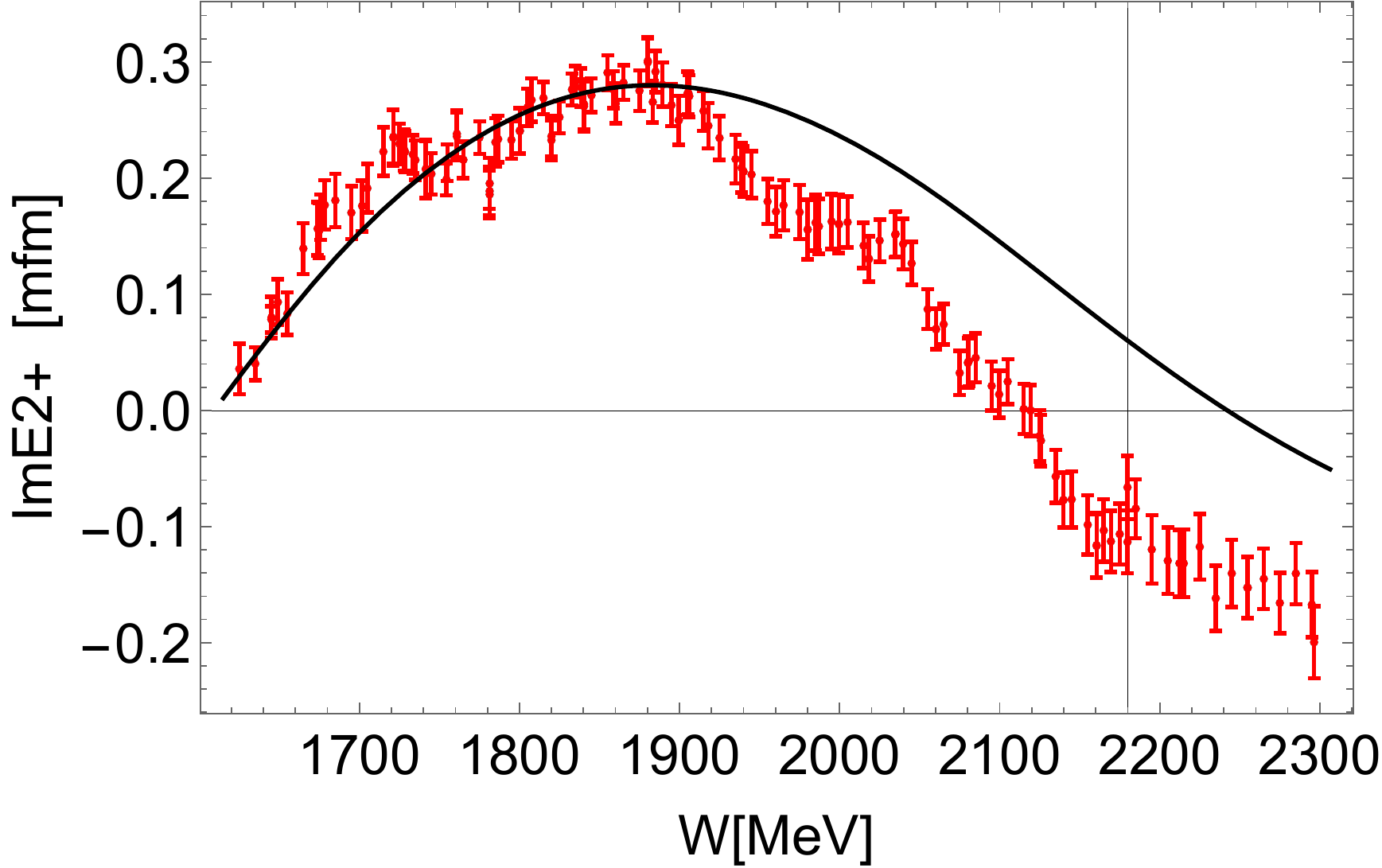}  \\
\includegraphics[width=0.37\textwidth]{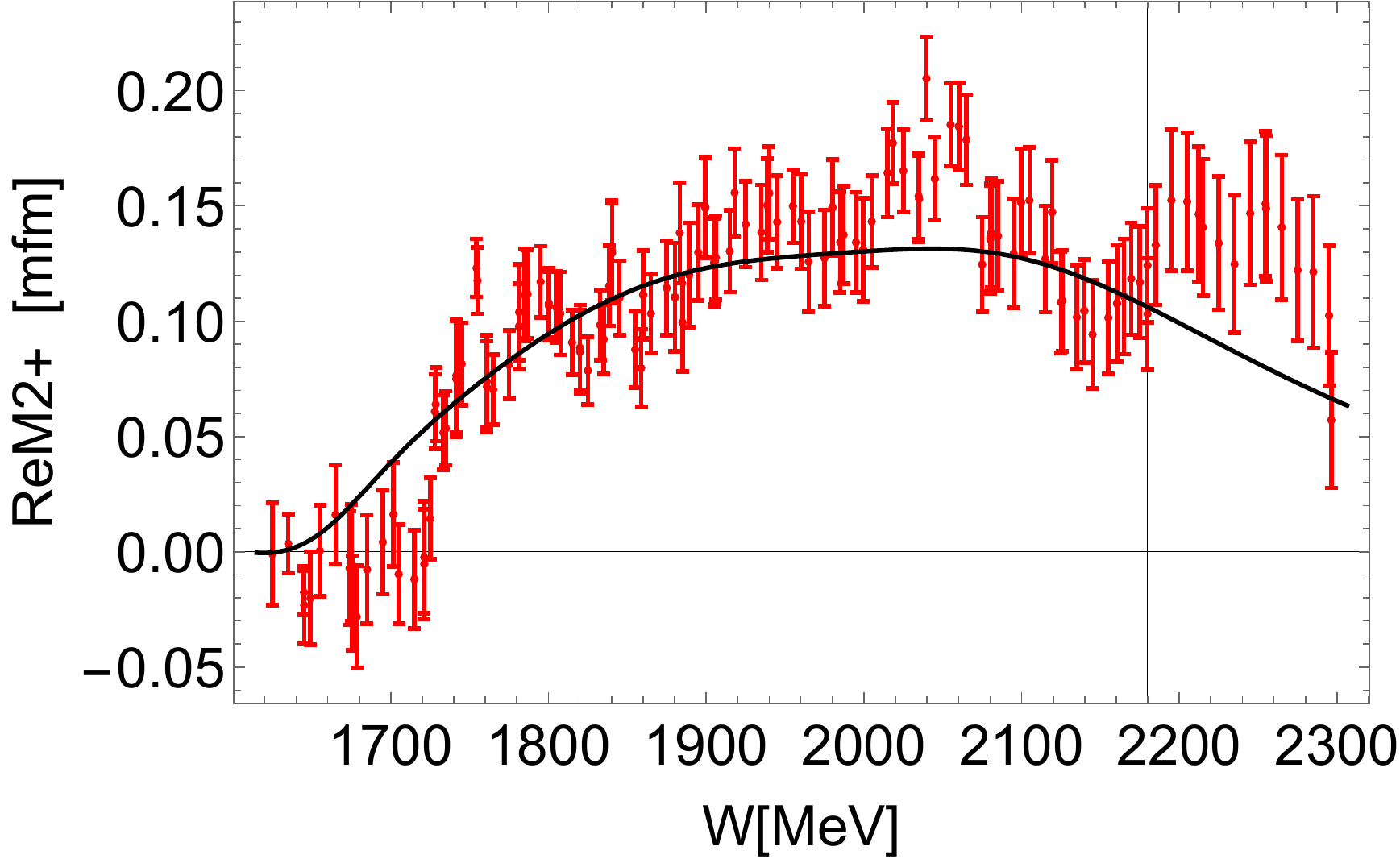} \hspace{0.5cm}
\includegraphics[width=0.37\textwidth]{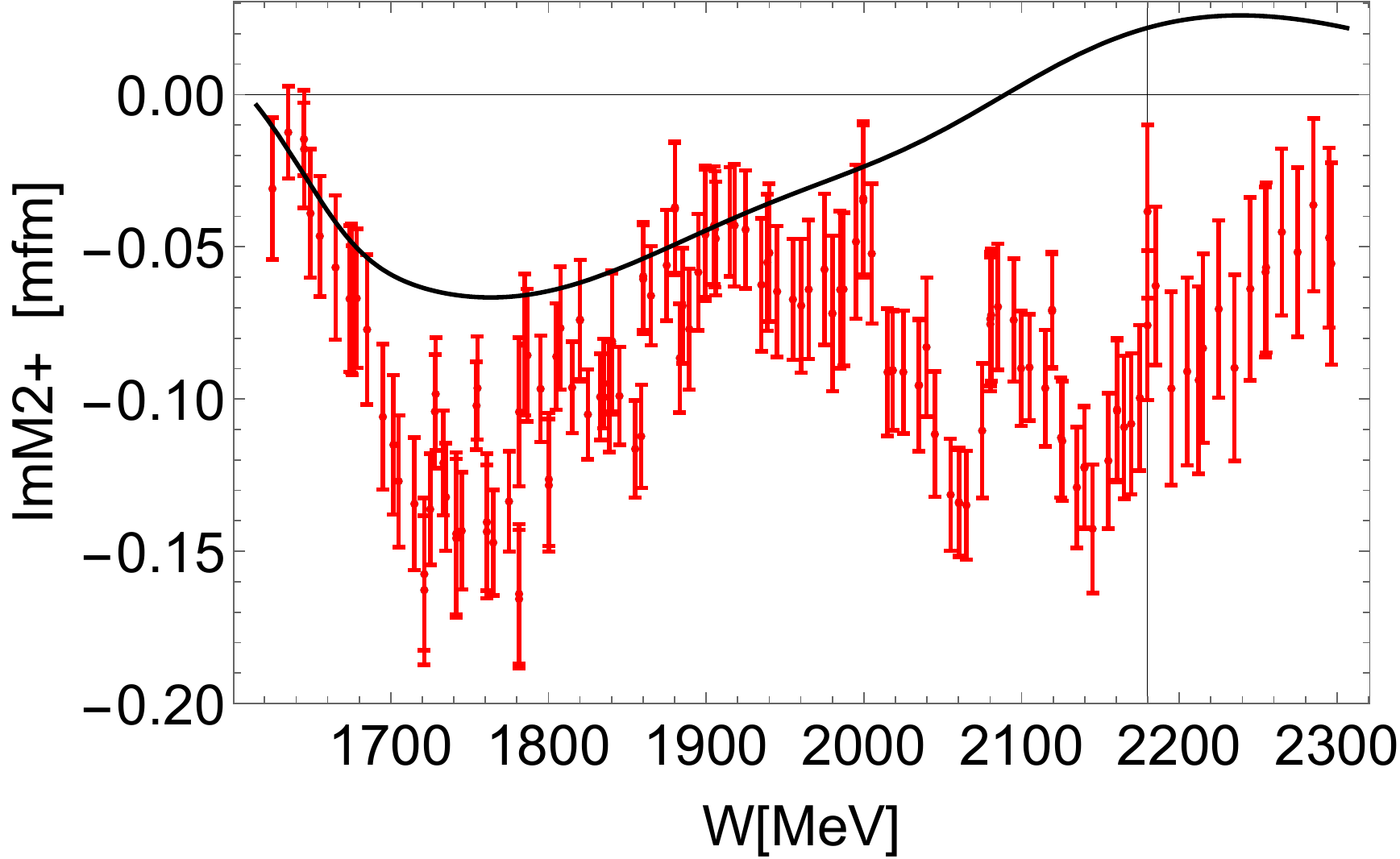}  \\
\includegraphics[width=0.37\textwidth]{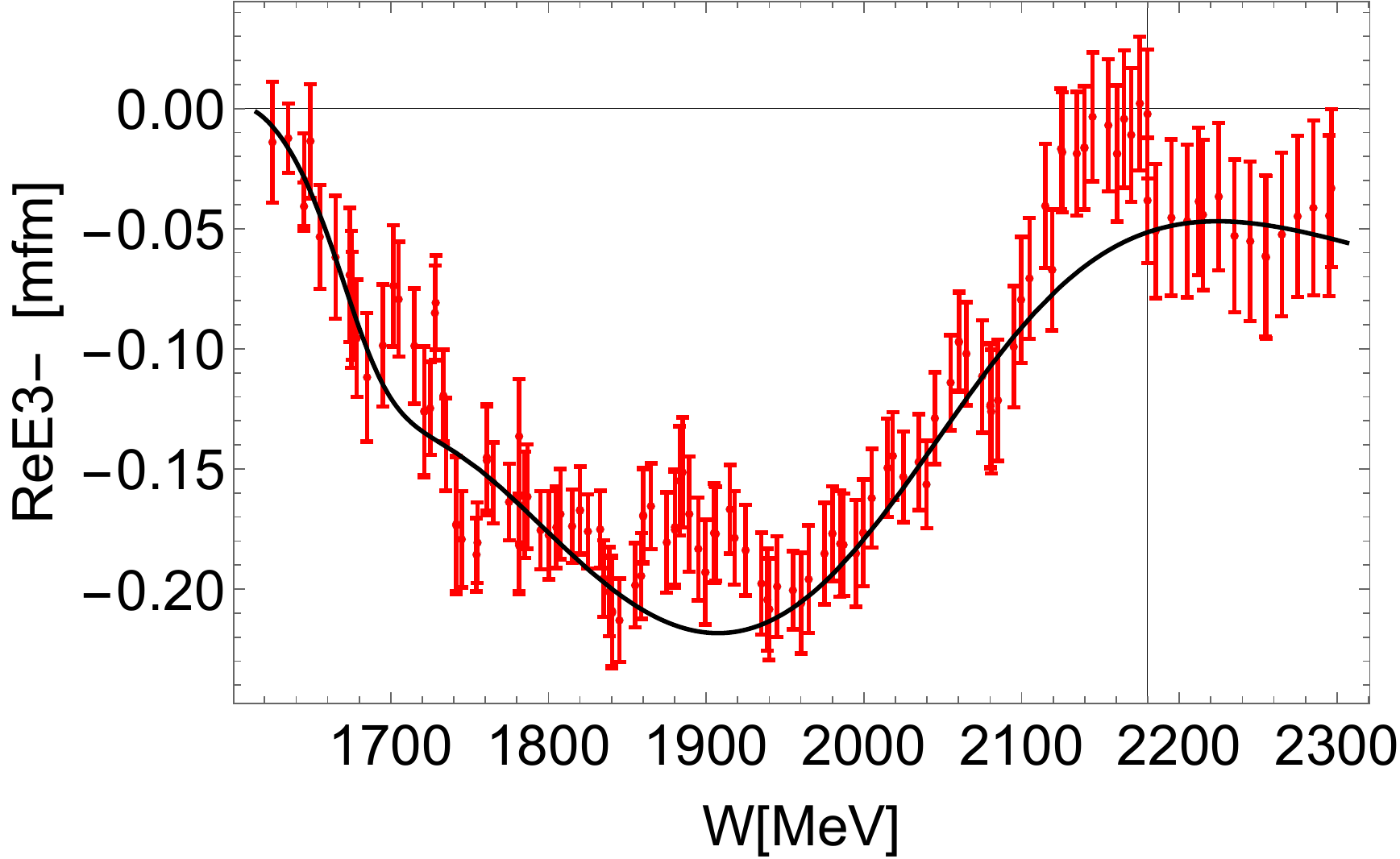} \hspace{0.5cm}
\includegraphics[width=0.37\textwidth]{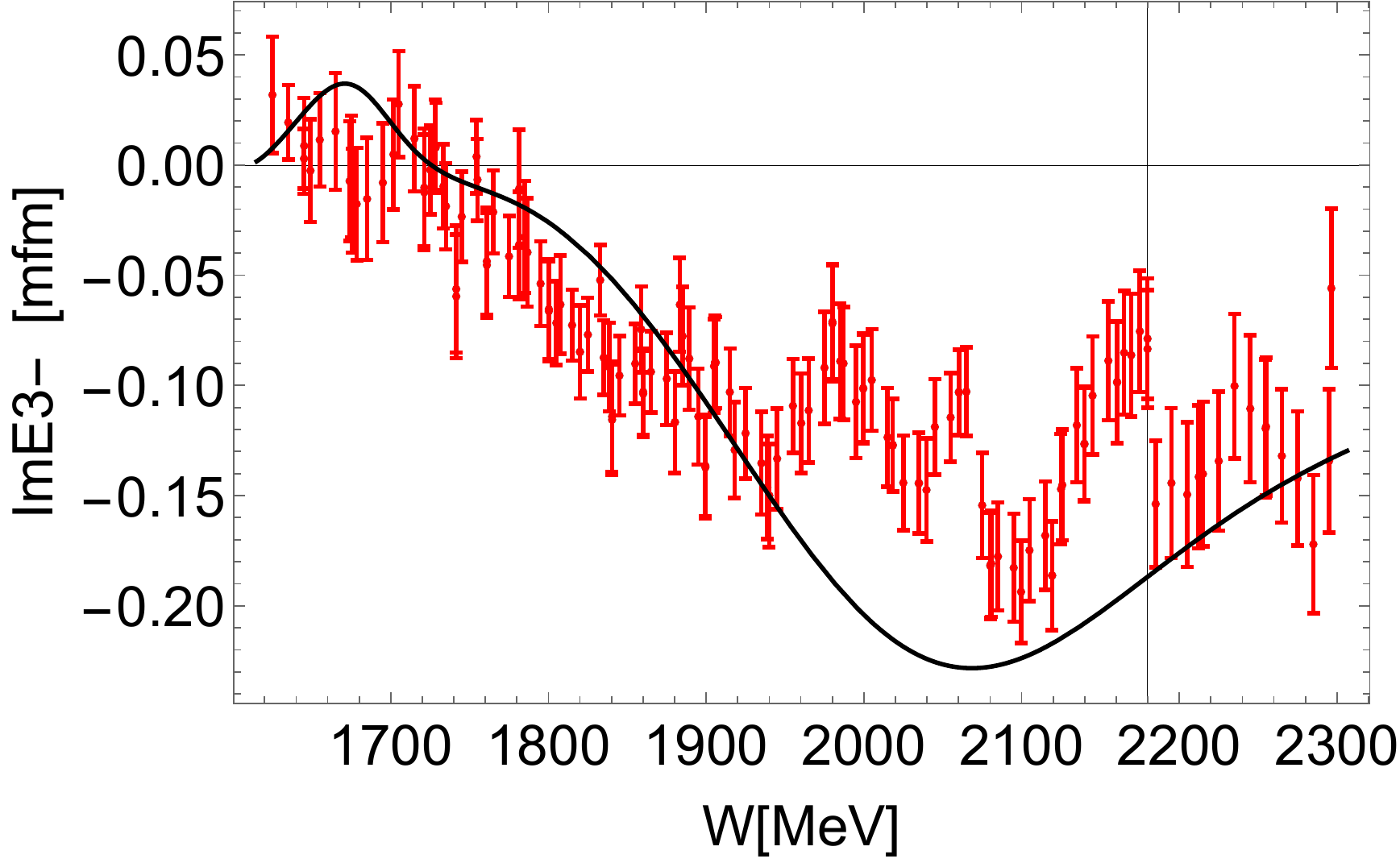}  \\
\includegraphics[width=0.37\textwidth]{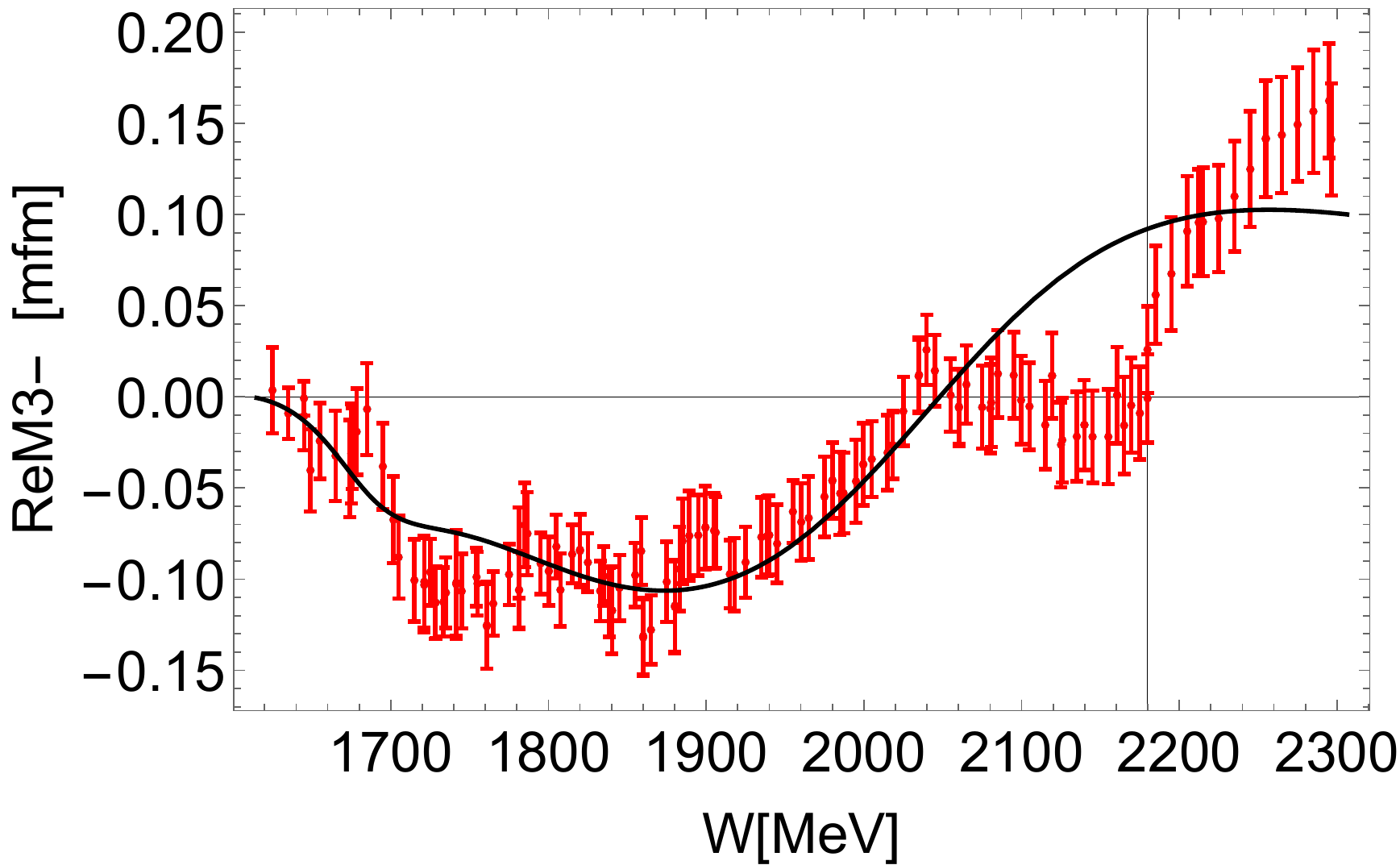} \hspace{0.5cm}
\includegraphics[width=0.37\textwidth]{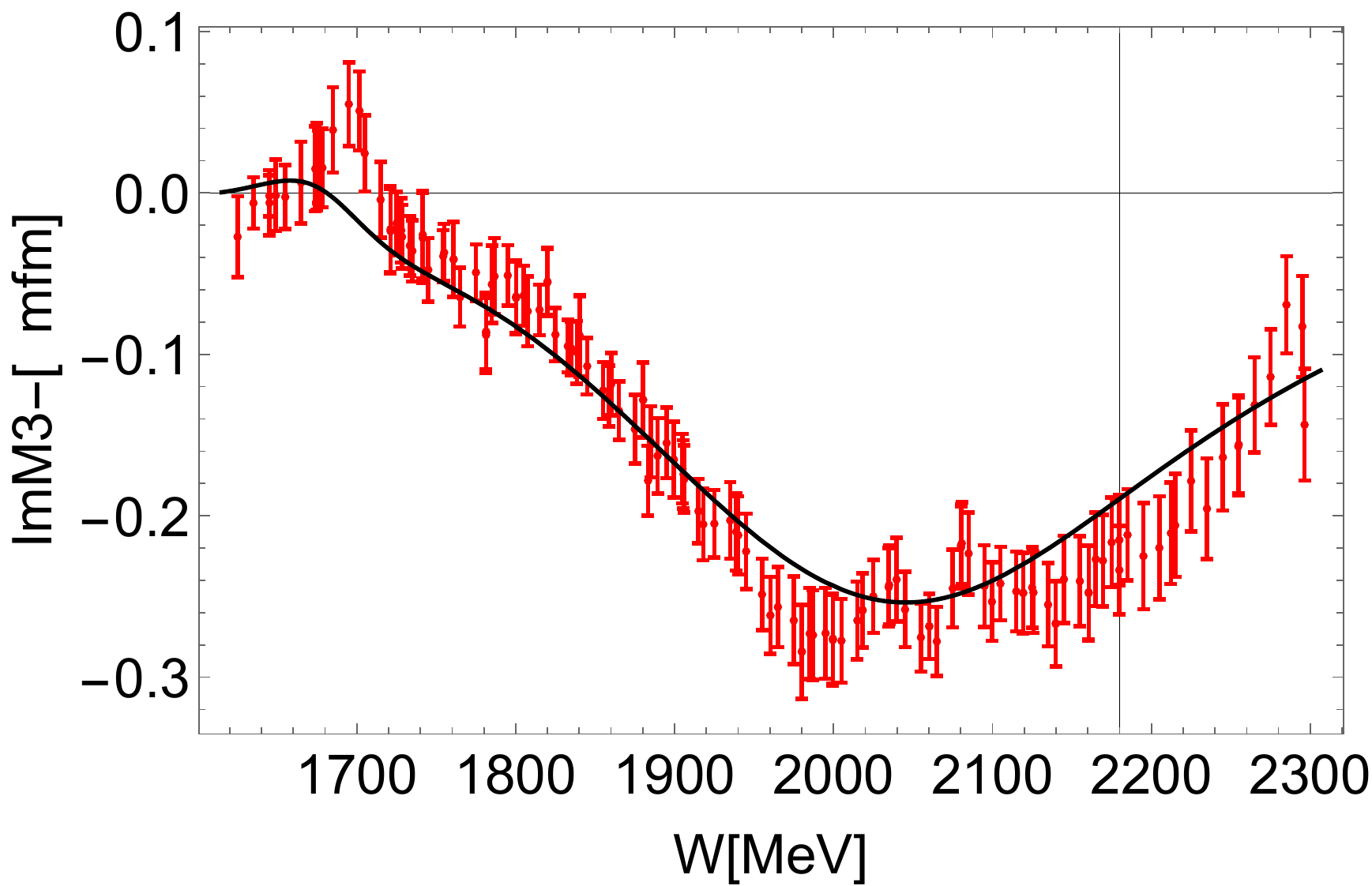}  \\
\caption{\label{Multipoles:b}(Color online) The multipoles for the $L=2$ and $3$ partial waves of our AA/PWA solution are shown.
Red discrete symbols correspond to our solution, and the black full line gives the BG2017 ED solution for comparison. The thin
vertical black line marks the energy where only 4 observables are measured instead of 8 (cf. Table~\ref{tab:expdata}).} \ec
\end{figure}

\begin{figure}[h!]
\bc
\includegraphics[width=0.35\textwidth]{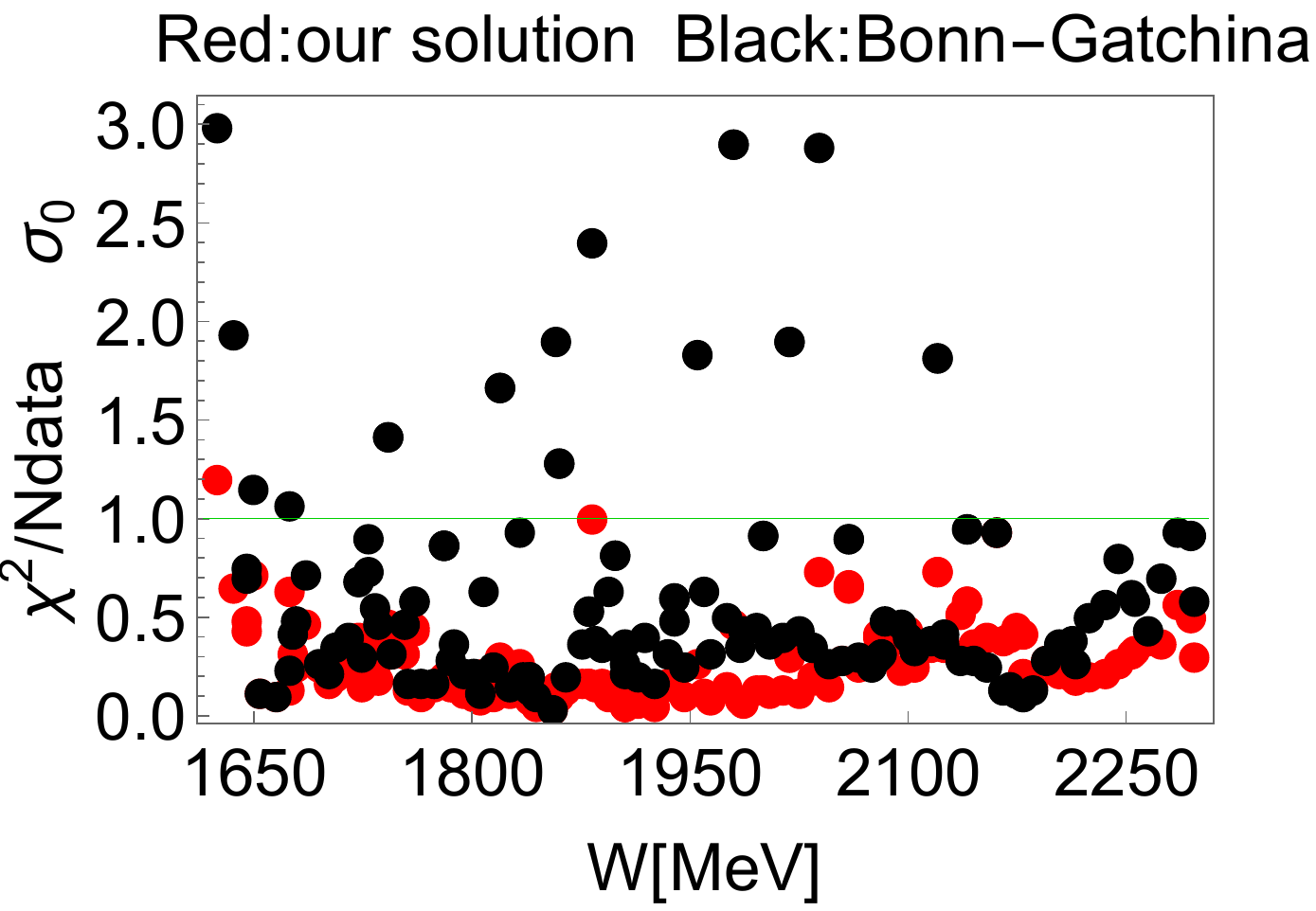} \hspace{0.5cm}
\includegraphics[width=0.35\textwidth]{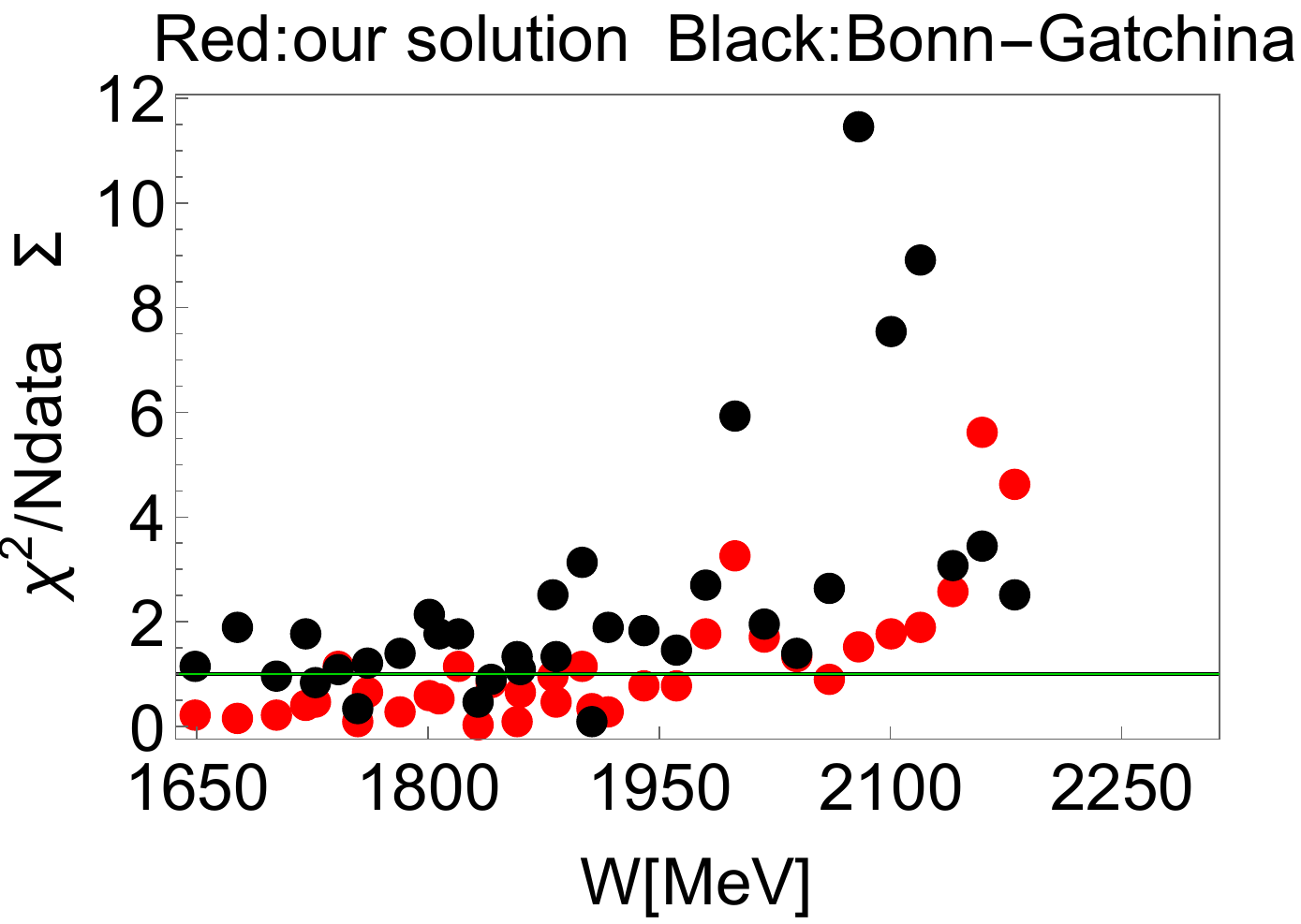} \\
\includegraphics[width=0.35\textwidth]{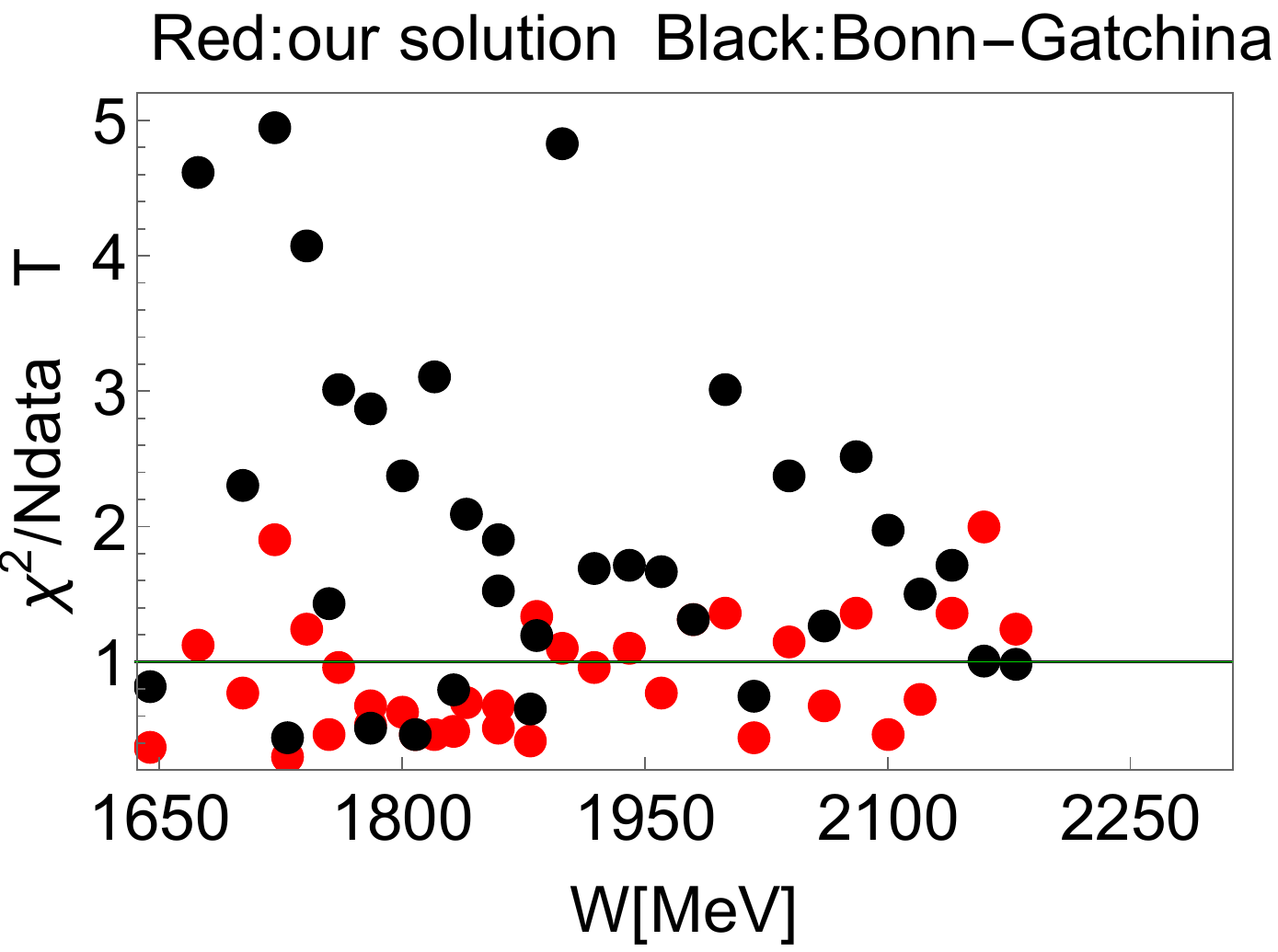} \hspace{0.5cm}
\includegraphics[width=0.35\textwidth]{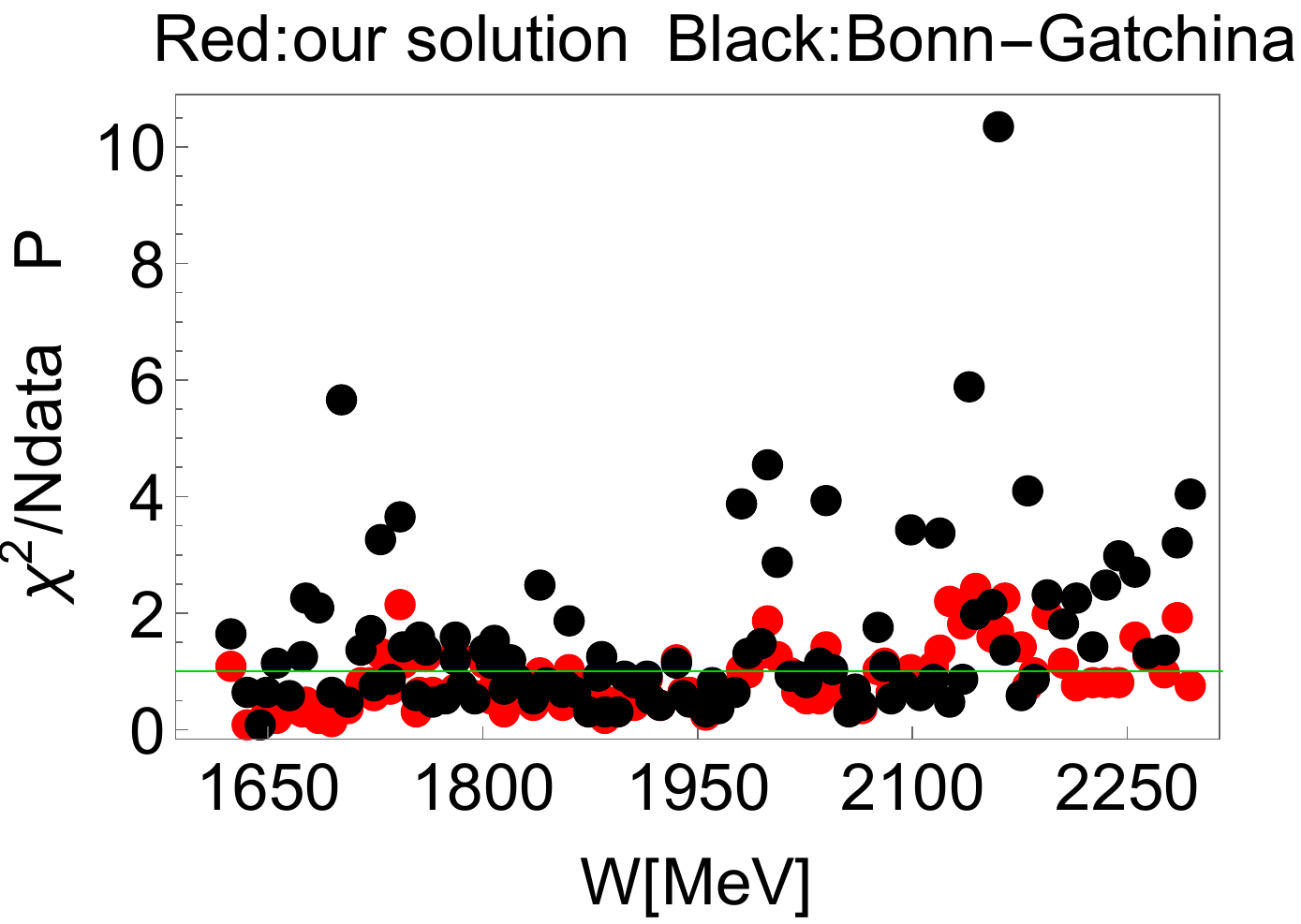} \\ \hspace{0.2cm}
\includegraphics[width=0.32\textwidth]{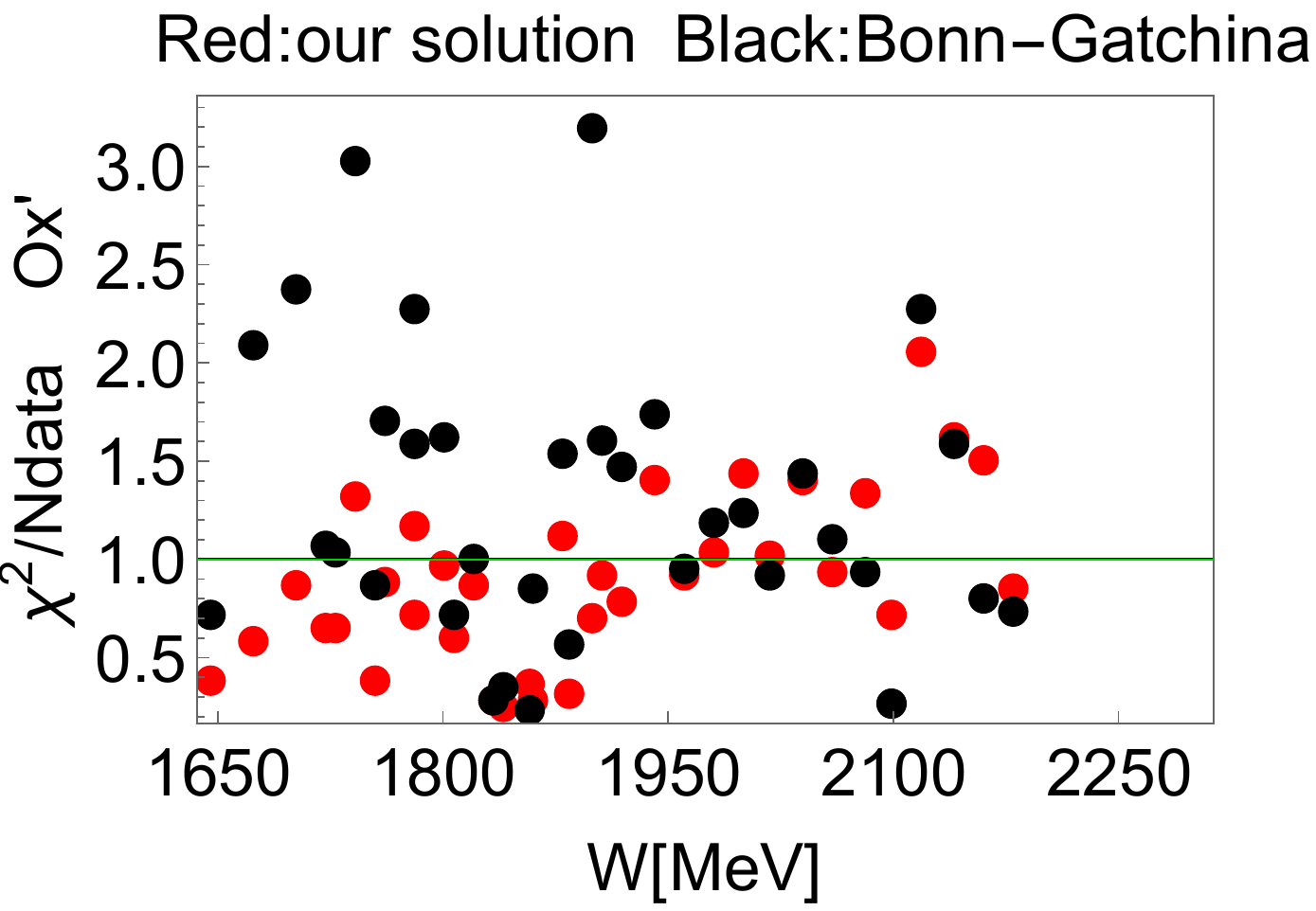} \hspace{0.8cm}
\includegraphics[width=0.35\textwidth]{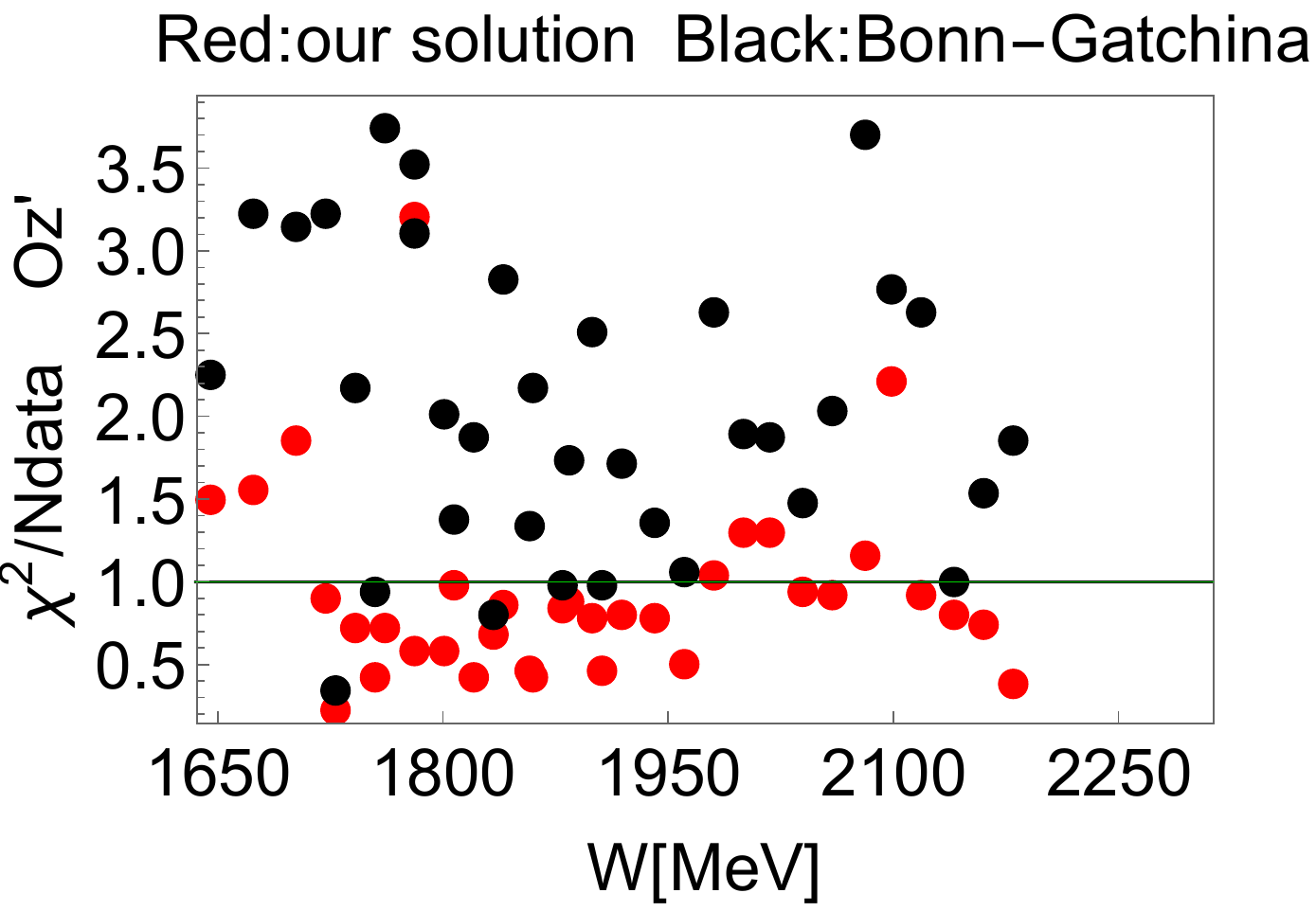} \\
\includegraphics[width=0.35\textwidth]{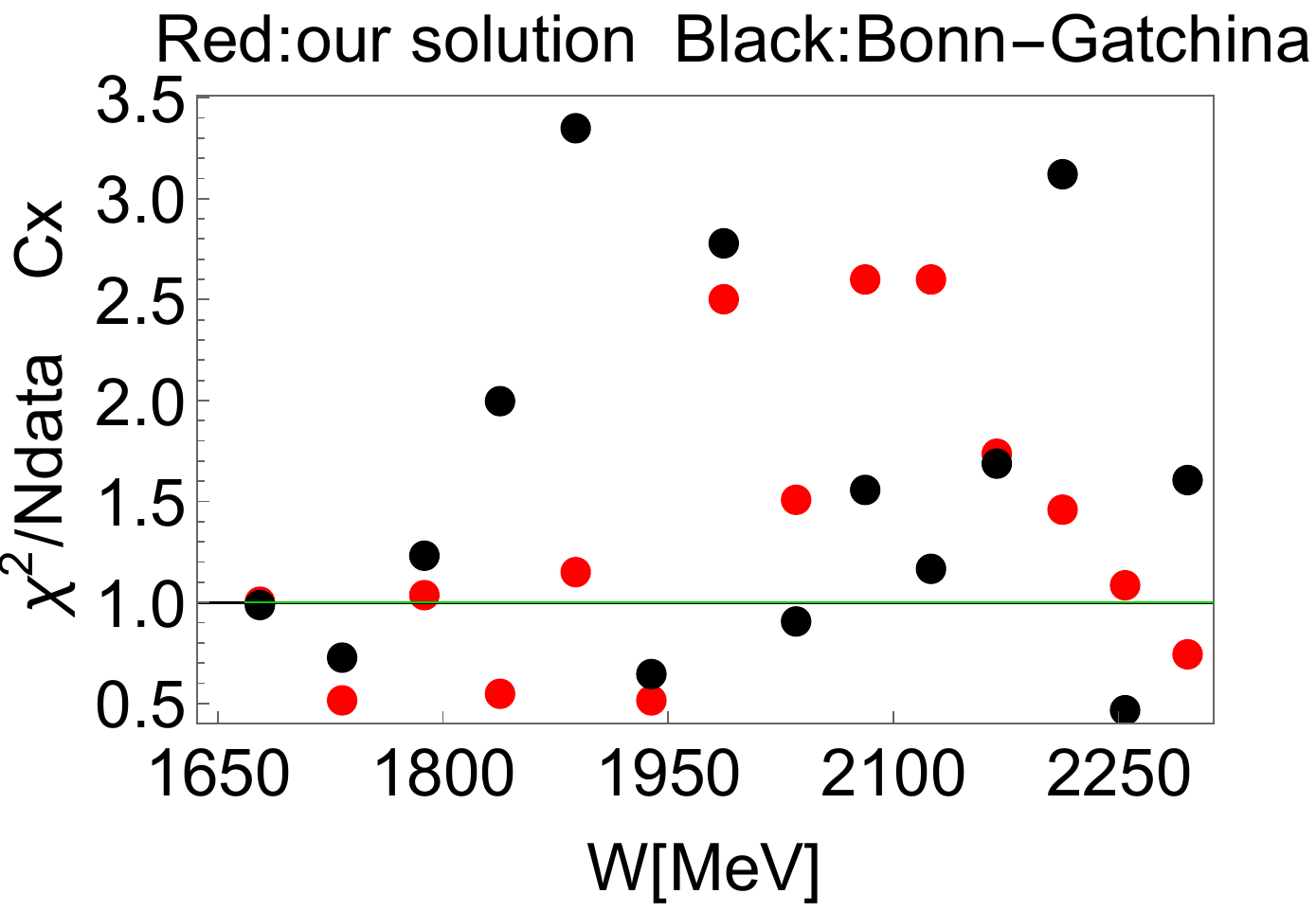} \hspace{0.5cm}
\includegraphics[width=0.35\textwidth]{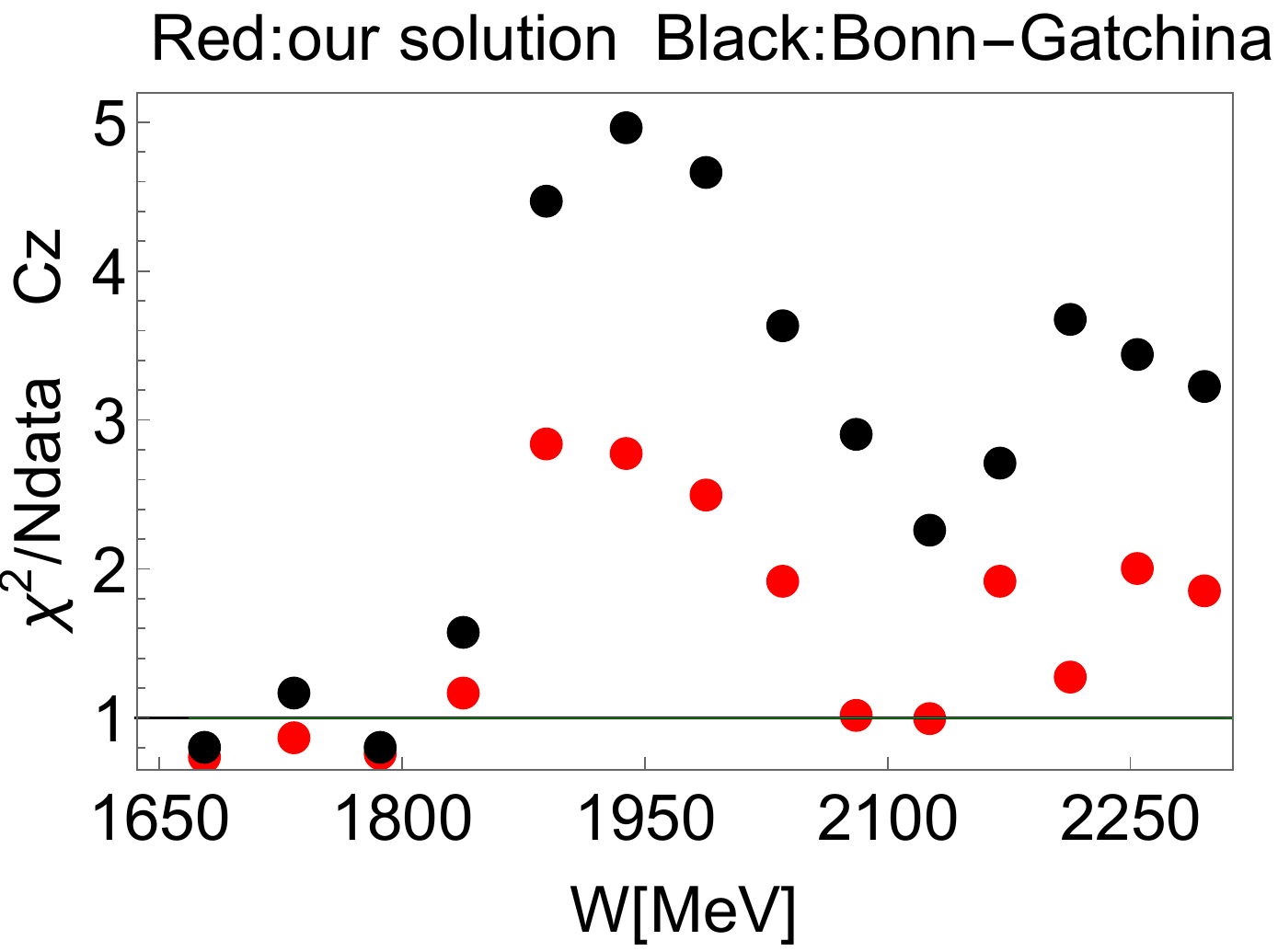} \\
\caption{\label{Chi2:Sol1}(Color online) The $\chi^2/N_{\text{data}}$ for individual observables, calculated on measured values
of energies and angles, is shown. Our AA/PWA solution is given with red symbols, and the same quantity evaluated for the ED
BG2017 solution~\cite{BG-web}  with black symbols.} \ec
\end{figure}

\begin{figure*}[h!]
\bc \hspace*{0.5cm} \includegraphics[width=1.\textwidth]{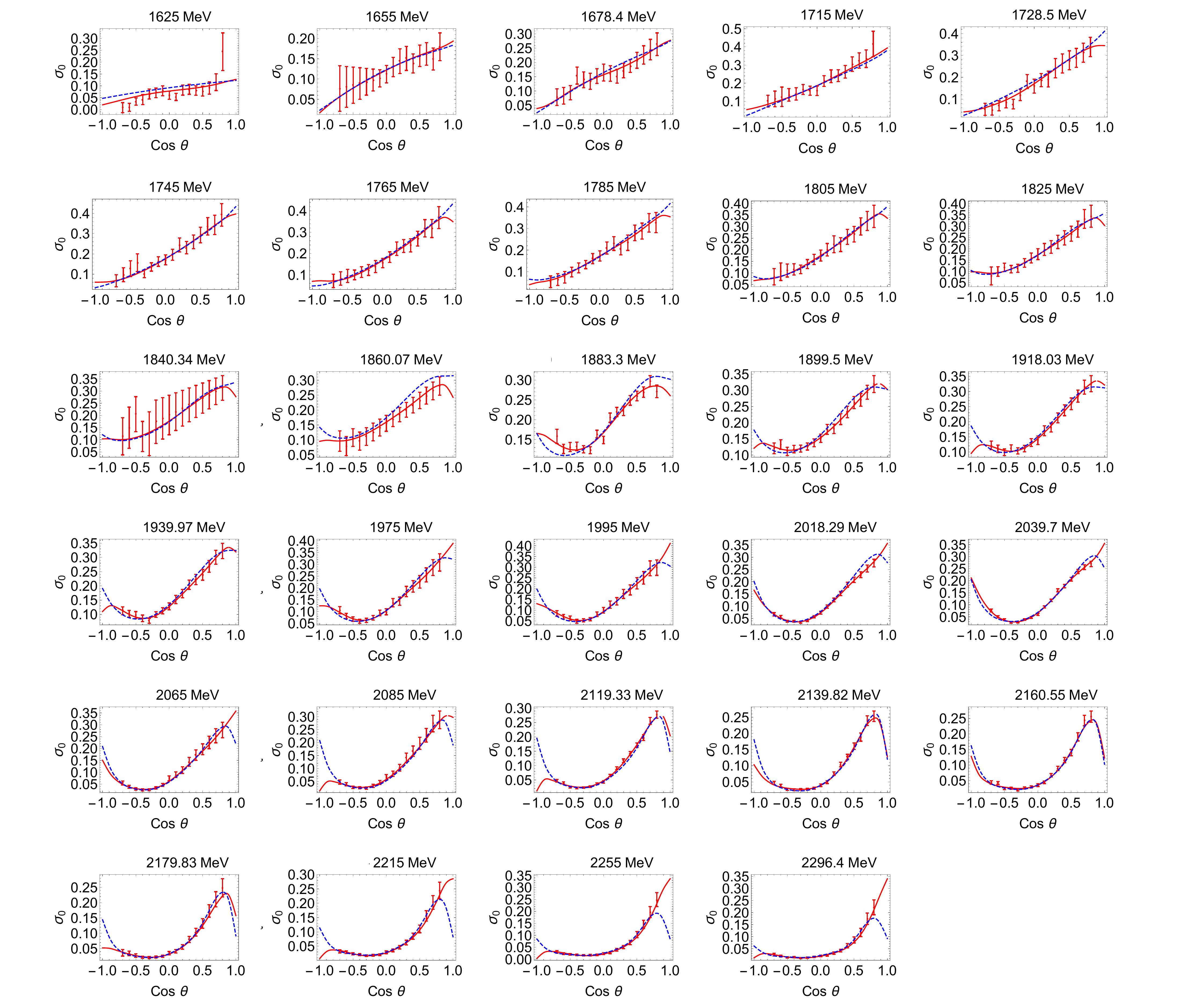} \caption{\label{Sol1:DCS}(Color online) Comparison of
experimental data for $\sigma_0$ (discrete symbols) with our results from AA/PWA (red full line) and the BG2017 fit (blue dashed
line) at representative energies.} \ec
\end{figure*}

\begin{figure*}[h!]
\bc \hspace*{0.5cm} \includegraphics[width=1.\textwidth]{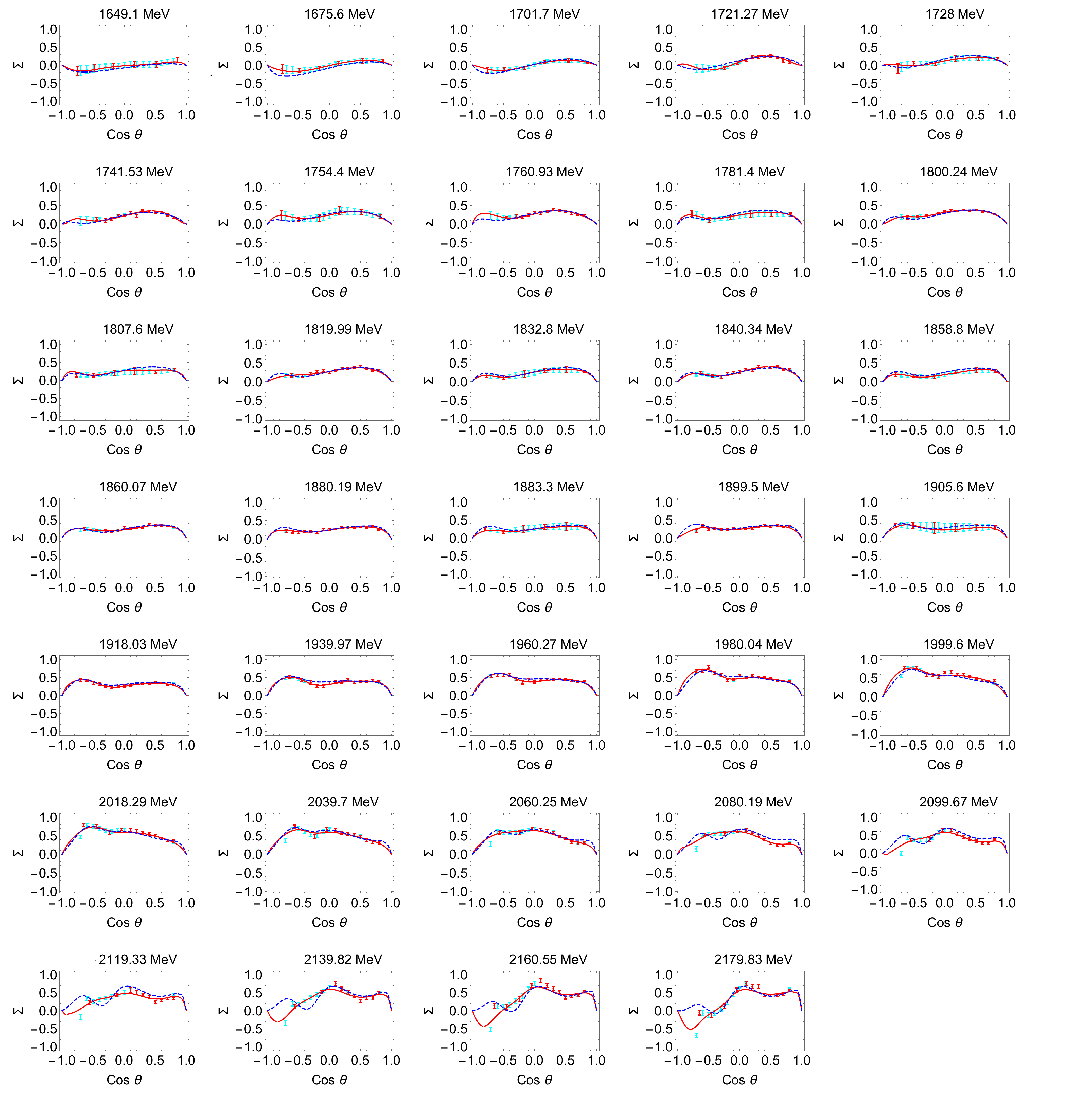} \caption{\label{Sol1:Sigma}(Color online) Comparison of
experimental data for $\Sigma$ (discrete symbols) and interpolated values (cyan symbols) with our results from AA/PWA (red full
line) and the BG2017 fit (blue dashed line) at representative energies.} \ec
\end{figure*}

\begin{figure*}[h!]
\bc \hspace*{0.5cm} \includegraphics[width=1.\textwidth]{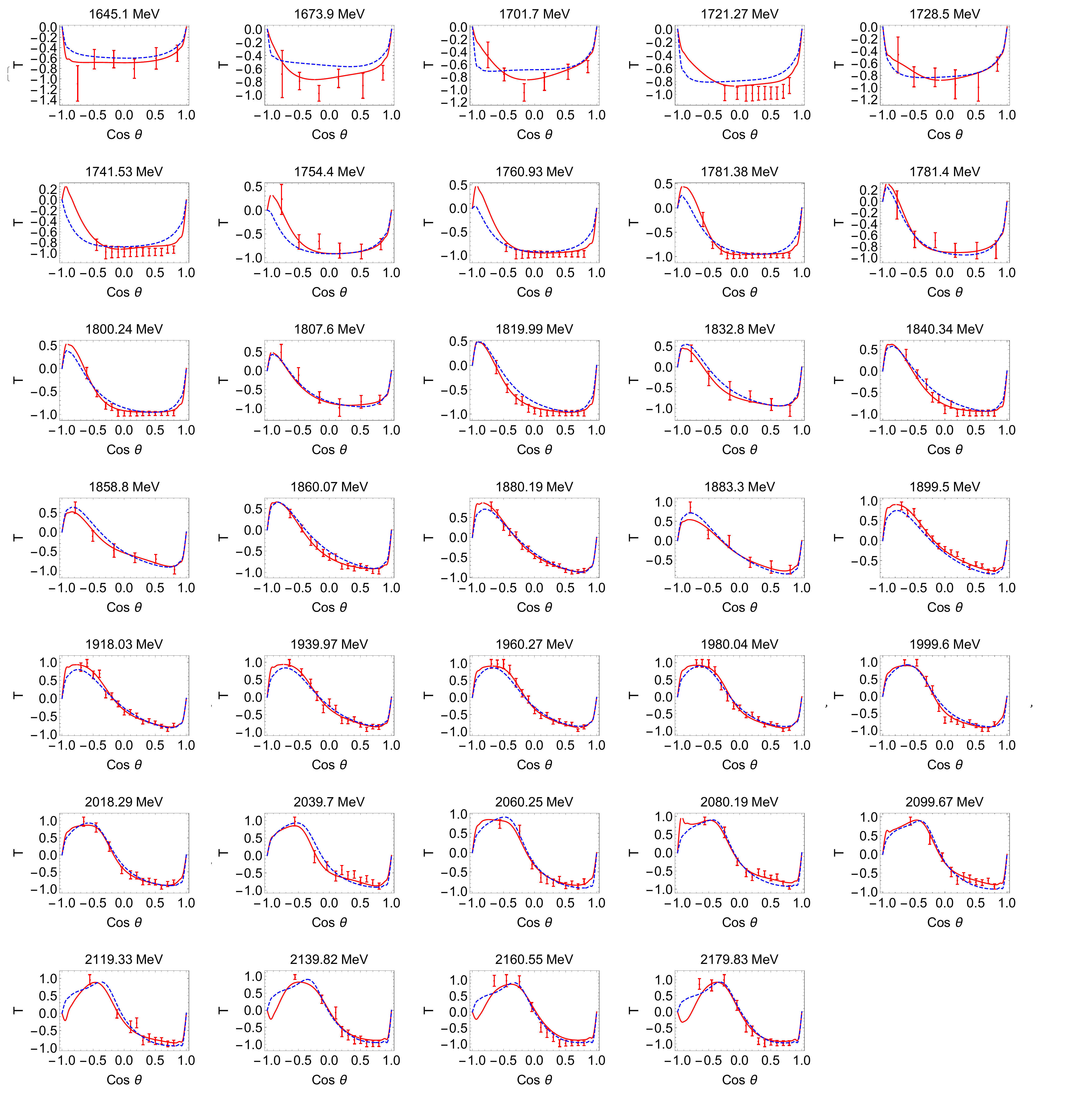} \caption{\label{Sol1:T}(Color online) Comparison of experimental
data for $T$ (discrete symbols) and interpolated values (cyan symbols) with our results from AA/PWA (red full line) and the
BG2017 fit (blue dashed line) at representative energies. } \ec
\end{figure*}

\begin{figure*}[h!]
\bc \hspace*{0.5cm} \includegraphics[width=1.\textwidth]{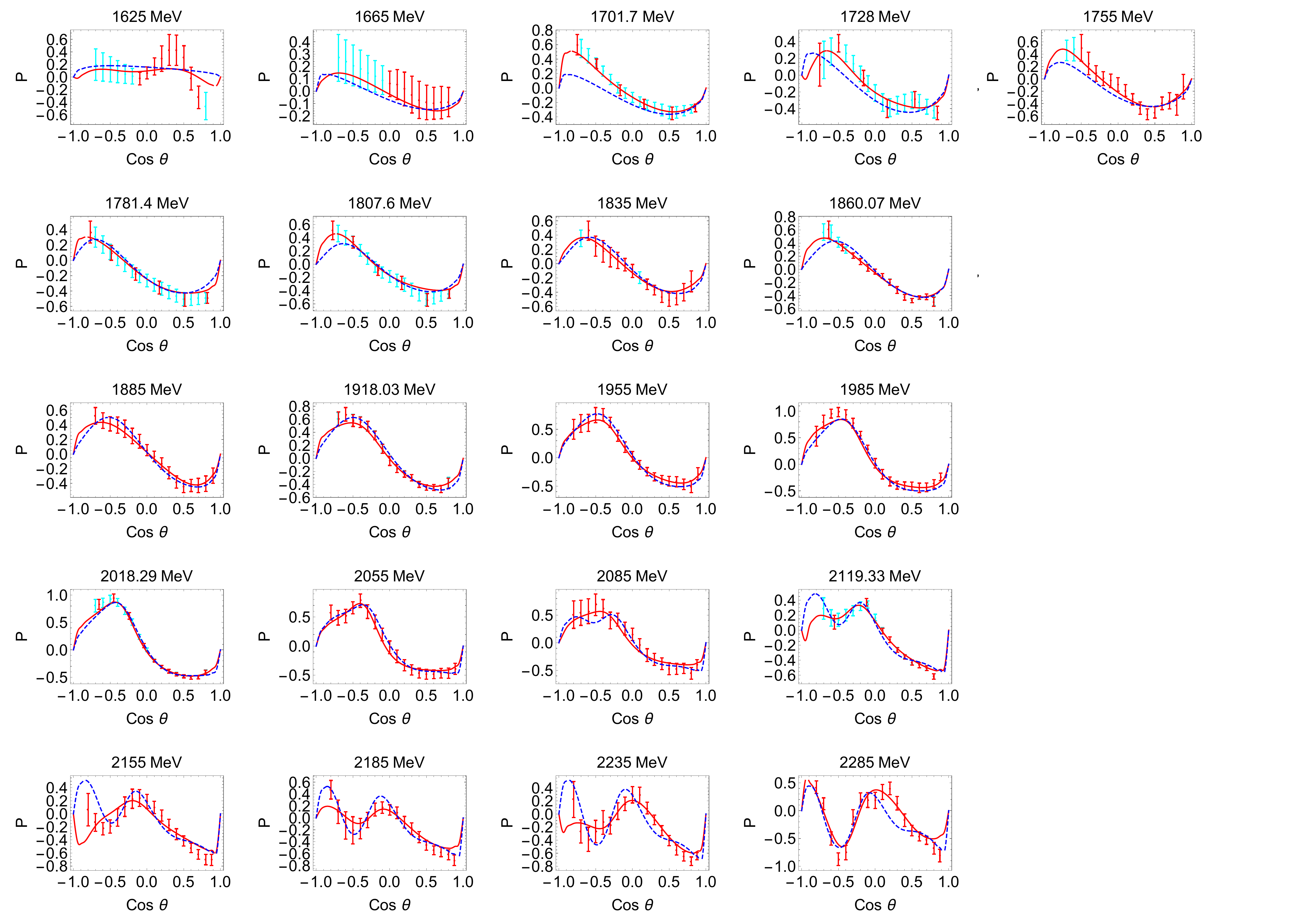} \caption{\label{Sol1:P}(Color online) Comparison of experimental
data for $P$ (discrete symbols) and interpolated values (cyan symbols) with our results from AA/PWA (red full line) and the
BG2017 fit (blue dashed line) at representative energies. } \ec
\end{figure*}

\begin{figure*}[h!]
\bc \hspace*{0.5cm} \includegraphics[width=1.\textwidth]{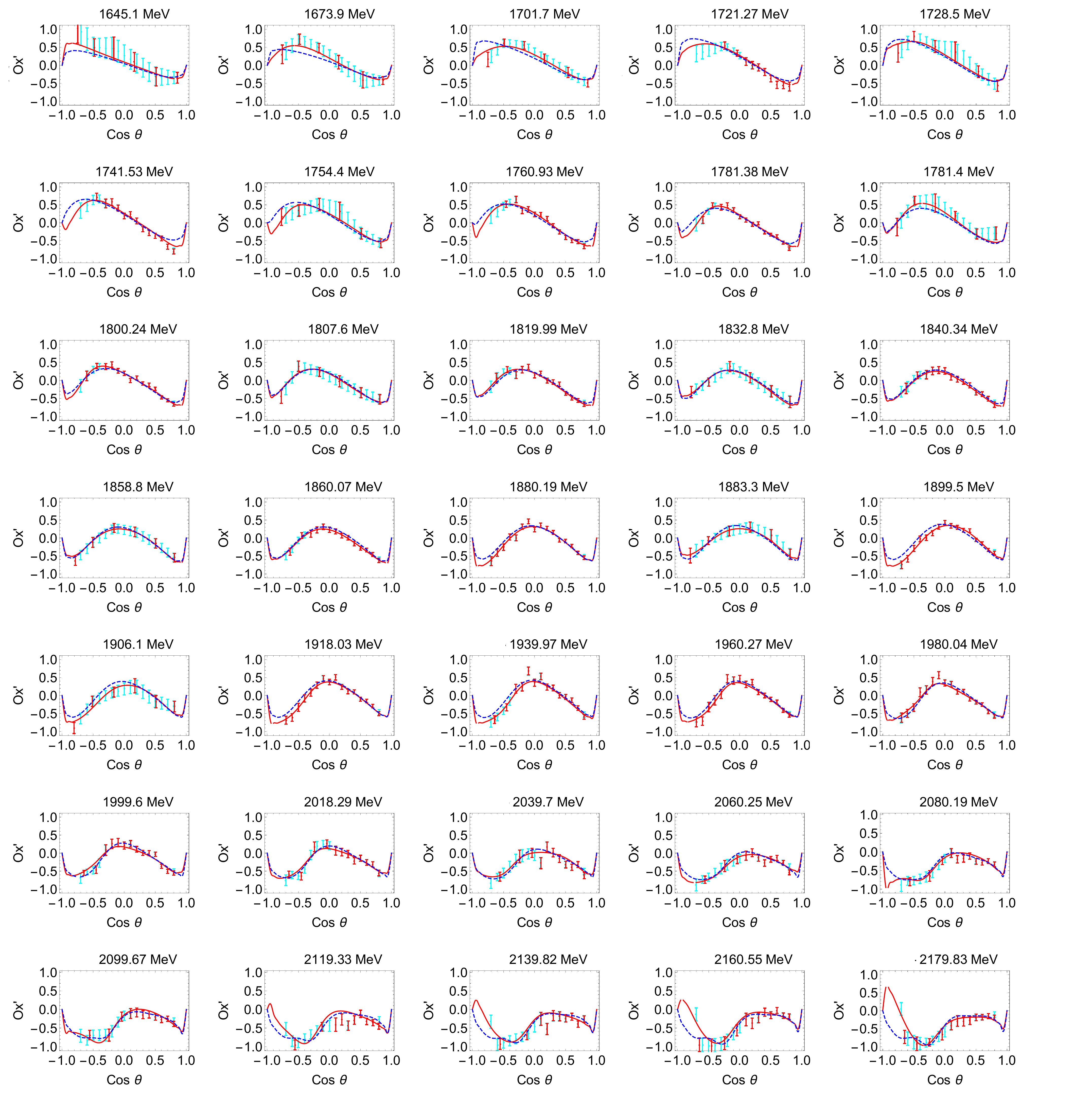} \caption{\label{Sol1:Oxprime}(Color online) Comparison of
experimental data for $O_{x'}$ (discrete symbols) and interpolated values (cyan symbols) with our results from AA/PWA (red full
line) and the BG2017 fit (blue dashed line) at representative energies.} \ec
\end{figure*}

\begin{figure*}[h!]
\bc \hspace*{0.5cm} \includegraphics[width=1.\textwidth]{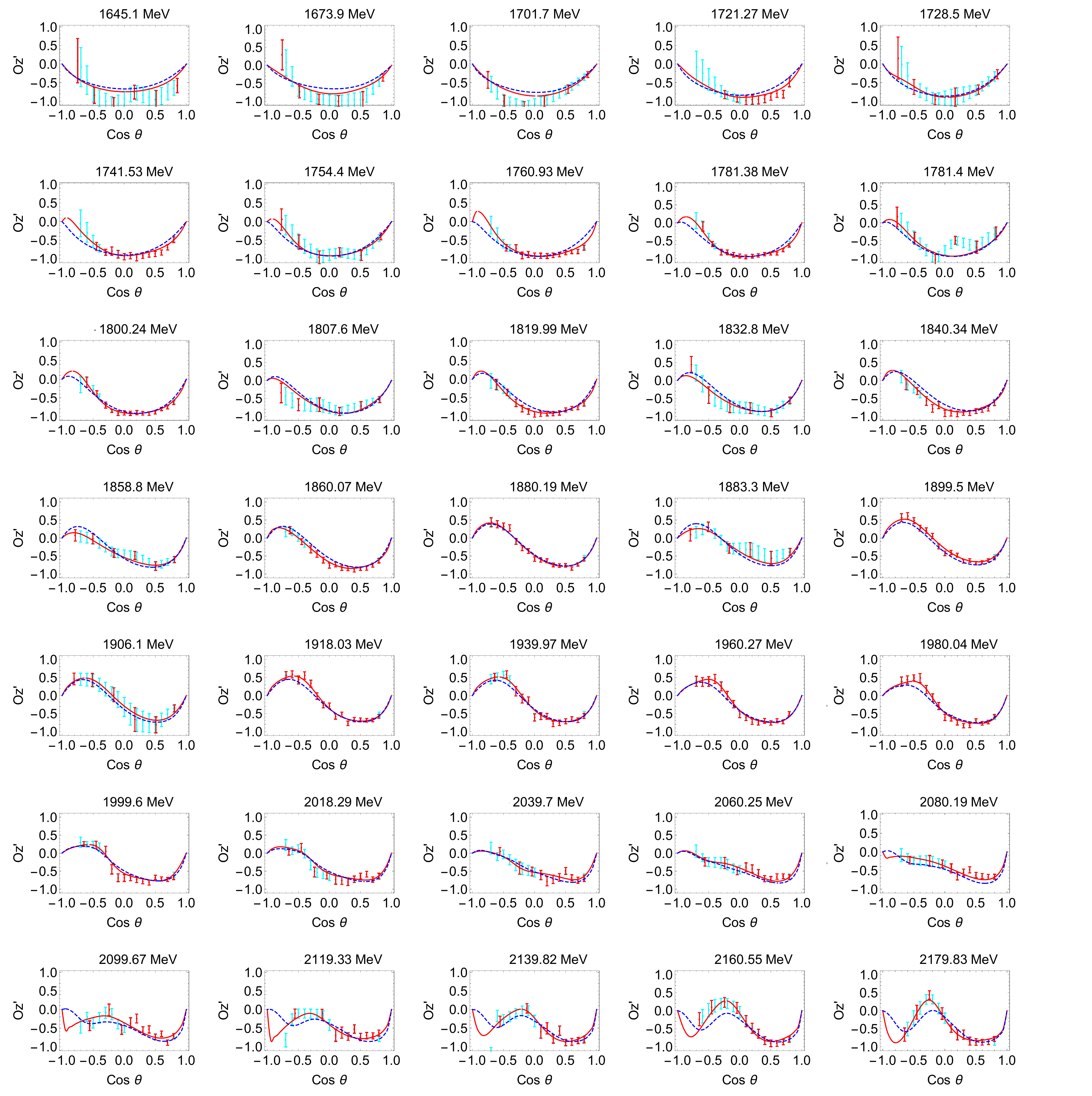} \caption{\label{Sol1:Ozprime}(Color online) Comparison of
experimental data for $O_{z'}$ (discrete symbols) and interpolated values (cyan symbols) with our results from AA/PWA (red full
line) and the BG2017 fit (blue dashed line) at representative energies.} \ec
\end{figure*}

\begin{figure*}[h!]
\bc \hspace*{0.5cm} \includegraphics[width=1.\textwidth]{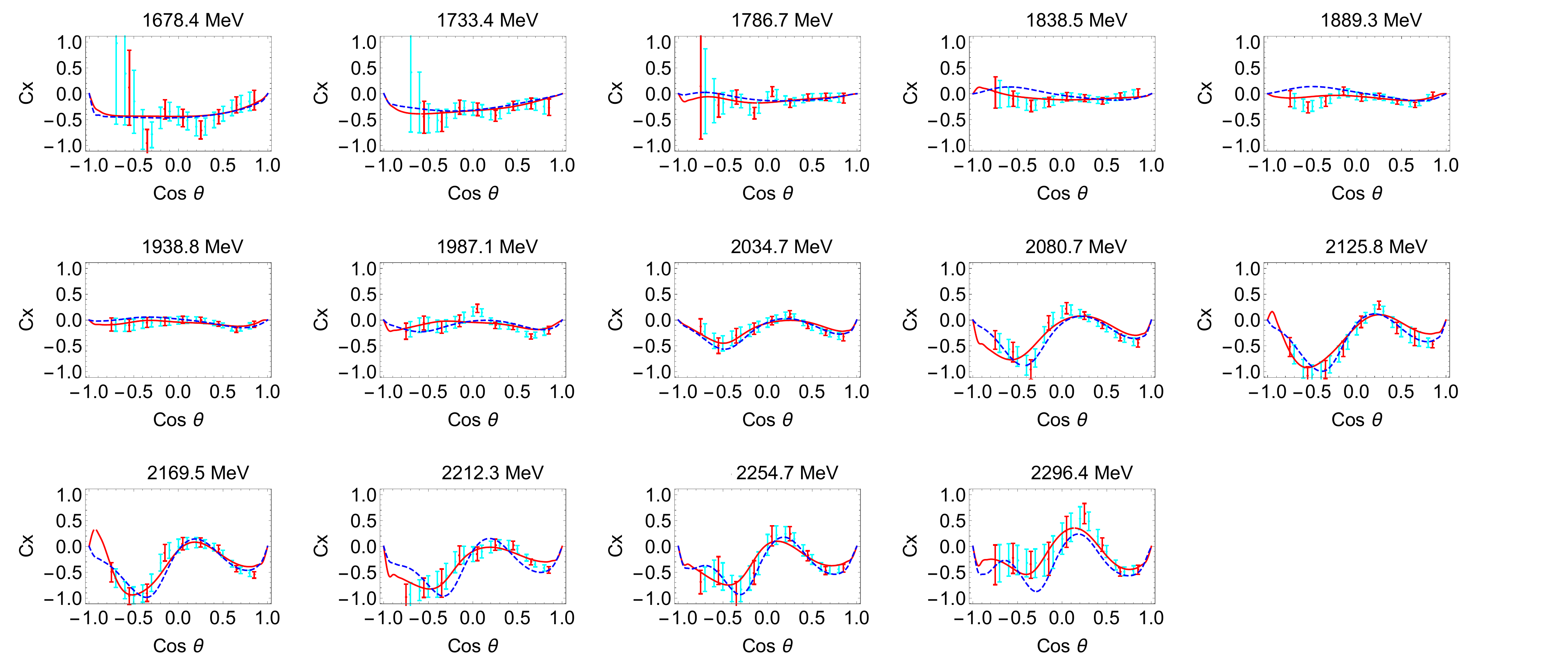} \caption{\label{Sol1:Cx}(Color online) Comparison of
experimental data for $C_{x}$ (discrete symbols) and interpolated values (cyan symbols) with our results from AA/PWA (red full
line) and the BG2017 fit (blue dashed line) at representative energies.} \ec
\end{figure*}

\begin{figure*}[h!]
\bc \hspace*{0.5cm} \includegraphics[width=1.\textwidth]{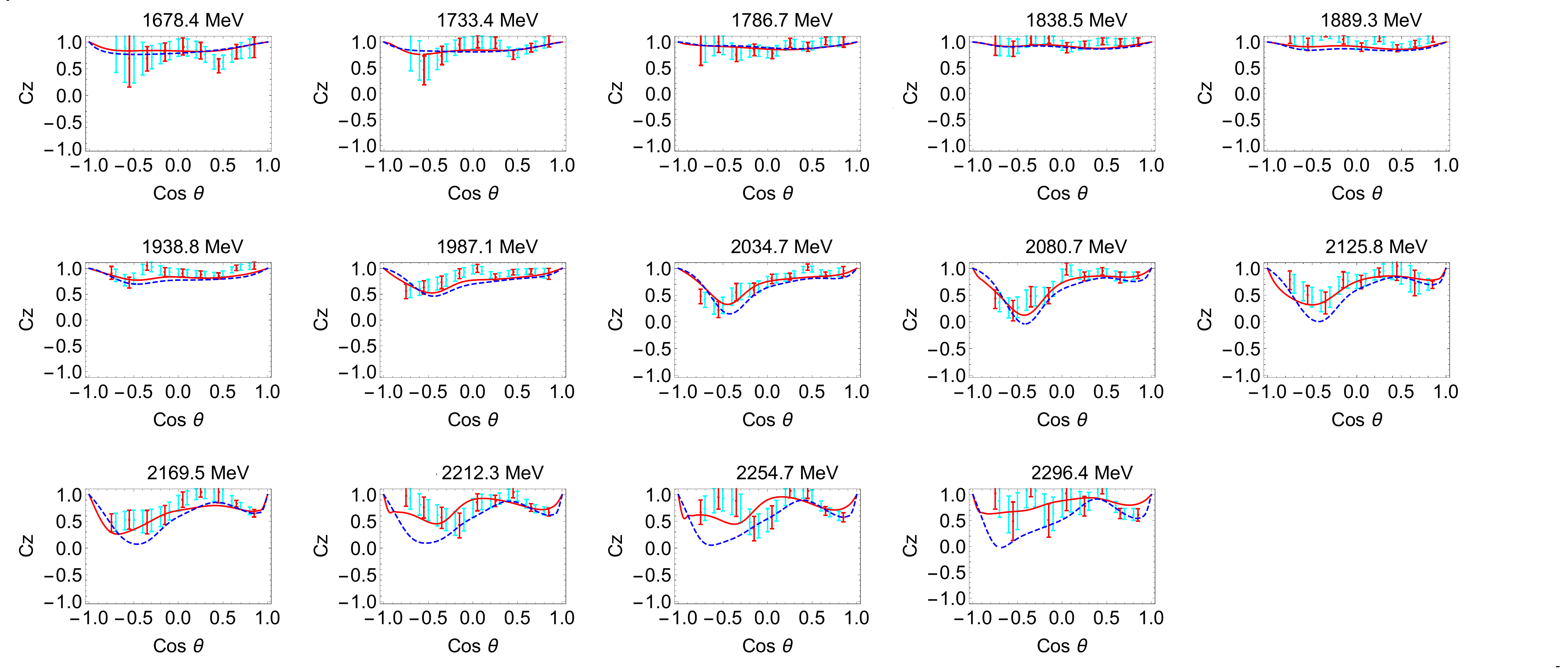} \caption{\label{Sol1:Cz}(Color online) Comparison of
experimental data for $C_{z}$ (discrete symbols) and interpolated values (cyan symbols) with our results from AA/PWA (red full
line) and the BG2017 fit (blue dashed line) at representative energies.} \ec
\end{figure*}

In Figs.~\ref{Multipoles:a} and \ref{Multipoles:b} we show our final results for multipoles on the full set of energies. The
penalty factor $\lambda$ is picked by hand, and is set to $\lambda=250$. Red symbols give the values of multipoles for our AA/PWA
solution, and the black full line gives the prediction of the BG2017 solution~\cite{BG-web}. In Fig.~\ref{Chi2:Sol1} we give the
$\chi^2$ per data point for each observable calculated on measured values of energies and angles as red symbols, and the same
quantity for the ED BG2017 solution~\cite{BG-web} as black symbols. In Figs.~\ref{Sol1:DCS} - \ref{Sol1:Cz} we give the fits to
measured observables resulting from our AA/PWA method (red full line) as well as predictions from the ED BG2017
solution~\cite{BG-web} (dashed blue line) at representative energies only. All further energies are available upon demand.

\newpage

\newpage

\clearpage We also give predictions resulting from our AA/PWA method for the unmeasured ${\cal BT}$ polarization observables $E$,
$F$, $G$, and $H$ at certain, representative energies in the full energy range.
 \begin{figure}[h!]
\bc
\hspace*{0.5cm} \hspace*{1.cm}\includegraphics[width=1.0\textwidth]{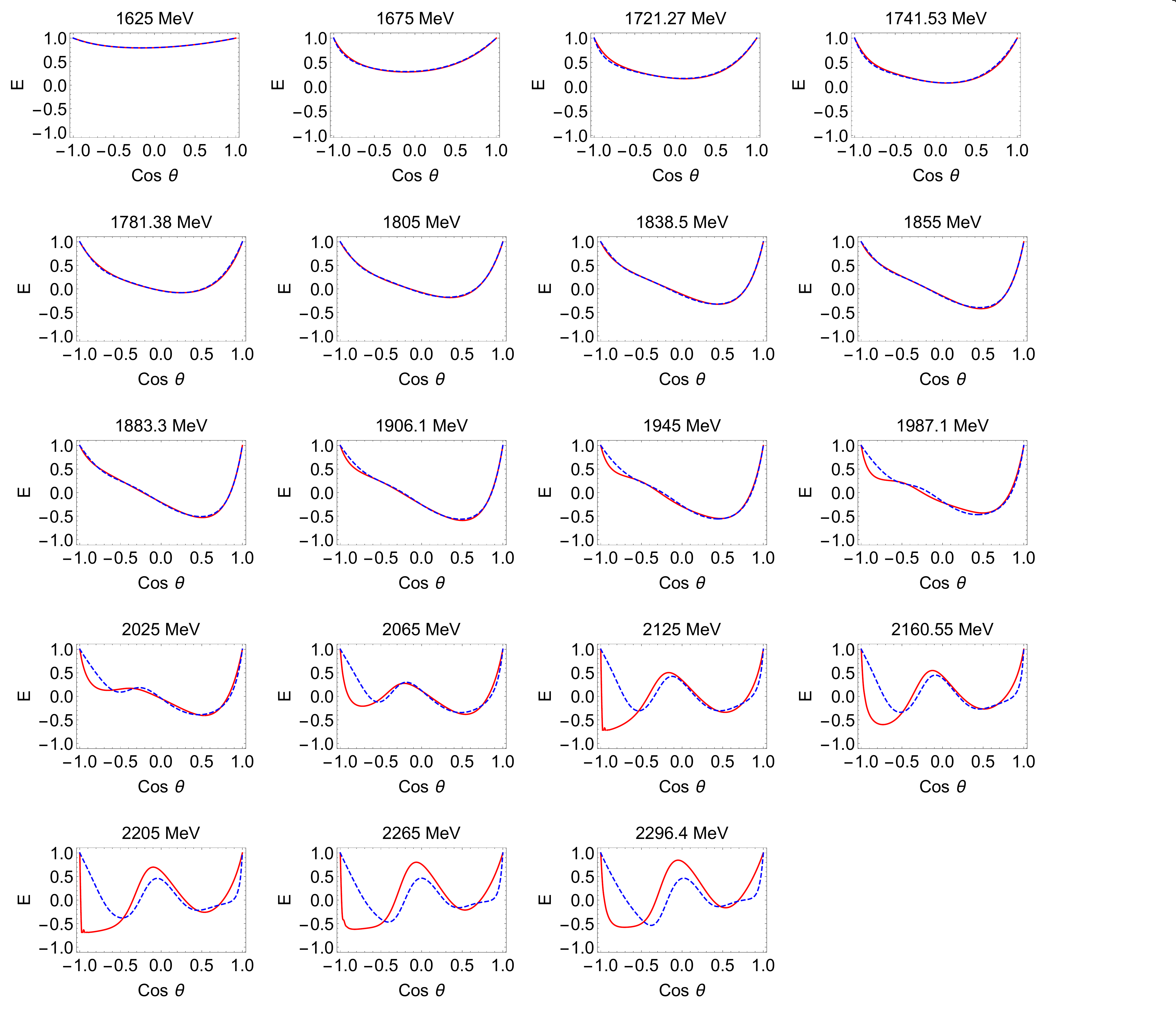} \\
\caption{\label{E-predictions} Prediction of our AA/PWA method (full red line) and the BG2017 fit (blue dashed line) for the $E$
spin observable at representative energies in the full energy range. } \ec
\end{figure}

 \begin{figure}[h!]
\bc
\hspace*{0.5cm}\includegraphics[width=1.0\textwidth]{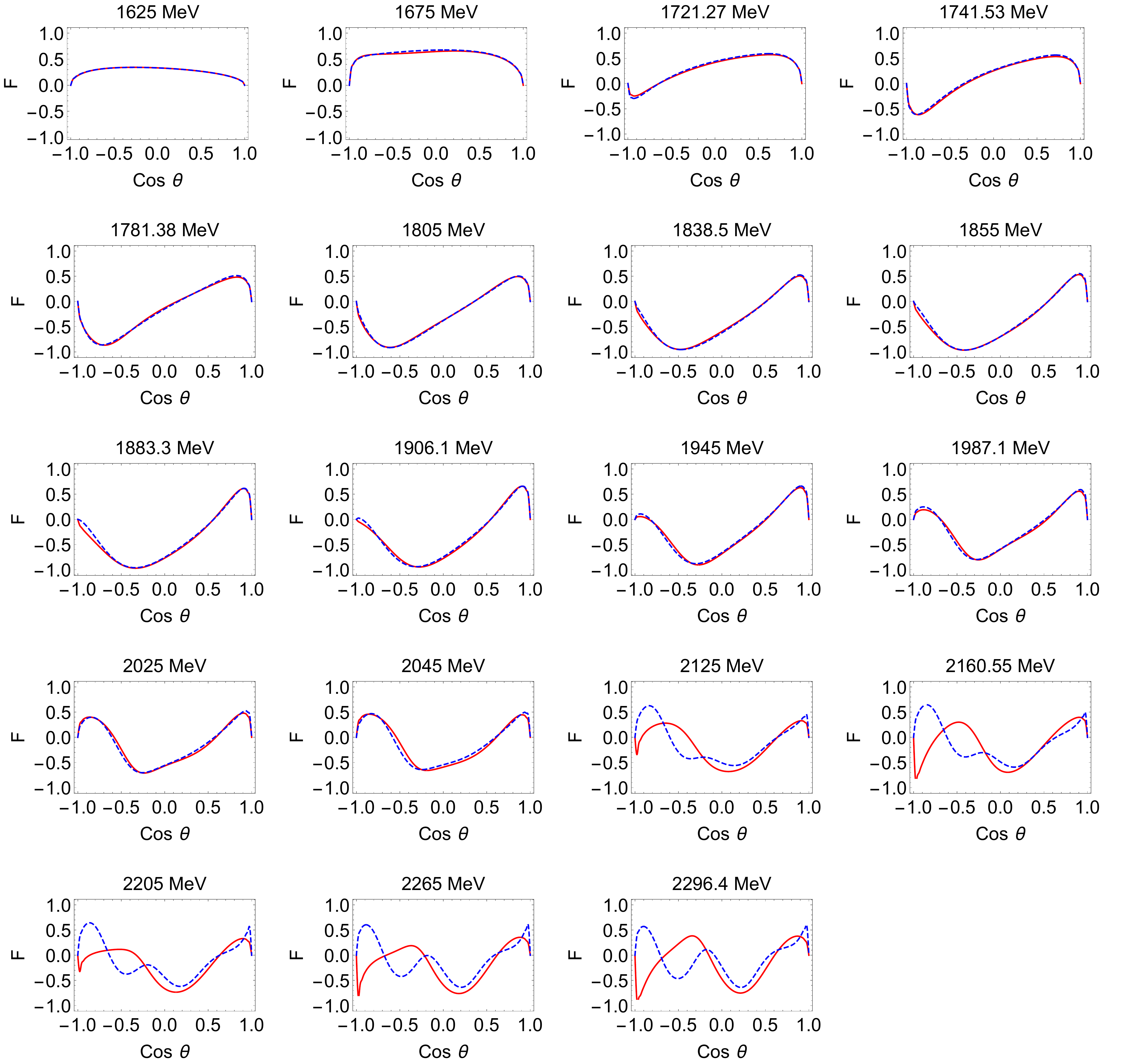} \\
\caption{\label{F-predictions} Prediction of our AA/PWA method (full red line) and the BG2017 fit (blue dashed line) for the $F$
spin observable at representative energies in the full energy range. } \ec
\end{figure}

 \begin{figure}[h!]
\bc
\hspace*{0.5cm}\includegraphics[width=1.0\textwidth]{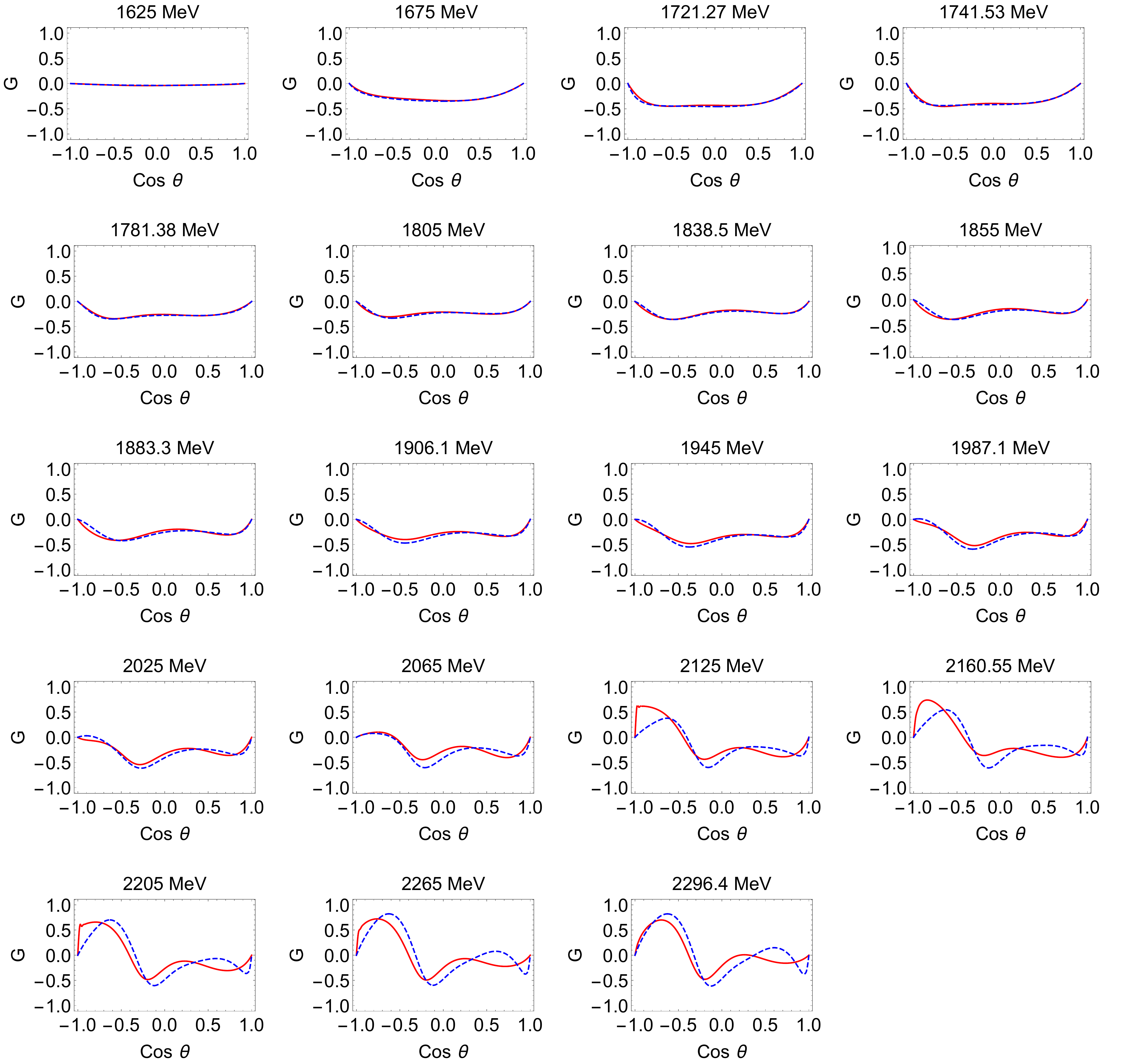} \\
\caption{\label{G-predictions} Prediction of our AA/PWA method (full red line) and the BG2017 fit (blue dashed line) for the $G$
spin observable at representative energies in the full energy range. } \ec
\end{figure}

 \begin{figure}[h!]
\bc
\hspace*{0.5cm}\includegraphics[width=1.0\textwidth]{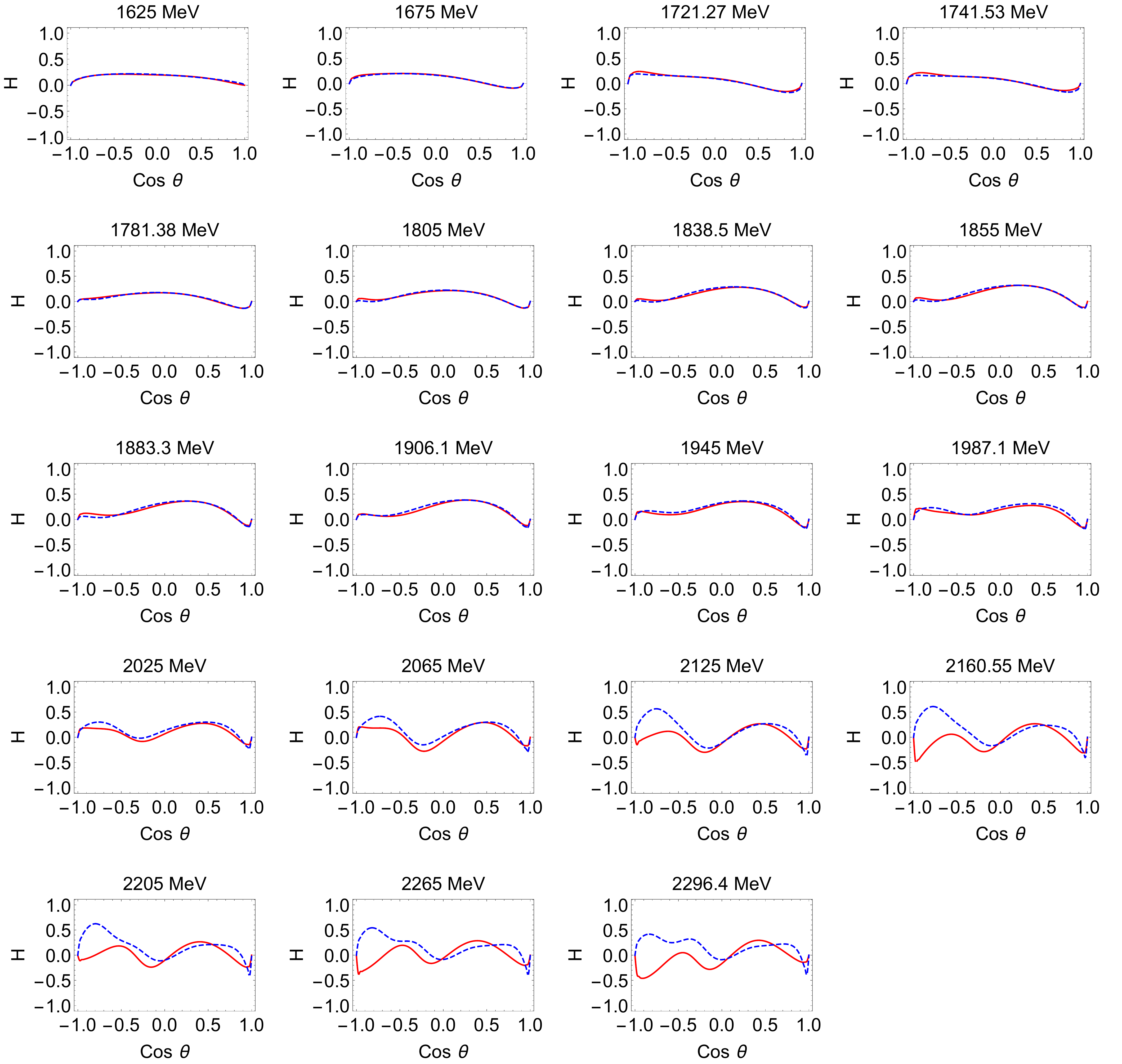} \\
\caption{\label{H-predictions} Prediction of our AA/PWA method (full red line) and the BG2017 fit (blue dashed line) for the $H$
spin observable at representative energies in the full energy range. } \ec
\end{figure}

\clearpage

\indent We have also given predictions for unmeasured ${\cal BT}$ observables $E$, $F$, $G$, and $H$. They show good agreement at
low energies and larger spread at the highest energies. This is, however, also already seen in the ${\cal BR}$ observables that
are fitted. Many observables shown in the paper, like $T$, show sharp structures near 0 and 180 degrees, especially for higher
energies. We consider those structures as natural, since most observables must vanish at those extreme angles, either as
$\sin(\Theta)$ or $\sin^2(\Theta)$. And at higher energies, when lots of multipoles can contribute, the bending towards zero
becomes quite sharp. In the experiment such structures are hard to see because of the small solid angles.

Up to about $2$ GeV the fitted data is practically complete, and further
additional polarization observables will hardly improve the PWA.
Above $2$ GeV and even more above $2.2$ GeV, the number of observables and
also the quality of the data is more limited and additional measurements
of ${\cal BT}$ observables can very well improve the PWA.

\clearpage
\subsection{Discussion} \label{sec:Discussion}
The obtained multipoles given in Figs.~\ref{Multipoles:a} and \ref{Multipoles:b} are very close to the values of the chosen
theoretical model BG2017. This demonstrates the stability of that model, however some additional resonant structures in all
multipoles are made more visible which is to expected as the AA/PWA is significantly improving the ED BG2017 fit for all
observables in the particular channel of $K \Lambda$ photoproduction, see Fig.~\ref{Chi2:Sol1}.
\\ \\ \indent
The AA/PWA multipoles are fairly smooth. However, we would like to warn the reader that we basically have two distinct energy
ranges: the lower one where 8 observables have been measured (1625 MeV $< W <$ 2179 MeV), and the higher one where only 4
observables have been measured (2179 MeV $< W <$ 2296 MeV). As fits are in principle done on individual energies one by one, they
are correlated only through the penalty function, so a change in multipoles due to the change of the number of observables might
be expected (cf. the solution theory discussed in appendix~\ref{sec:SolutionTheory}). The crucial energy where the transition
happens is indicated by the vertical black line at 2179 MeV in all figures. The fits in the lower energy range should tend to be
smoother, and more constrained, while some visible changes might occur at higher energies. This produces discontinuities. As the
AA/PWA method gives a set of smooth multipoles for a self-consistent and complete dataset by forcing the reaction-amplitude
phases to be smooth (this has been shown in ref.~\cite{Svarc2018}), this indicates that the remaining discontinuities are the
result of an inconsistency of the database. So, the AA/PWA method offers a possibility to test the self-consistency of the
experimental database. However, this discussion can be reliably performed only when a confident method for pole detection is
used, so one should in principle get some answers with the use of Laurent plus Pietarinen (L+P)
formalism~\cite{L+P2015,Svarc2013,Svarc2014,Svarc2014a}.

\subsection{Comparison of AA/PWA with BG SC-SE-PWA of ref.~\cite{Anisovich2017}} \label{sec:Comparison-with-BG}
\begin{figure}[h!]
\bc
\includegraphics[width=0.45\textwidth]{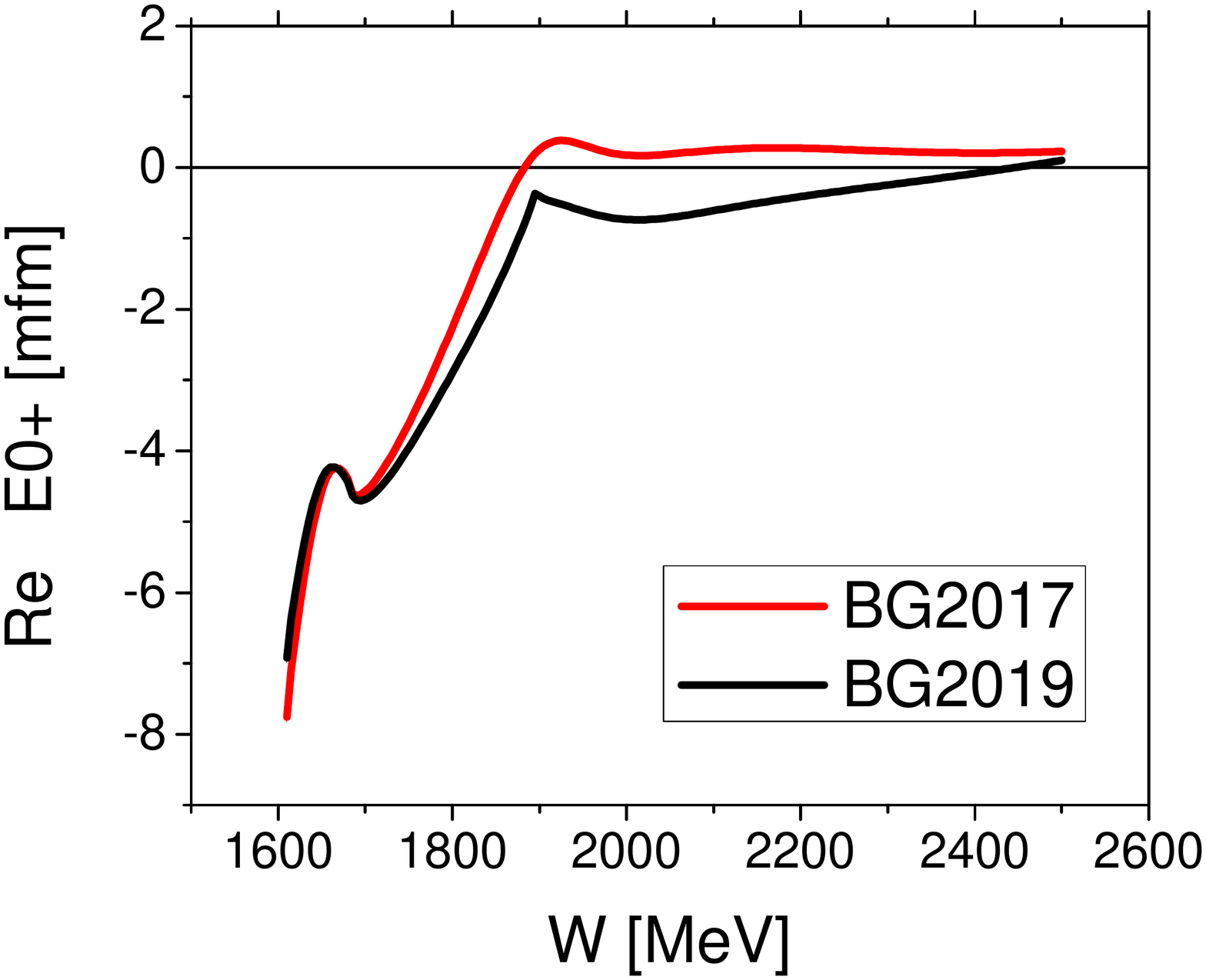} \hspace{0.5cm}
\includegraphics[width=0.45\textwidth]{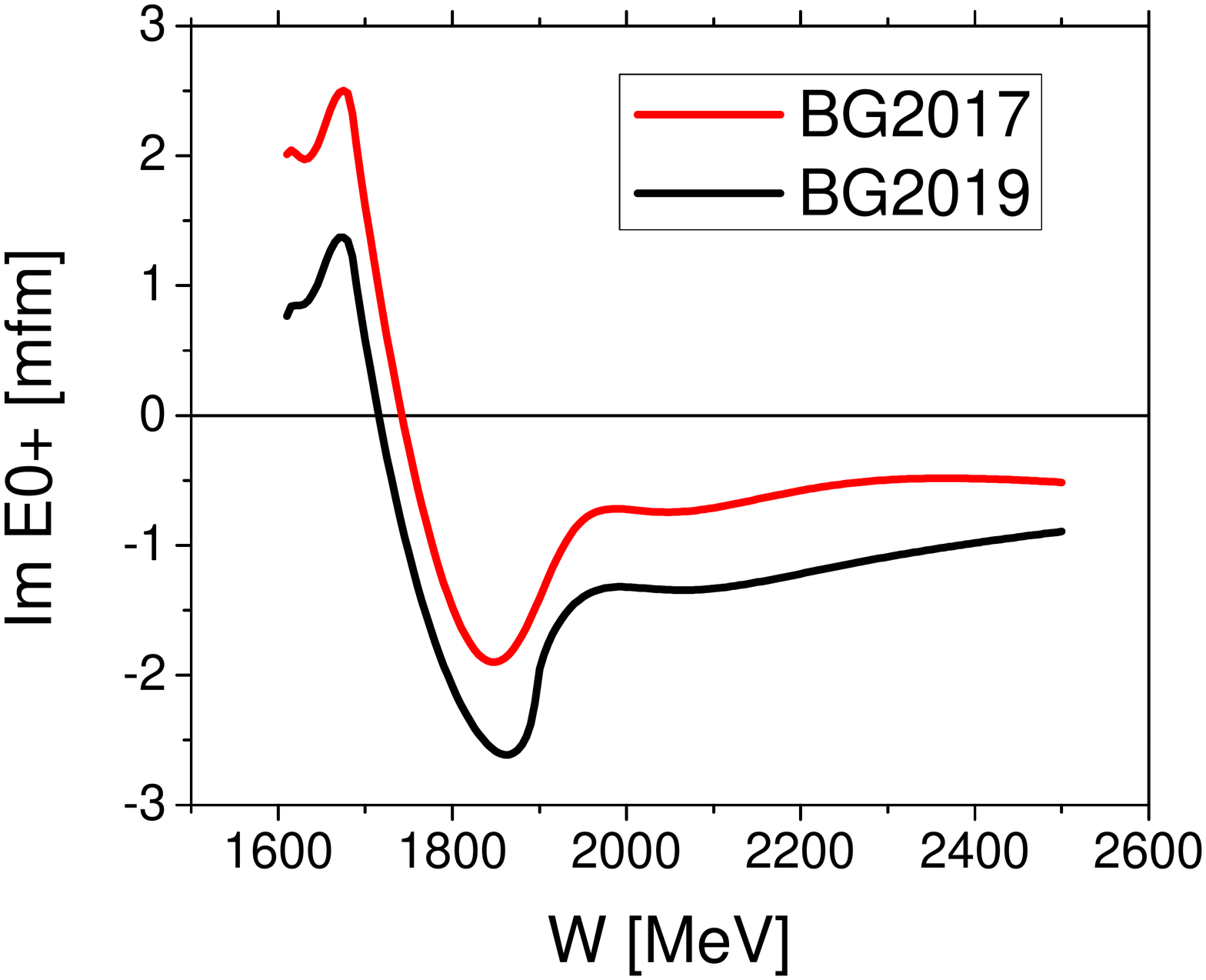} \hspace{0.5cm}
\caption{\label{E0pComparison}(Color online) A comparison of $E_{0+}$ BG ED solution BG2017 from ref.~\cite{Anisovich2017} (red
line) which was used in this publication, and the BG2019 solution given in BG2019~\cite{BG-web} (black line). } \ec
\end{figure}

For relatively new methods such as AA/PWA, a comparison with old, double-checked and worldly accepted models is crucial. Such an
opportunity is offered to us by the Bonn-Gatchina group. They have performed SE-PWA and used the standard constrained PWA method
(letting lower multipoles free, and strongly penalizing higher ones to theoretical ED values, cf.
section~\ref{sec:SummaryPWAAA}). This enabled us to make a direct comparison of our results with theirs. As we mentioned before,
their constraining ED solution was, unfortunately, not made public, but differed notably from the values of the published old
BG2014-02 and new BG2019 solutions~\cite{BG-web} (we show the difference for the $E_{0+}$  multipole in
Fig.~\ref{E0pComparison}). However, upon request, they provided us with the exact numbers~\cite{Anisovich-priv-com} for all
multipoles. So, as our AA/PWA method required exactly the same input (data base and theoretical ED multipoles), and we have
achieved that, we used the chance to compare the results directly.

\begin{figure}[h!]
\bc
\includegraphics[width=0.37\textwidth]{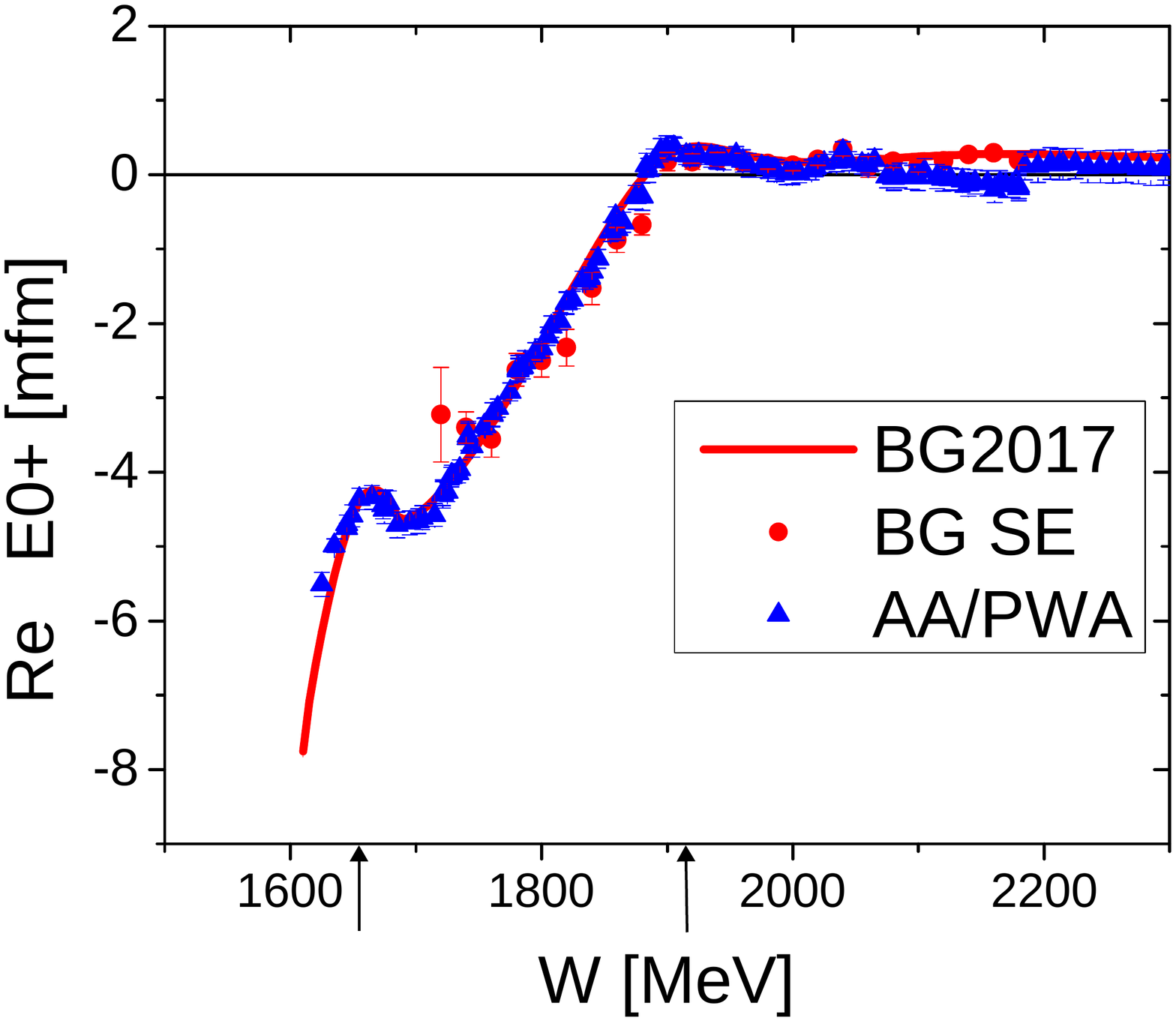} \hspace{0.5cm}
\includegraphics[width=0.37\textwidth]{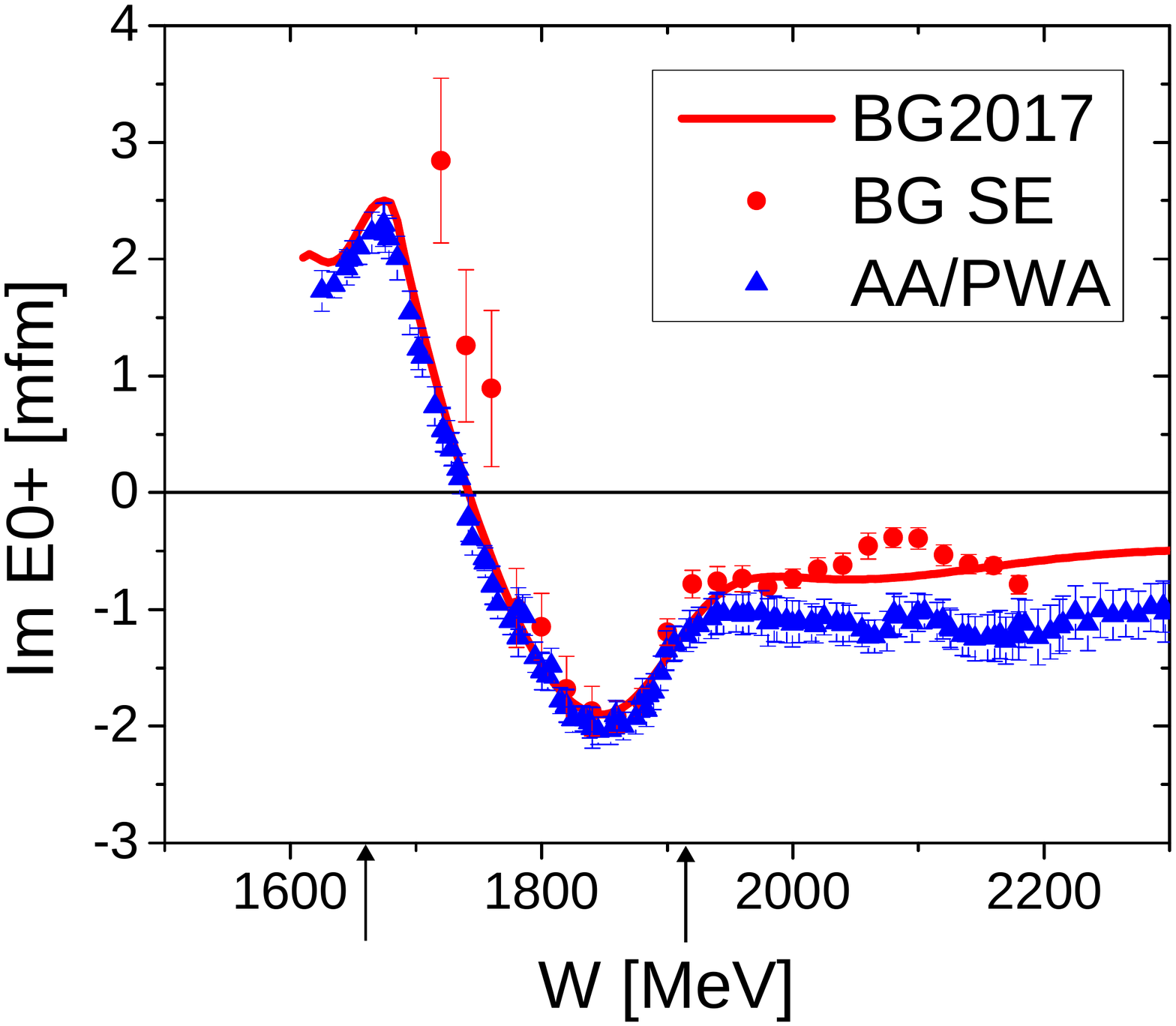}  \\
\includegraphics[width=0.37\textwidth]{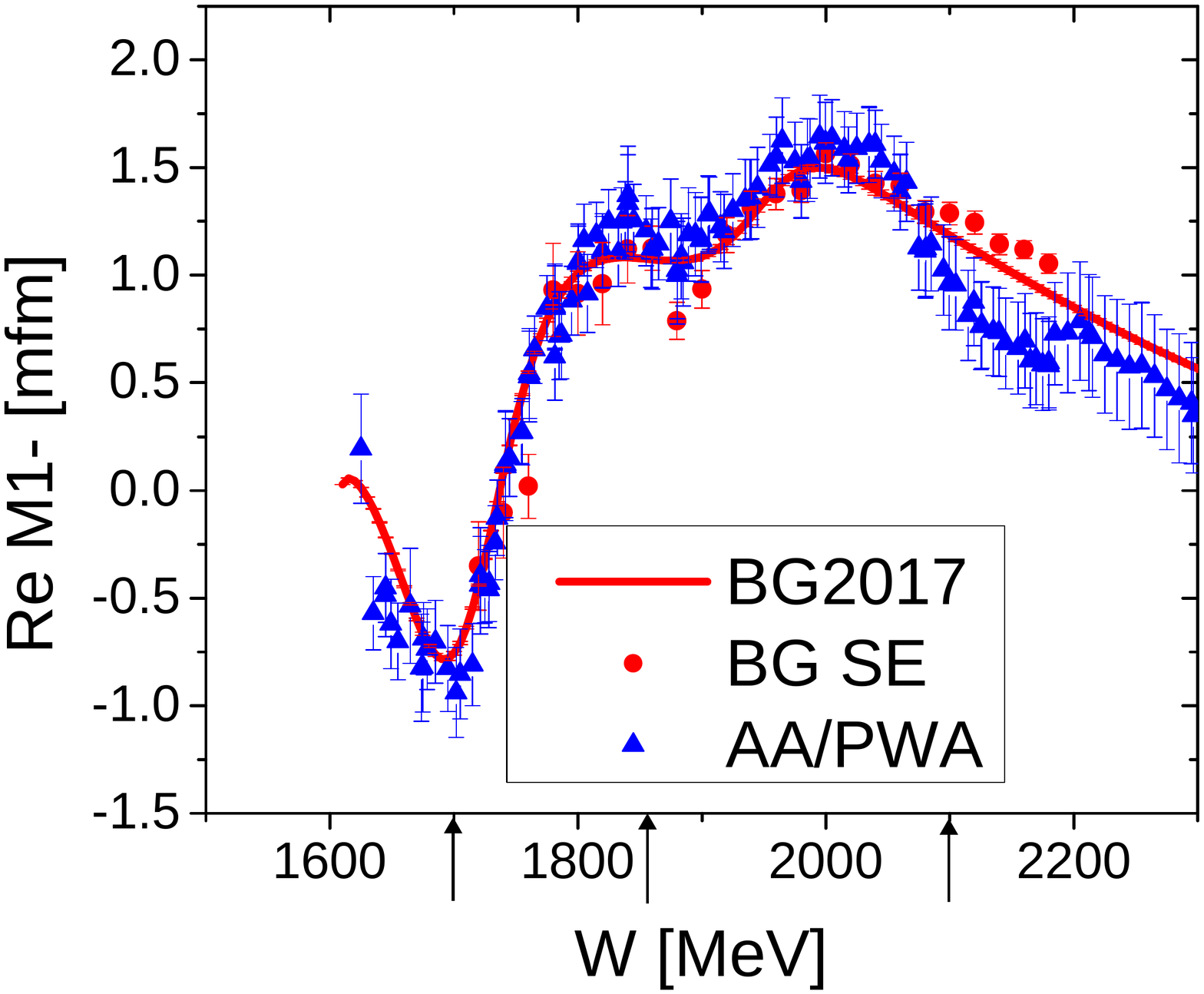} \hspace{0.5cm}
\includegraphics[width=0.37\textwidth]{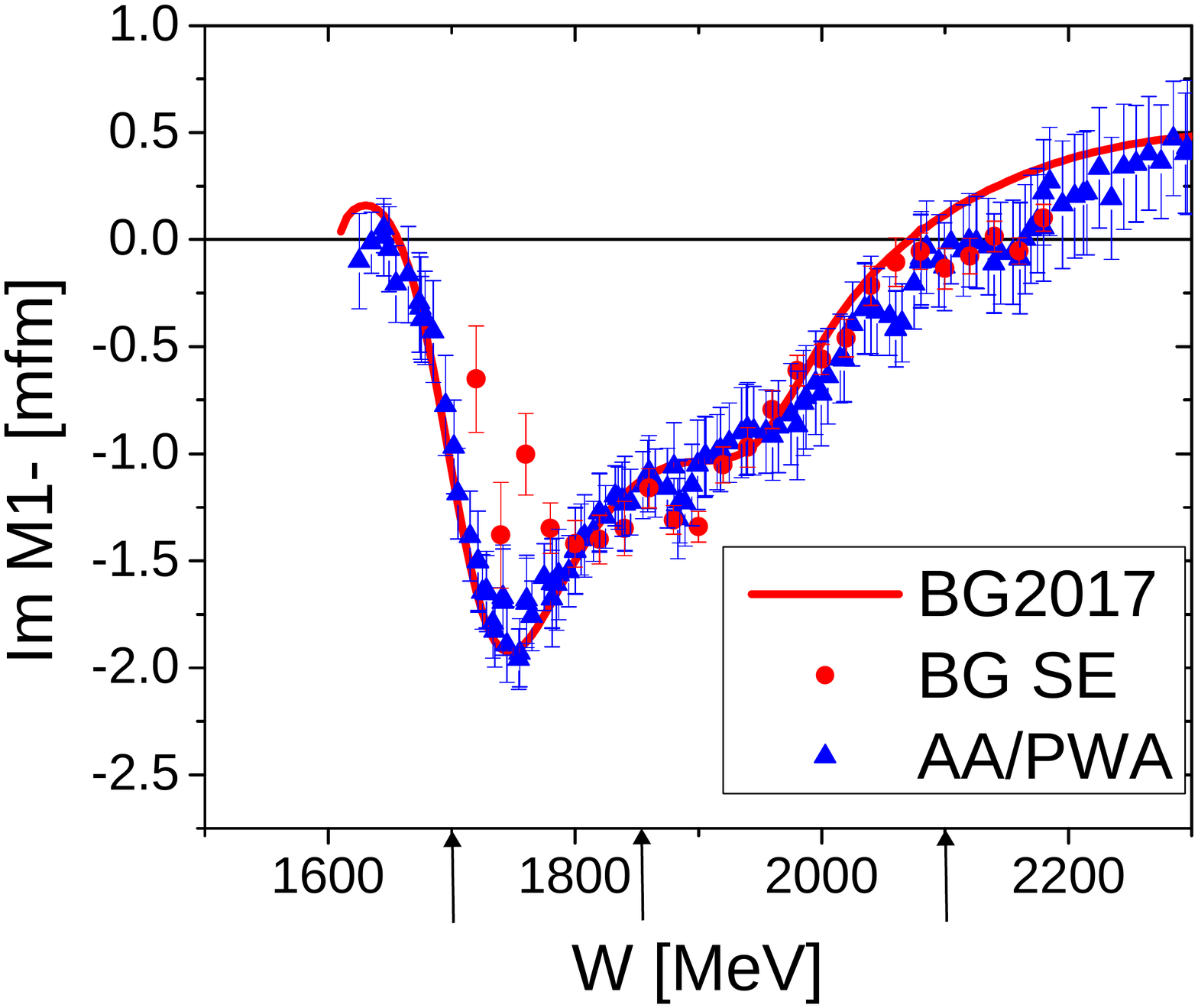}
\includegraphics[width=0.37\textwidth]{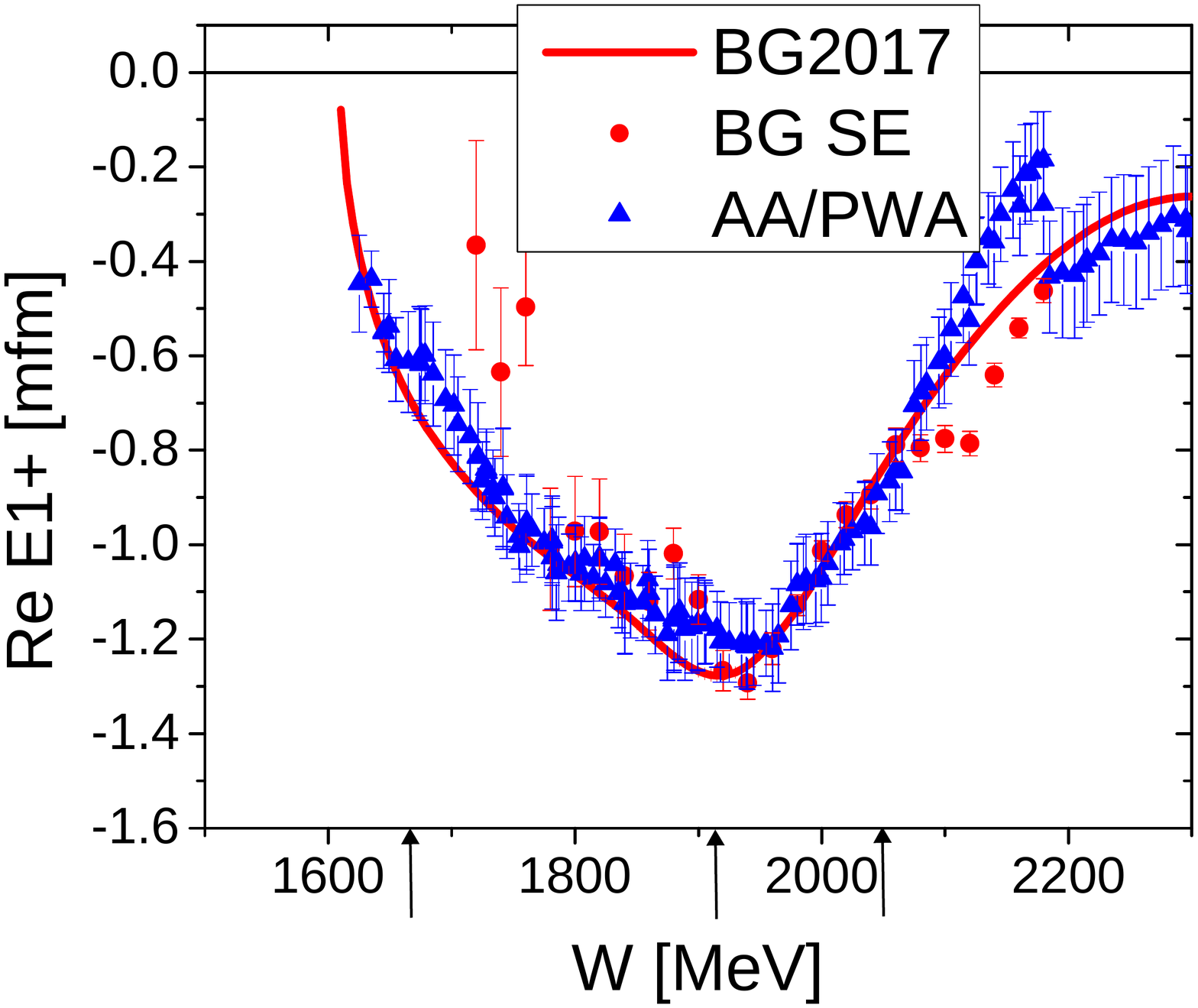} \hspace{0.5cm}
\includegraphics[width=0.37\textwidth]{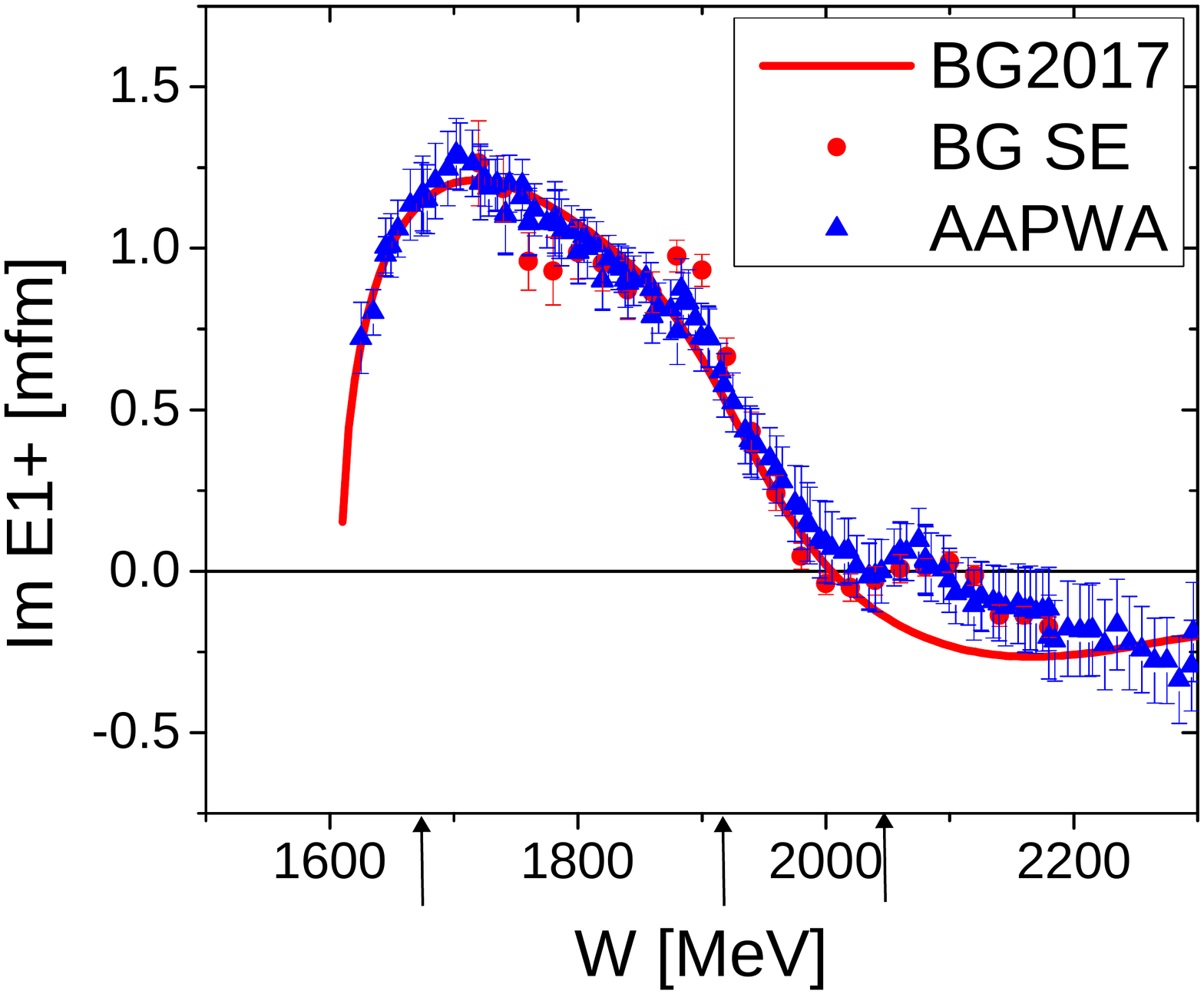}  \\
\includegraphics[width=0.37\textwidth]{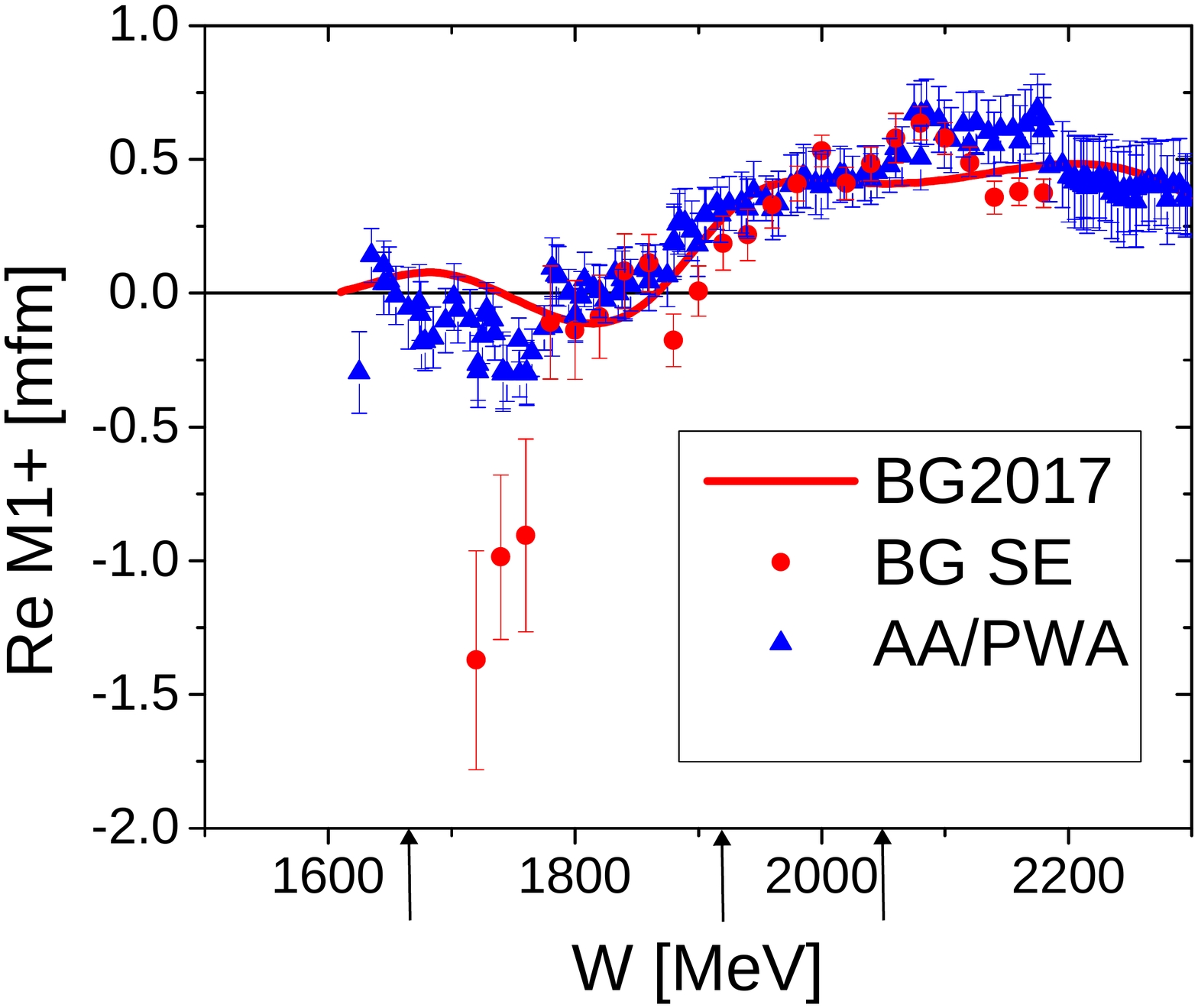} \hspace{0.5cm}
\includegraphics[width=0.37\textwidth]{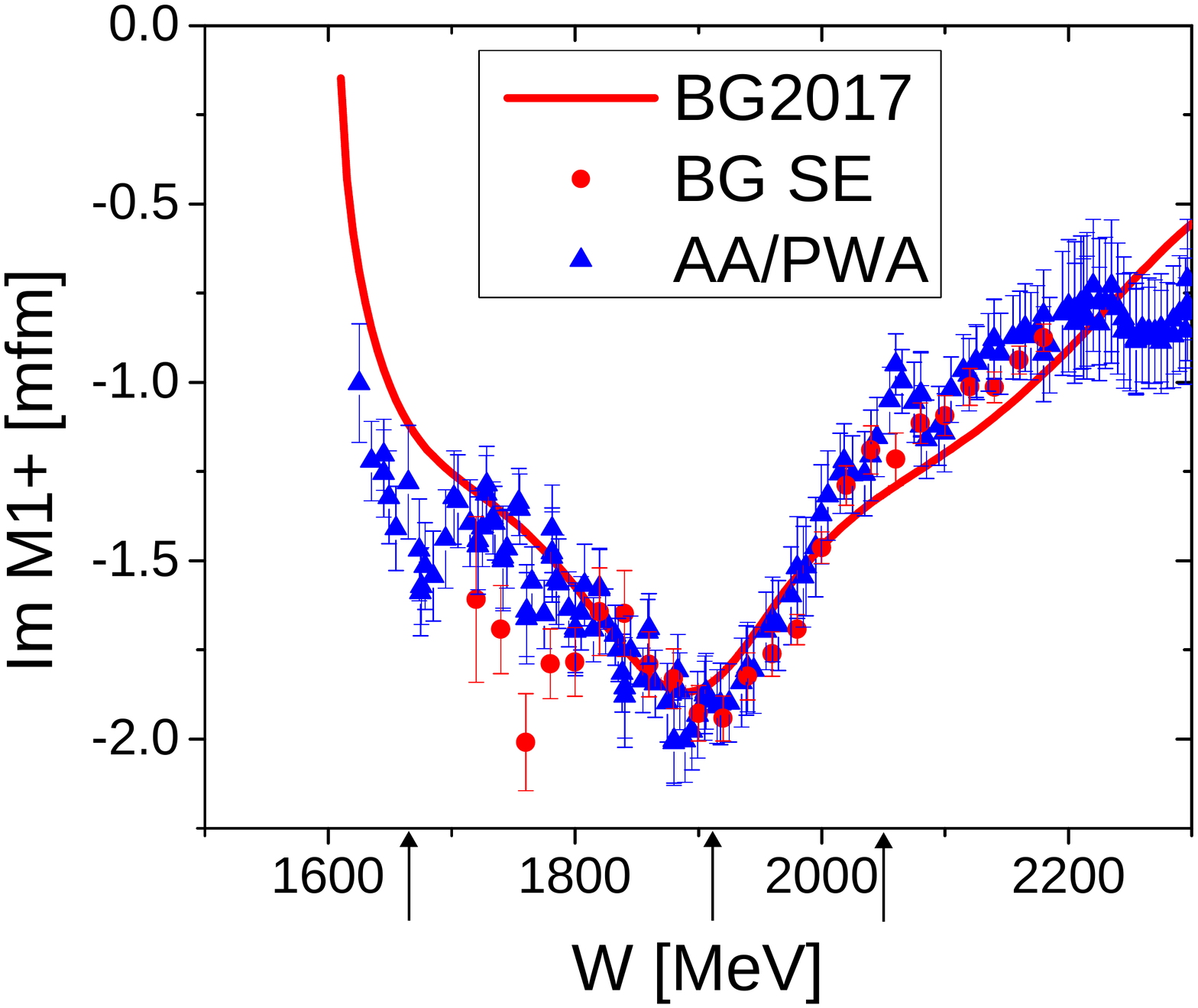}  \\
\caption{\label{MultipolesComparison}(Color online)  Comparison of $E_{0+}$, $M_{1-}$, $E_{1+}$ and $M_{1+}$ SC-SE-BG PWA
solutions presented in ref.~\cite{Anisovich2017} (red circles) and  the same multipoles obtained using the AA/PWA method in this
publication (blue triangles). The red line is the BG2017 ED solution used in the SC-SE-BG PWA publication, see
ref.~\cite{BG-web}. Black arrows on the horizontal axis mark the pole masses of nucleon resonances as given by PDG~\cite{PDG}. }
\ec
\end{figure}

We show in Fig.~\ref{MultipolesComparison} the result for the four lower multipoles, where the SE results of
ref.~\cite{Anisovich2017} only exist.

\clearpage We see a very good agreement between both solutions regarding the overall absolute values and functional shape. Both
solutions are also quite close to the constraining BG2017 ED solution.
\\ \\ \indent
We see that the AA/PWA method results in many more points. This is a result of the data-preparation philosophy as the AA/PWA
method uses data interpolation instead of data binning as the Bonn-Gatchina SE-PWA does.
\\ \\ \indent
We see some discrepancy in the absolute values of some multipoles at lower energies, but our values are closer to the BG2017
values. We conclude that both SE methods are in full agreement.
\\ \\ \indent
The advantage of the AA/PWA method is obvious: first we generate much many more data points in reconstruction multipoles, and
second we generate all multipoles, and not only the lowest ones.
\\ \\ \indent
The main issue is to answer the question whether we reproduce the structures in the $M_{1-}$ multipole around 1890 MeV, which
were interpreted in ref.~\cite{Anisovich2017} as a confirmation of the $N(1880) \frac{1}{2}^{+}$ resonance. The answer is
definitely affirmative: Yes, we do. However, the "size" of the structure is not so pronounced, so we need a detailed L+P analysis
to confirm or dismiss this interpretation. This will be done in the forthcoming chapter of this paper.

\section{Extraction of resonance parameters from the single-energy multipoles} \label{eq:PhysicalResultsFromPWAAA}
Extracting poles directly from SE partial-wave solutions is very difficult, and the only method which showed quite some
flexibility and confidence is the L+P method~\cite{PDG,L+P2015,Svarc2013,Svarc2014,Svarc2014a}. So, we use this method to analyze
the crucial multipole $M_{1-}$. However, for the convenience of the reader, we repeat the essence of the method in section
\ref{sec:SummaryLPlusP}.

\subsection{The L+P method} \label{sec:SummaryLPlusP}
The driving concept behind the Laurent-Pietarinen expansion (L+P) was the aim to replace elaborate theoretical models by a local
power-series representation of partial wave amplitudes~\cite{Svarc2013}. The complexity of any reaction-theoretical model which
leads to partial waves is thus replaced by a much simpler model-independent expansion which just exploits analyticity. The L+P
approach separates the pole- and regular parts in the form of a Laurent expansion, and instead of modeling the regular part by
some physical model it uses a conformal mapping to expand it into a rapidly converging power series with well defined analytic
properties. In such an approach the model dependence is minimized, and is reduced to the choice of the number and location of
branch points used in the L+P expansion.

The L+P expansion is based on the Pietarinen expansion used in some older papers in the analysis of pion-nucleon scattering data
\cite{Ciulli,CiulliFisher,Pietarinen,Pietarinen1}, but for the L+P model the Pietarinen expansion is applied in a different
manner. It exploits the Mittag-Leffler theorem\footnote{Mittag-Leffler theorem \cite{Mittag-Leffler}: This theorem provides the
generalization of a Laurent expansion to a more-than-one pole situation. For simplicity, we will simply refer to this as a
Laurent expansion.} of partial wave amplitudes near the real energy axis, representing the regular, but unknown, background term
by a conformal-mapping-gener\-ated, rapidly converging power series called a Pietarinen expansion\footnote{A conformal-mapping
expansion of this particular type was introduced by Ciulli and Fisher \cite{Ciulli,CiulliFisher}. It was described in detail and
used in pion-nucleon scattering by Esco Pietarinen \cite{Pietarinen,Pietarinen1}. The procedure was denoted as a Pietarinen
expansion by G. H\"{o}hler in \cite{Hoehler84}.}. The method was used successfully in several few-body reactions
\cite{Svarc2014,Svarc2014a,L+P2015}, and recently generalized to the multi-channel case \cite{Svarc2016}. The formulae used in
the L+P approach are collected in Table~\ref{LplusP}.

\begin{table*}[h!]
\caption{\label{LplusP}Formulae defining the Laurent plus Pietarinen (L+P) expansion (see
ref.~\cite{Anisovich2017a}).\vspace{-5mm}} {\footnotesize
\begin{align*}
\label{eq:Laurent-Pietarinen}
T^a(W) &=& \sum _{j=1}^{{N}_{pole}} \frac{x^{a}_{j} + \imath \, \, y^{a}_{j}  }{W_j-W} +
      \sum _{k=0}^{K^a}  c^a_k \, X^a (W)^k  +  \sum _{l=0}^{L^a} d_l^a \, Y^a (W)^l +  \sum
_{m=0}^{M^a} e_m^a \, Z^a (W)^m \nonumber\hspace{50mm} \\
X^a (W )&=&  \frac{\alpha^a-\sqrt{x_P^a-W}}{\alpha^a+\sqrt{x_P^a - W }}; \, \, \, \, \,   Y^a(W ) =  \frac{\beta^a-\sqrt{x_Q^a-W }}{\beta^a+\sqrt{x_Q^a-W }};  \, \, \, \, \,
Z^a(W ) =  \frac{\gamma^a-\sqrt{x_R^a-W}}{\gamma^a+\sqrt{x_R^a-W }}
 \nonumber \hspace{42mm}\\
  D_{dp}^a \ \ \ \ \ & = & \frac{1}{2 \, N_{W}^a - N_{par}^a} \, \, \sum_{i=1}^{N_{W}^a} \left\{ \left[ \frac{{\rm Re} \,T^{a}(W^{(i)})-{\rm Re} \, T^{a,exp}(W^{(i)})}{ Err_{i,a}^{\rm Re}}  \right]^2 + \right.
            \left. \left[ \frac{{\rm Im} \, T^{a}(W^{(i)})-{\rm Im} \, T^{a,exp}(W^{(i)})}{ Err_{i,a}^{\rm Im}} \right]^2 \right\}  + {\cal P}^a  \nonumber\hspace{3mm} \\
{\cal P}^{a}\ \ \ \ \  &=& \lambda_c^a \, \sum _{k=1}^{K^a} (c^a_k)^2 \, k^3 +
\lambda_d^a \, \sum _{l=1}^{L^a} (d^a_l)^2 \, l^3 +  \lambda_e^a \,
\sum _{m=1}^{M^a} (e^a_m)^2 \, m^3 \qquad\qquad
 D_{dp}  =  \sum _{a}^{all}D_{dp}^a \nonumber \hspace{40mm}\\
 &&a\, \, .....   \, \,    {\rm  channel \, \,  index} \qquad\qquad
  N_{pole}\; .....\;      {\rm number\; of\; poles} \qquad\qquad
  W_j,W \in \mathbb{C}  \nonumber \hspace{42mm}\\
&& x_i^a, \, y_i^a, \, c_k^a, \, d_l^a, \, e_m^a, \, \alpha^a, \, \beta^a, \, \gamma^a ... \in  \mathbb{R}  \, \,
 \nonumber  \hspace{96mm}\\
 &&  K^a, \, L^a, \, M^a \, ... \, \in  \mathbb{N} \, \, \, {\rm number \, \, of \, \, Pietarinen \, \, coefficients \, \, in \, \, channel \, \, \mathit{a} }.
 \nonumber \hspace{53mm}\\
 && D_{dp}^a \; .....  \; {\rm discrepancy \; function \; in\; channel \; }a \qquad\qquad \hspace{12mm}
N_{W}^a \; .....  \; {\rm number\; of\; energies\; in\; channel \; }a \nonumber \hspace{8mm}\\
 && N_{par}^a \; .....  \; {\rm number\; of\; fitting \; parameters \; in\; channel \; }a
\qquad\qquad {\cal P}^a \, \,  .....   \, \, {\rm Pietarinen  \, \, penalty \, \, function}
\nonumber \hspace{14mm}\\
 && \lambda_c^a, \, \lambda_d^a, \,\lambda_e^a \, \, \,  .....   \, \, {\rm Pietarinen  \, \,  weighting \, \, factors} \nonumber \qquad\qquad\hspace{10mm} x_P^a, \, x_Q^a, \, x_R^a  \in \mathbb{R}
 \, \, \, \,( {\rm  or} \, \, \in \mathbb{C}). \nonumber  \hspace{26mm}
\\
&& Err_{i,a}^{\rm Re, \, Im} ..... {\rm \, \, minimization \, \, error\, \, of \, \, real \,\, and \, \, imaginary \, \, part \, \, respectively.} \nonumber \hspace{45mm}
\end{align*}
}%
\end{table*}

In the fits, the regular background part is represented by three Pietarinen series and all free parameters are fitted. The first
Pietarinen expansion with branch point $x_P$ is restricted to an unphysical energy range and collectively represents all
left-hand singularities. The next two Pietarinen expansions describe the background in the physical energy-range with branch
points $x_Q$ and $x_R$ respecting the analytic properties of the analyzed partial wave. The second branch point is in most cases
fixed to the elastic channel branch point, the third one is either fixed to the dominant inelastic threshold, or left free. Thus,
only rather general physical assumptions about the analytic properties are made like the number of poles and the number and the
position of branch points, and the simplest analytic function with the correct set of poles and branch points is constructed. The
method is applicable to both, theoretical and experimental input, and represents the first reliable procedure to extract pole
positions from experimental data, with minimal model bias.

The generalization of the L+P method to a multichannel L+P method is performed in the following way: i)~separate Laurent
expansions are made for each channel; ii)~pole positions are fixed for all channels, iii)~residues and Pietarinen coefficients
are varied freely; iv)~branch points are chosen as for the single-channel model; v)~the single-channel discrepancy function
$D_{dp}^a$ (see Eq. (5) in ref. \cite{L+P2015}) which quantifies the deviation of the fitted function from the input is
generalized to a multi-channel quantity $D_{dp}$ by summing up all single-channel contributions, and vi)~the minimization is
performed for all channels in order to obtain the final solution.

\subsection{Detailed analysis of the $M_{1-}$ multipole using the L+P method} \label{sec:EnergyDependentFitExampleMultipole}

\begin{figure}[h!]
\hspace*{0.5cm}
 \includegraphics[width=0.1\textwidth]{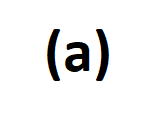} \hspace*{4.2cm}
\includegraphics[width=0.1\textwidth]{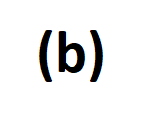} \hspace*{4.2cm}
\includegraphics[width=0.1\textwidth]{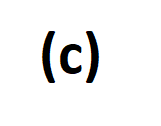} \\
\bc
\includegraphics[width=0.3\textwidth]{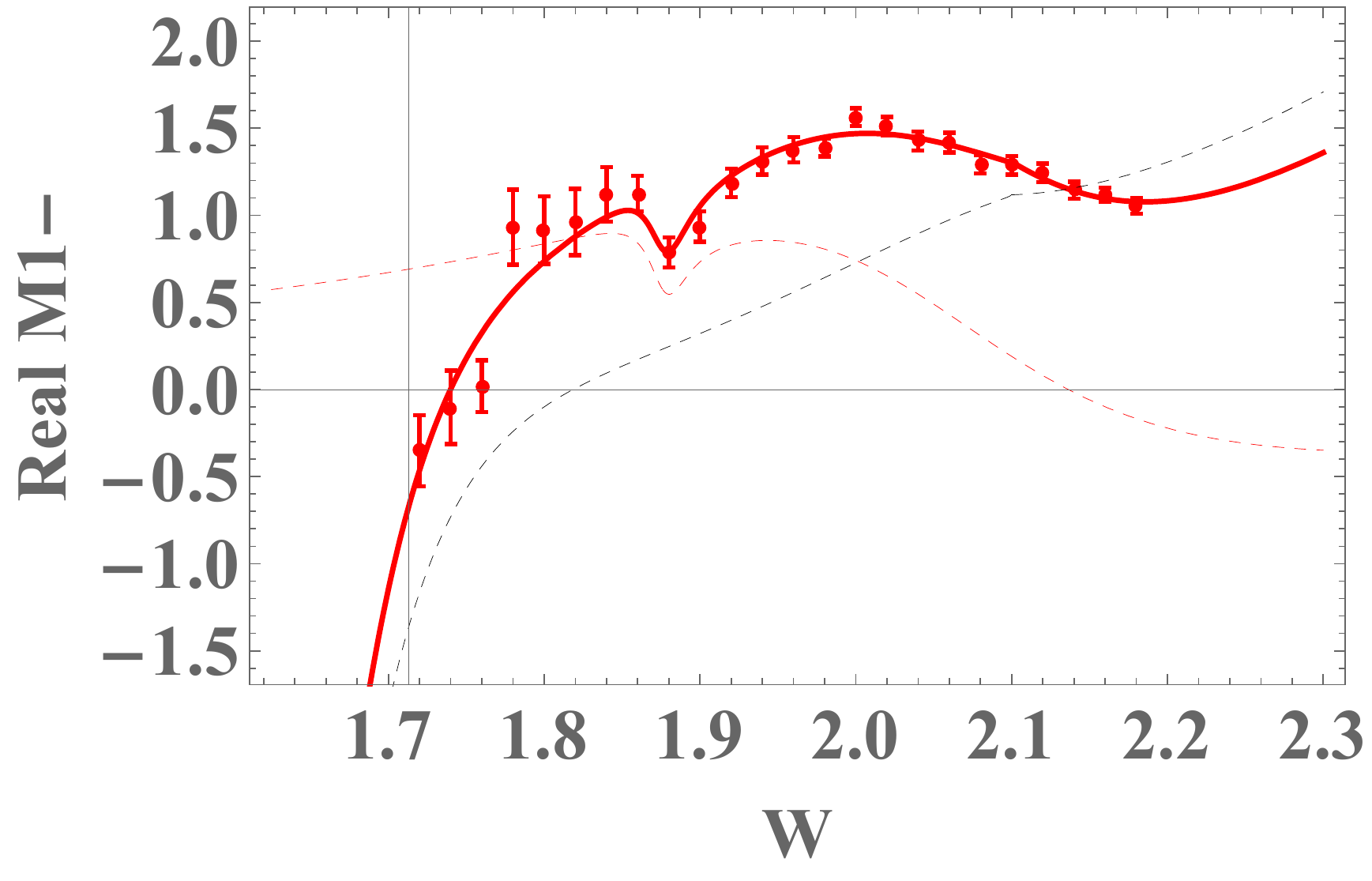} \hspace{0.5cm}
\includegraphics[width=0.3\textwidth]{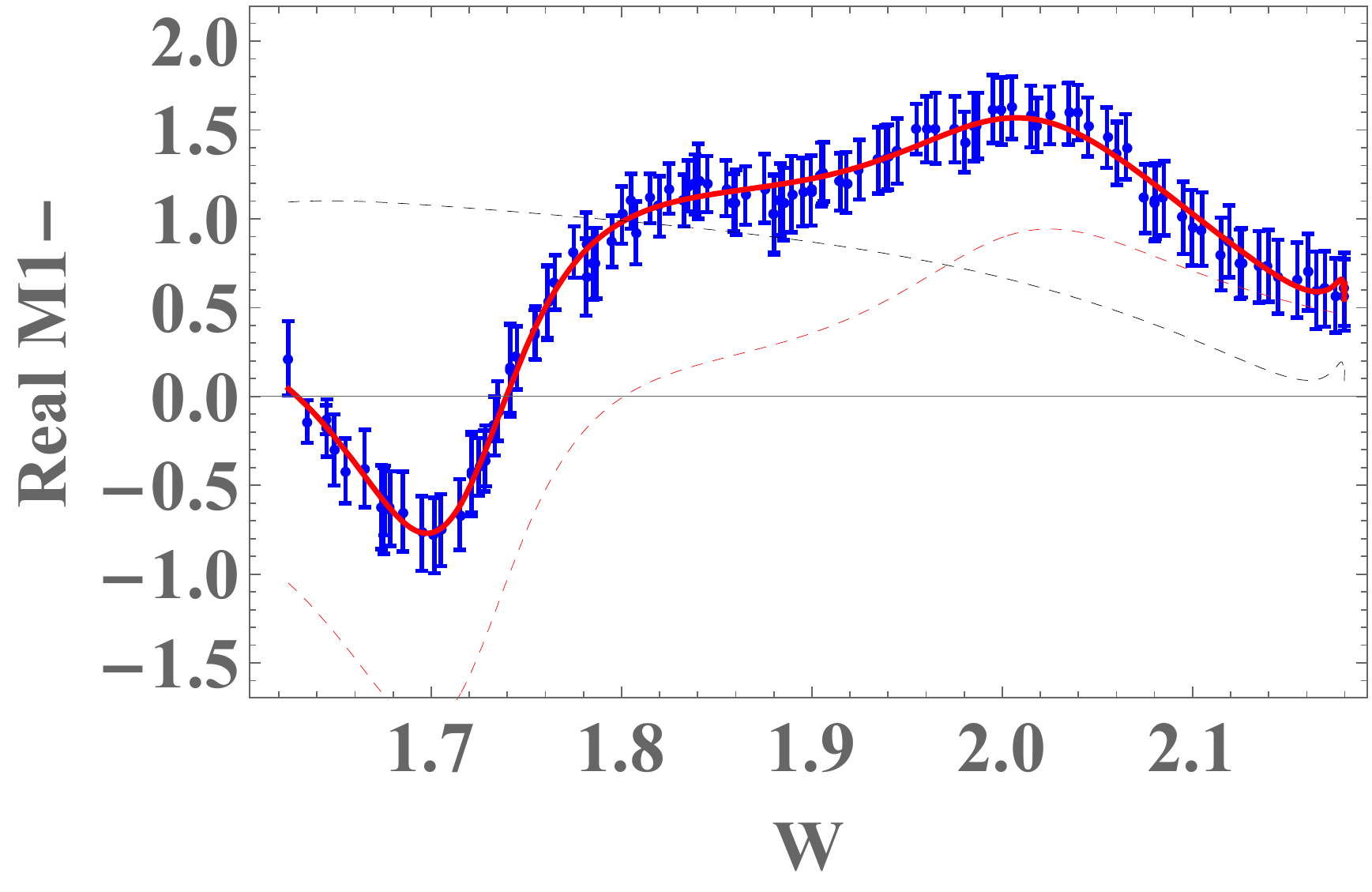} \hspace{0.5cm}
\includegraphics[width=0.3\textwidth]{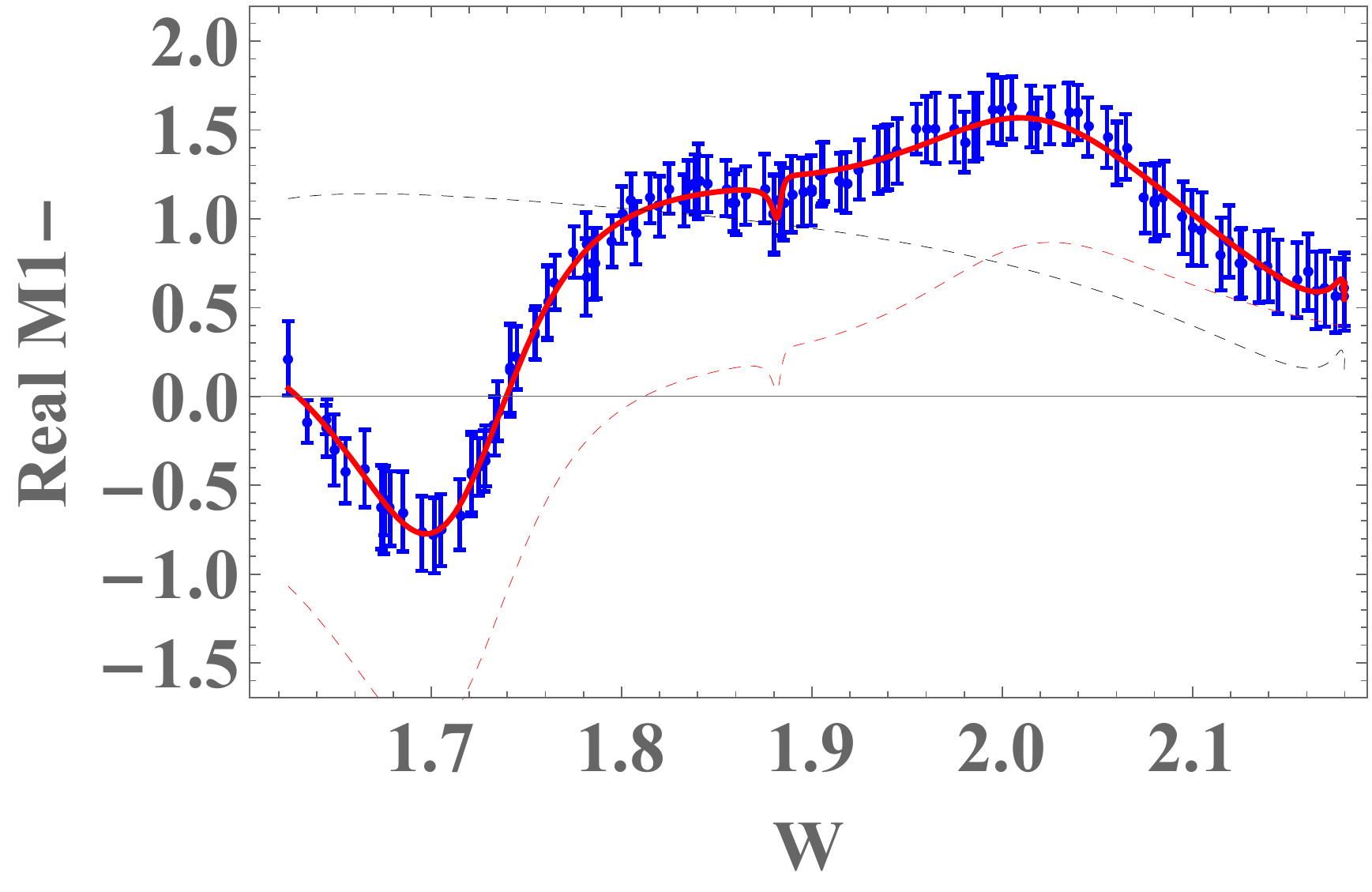} \\
\includegraphics[width=0.3\textwidth]{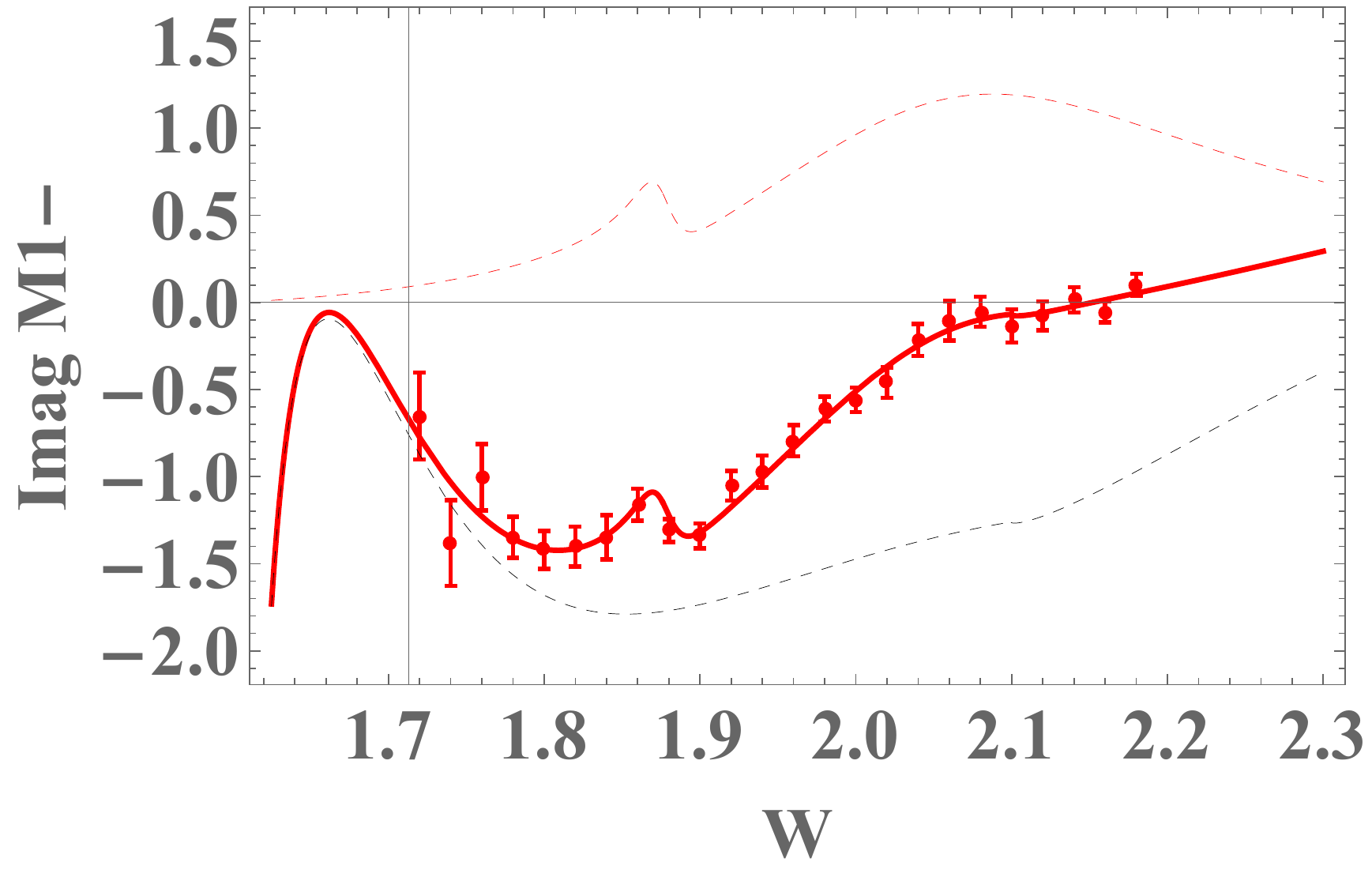} \hspace{0.5cm}
\includegraphics[width=0.3\textwidth]{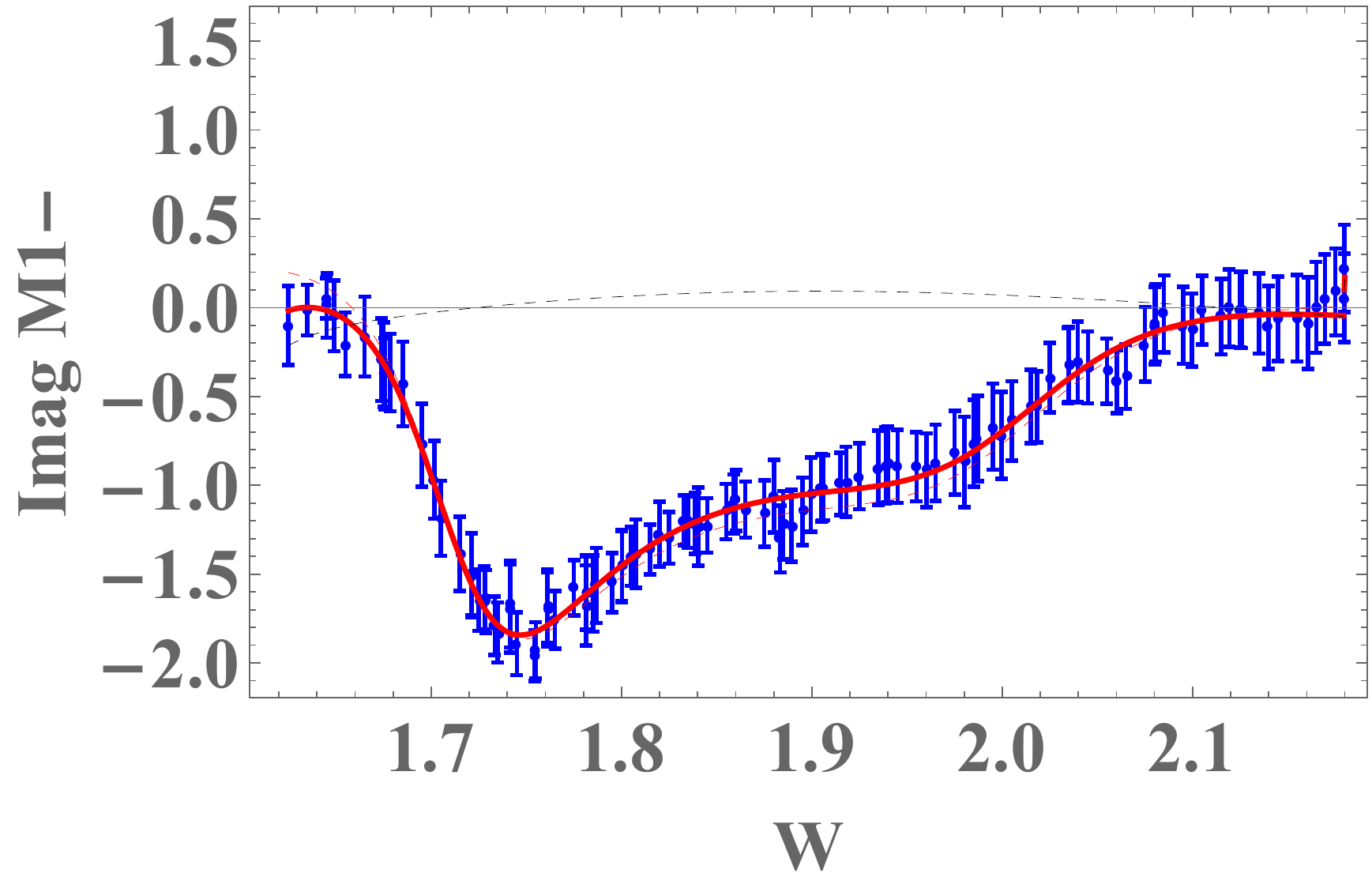} \hspace{0.5cm}
\includegraphics[width=0.3\textwidth]{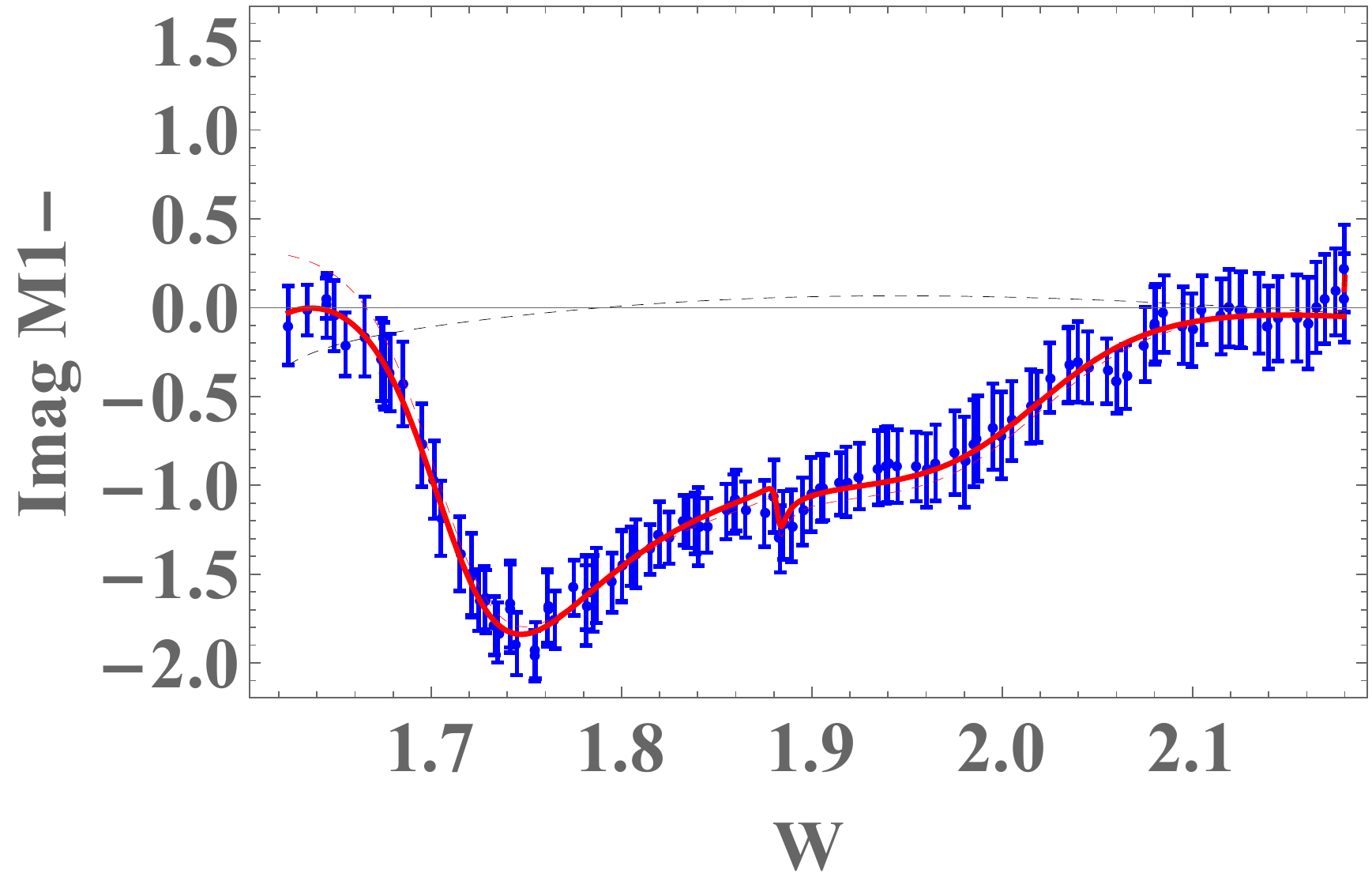}
\caption{\label{L+P}(Color online) The L+P fit of the $M_{1-}$ multipole. In Fig.~(a) we show the result of the single-channel
fit of BG-SE data, in Fig.~(b) we show the result of the 2-pole fit of our AA/PWA solution, and in Fig.~(c) we show the result of
its 3-pole fit. Red and blue symbols are single-energy BG-SE and AA/PWA solutions, the thin dashed red line is the resonant
contribution, thin dashed black line is the background part, and thick red line is the result of the full fit.} \ec
\end{figure}

The only reliable way to establish whether the structure seen in the $M_{1-}$ multipole corresponds to a resonance pole or not is
to use the L+P formalism, and try to find any analytic function with an explicit pole and realistic background which fits the
data. If we find it, then we can claim that the observed structure is at least consistent with a function having a pole. Very
often, especially for narrow or small resonances, such a function cannot be found, so this is an indication that the observed
structure is originating not from a pole, but from some other effect possible in the method.
\\ \\ \indent
In ref.~\cite{Anisovich2017} the L+P formalism was used to analyze the obtained result, and it was found that a function
containing a pole of mass $M=1876$ MeV and a width of $\Gamma=33$ MeV can describe the data well, so in spite of the fact that
the width was rather narrow for the formerly found state $N(1880) \frac{1}{2}^{+}$ (33(19) MeV corresponding to previously
established 230(50) MeV) it was interpreted as a signal of a new resonant state. Unfortunately, in the present paper we do not
confirm this finding.
\\ \\ \indent
In this paper we have performed an L+P analysis of our AA/PWA solution. Instead of using the whole energy range up to 2295 MeV,
from our analysis we have omitted the high-energy part above 2179 MeV, where four spin observables $\Sigma$, $T$, $O_{x'}$ and
$O_{z'}$ are not measured (cf. Table~\ref{tab:expdata}), as we anticipate discontinuities in our solution in that energy range.
Observe that this is also the energy range analyzed in refs.~\cite{Anisovich2017,Anisovich2017a}. The fit has been performed with
two and three poles, and our solutions are documented in Fig.~\ref{L+P} and Table~\ref{Table1}.
\\ \\ \indent
\begin{table}[htb]
\caption{\label{Table1} The pole parameters for the $M_{1-}$ multipole extracted using the single-channel L+P formalism of
ref.~\cite{L+P2015,Svarc2013,Svarc2014,Svarc2014a} are given. BG-SE denotes the solution of
refs.~\cite{Anisovich2017,Anisovich2017a} with one pole, and AA/PWA denotes two solutions for AA/PWA determined in this paper,
with two and three poles respectively.  $M_i$, $\Gamma_i$, $r_i$ and $\Theta_i$, $i=1,\ldots,3$ are pole masses, widths as well
as absolute values of the residue and its phase, while $\chi^2$ is the value of the total chi-squared.  Values given by the
PDG~\cite{PDG} for these resonances are given in bold letters for comparison. }
\bigskip
%\begin{ruledtabular}
\scalebox{0.95}{
\begin{tabular}{cccccccccccccc} \hline \hline  \\[-2ex]
     Model    & $M_1$ & $\Gamma _1$ & $|a_1|$  & $\Theta_1$ & $M_2$ & $\Gamma _2$ & $|a_2|$  & $\Theta_2$ & $M_3$ & $\Gamma _3$ & $|a_3|$  & $\Theta_3$ & $\chi^2$   \\[0.5ex]
\hline  \\[-2ex]
PDG    &  \textbf{1700(40)}  &     \textbf{ 120(40)}      &     -     &      -       &  \textbf{1860(40)}     &    \textbf{230(50)}          &    -      &   -          &     \textbf{2100(50)}     &  \textbf{290(50)} & -   &  - &\\ [2ex]
BG-SE    &  -     &  -    &  -    &   -  &  1876(11)    &  31(19)    &  6(2)     &   57(14)   &  - &  -  &       &   -     &   36      \\ [2ex]
AA/PWA$^{\rm 2 poles}$    &  1715(12)     &  118(31)    &  117(62)    &   12454(35)  &  -    &  -    &  -     &   -&   2002(59) &  201(100)  &   84(103)      &   -116(92)     &   32.7       \\ [2ex]
 AA/PWA$^{\rm 3 poles}$        &  1714(19)     &  120(33)    &  122(98)    &   121(45)  &  1882(3)    &  6(6)    &  0.6(0.6)     &
123(54)   &   2007(70) &  187(102)  &   71(120)      &   -113(108)     &   29.2       \\
 \hline \hline
 \end{tabular} }
\end{table}
\newpage
In spite of obtaining a suspiciously narrow width of 33 MeV, the BG-SE solution is completely consistent with a $N(1880)
\frac{1}{2}^{+}$ resonant state, and is in refs.~\cite{Anisovich2017,Anisovich2017a} it has been interpreted as such. However,
both our fits depicted in Fig.~(\ref{L+P}) (b) and (c) do not support this conclusion. While the BG-SE fit identifies only a
$N(1880) \frac{1}{2}^{+}$ state, our model clearly confirms the existence of $N(1710) \frac{1}{2}^{+}$ and $N(2100)
\frac{1}{2}^{+}$ states too. In Fig.~(\ref{L+P}) (b) we show the fit of the data with two poles only. The fit is smooth and
reliable, with $\chi^2=32$, and covers all data in the whole energy range very reliably. Some deviation from the data is only
seen in the energy range of the $N(1880)\frac{1}{2}^{+}$ resonance, but it is small. In Fig.~(\ref{L+P}) (c) we show the
three-pole fit to the data. The $\chi^2$ is slightly improved from  32 to 29.2, but the obtained resonance, in spite of being in
the right place at 1882 MeV, has an extremely narrow width of 6 MeV. \clearpage

\noindent As such a narrow state of 6 MeV is extremely unlikely, and we do not see the mechanism how a wider state could
influence data in such a narrow energy range, we do not confirm the existence of $N(1880)\frac{1}{2}^{+}$ state contrary to
Bonn-Gatchina result. We confirm that we do have a "one- or two-points structure", but as the energy step in BG-SE case is much
wider (20 MeV) the width of the disturbance raises to ca 30 MeV. In our case our energy step is much lower (ca 2-5 MeV), a
"two-point" structure remains, but the width of a corresponding resonance becomes much narrower.
\\ \\ \noindent
We conclude that L+P analysis indicates that the narrow disturbance in this energy range is more consistent with the instability
due to violation of data consistency at this narrow energy range than to the existence of $N(1880)\frac{1}{2}^{+}$ resonance.

\section{Conclusions and Outlook} \label{sec:ConclusionsAndOutlook}
The new single-channel, single-energy data analysis method AA/PWA, fully explained and demonstrated for $\eta$ photoproduction in
ref.~\cite{Svarc2020}, has been applied to the world collection of data for $K \Lambda$ photoproduction. It is shown that a the
precise set of multipoles which improves the agreement with the data compared to the theoretical ED Bonn-Gatchina
model~\cite{Anisovich-priv-com} was obtained. Some discontinuities in the obtained multipoles are observed, and this is explained
by the violation of self-consistency of measured data, as well as the incompleteness of existing data sets. In order to overcome
these difficulties a stronger constraint on the penalizing functions is required. Let us observe that after Step 1, the
constraining amplitudes can be discontinuous because their continuity is in the present AA/PWA method guaranteed only by data
consistency and completeness of the data set. As both requirements are not met in most measured processes, additional conditions
for the achievement of continuity are needed. One natural way is to impose fixed-$t$
analyticity~\cite{Hoehler84,Osmanovic2019,Osmanovic2021}.
\\ \\ \indent
The obtained results were compared to the already published classic SE-PWA made in refs.~\cite{Anisovich2017,Anisovich2017a}, and
it has been shown that our results qualitatively and quantitatively agree with the results of  these references. The mutual
agreement of both approaches improves the probability that they are correct.
\\ \\ \indent
We confirm that the $M_{1-}$ multipole in our model reproduces a narrow structure around 1880 MeV, but our L+P analysis fails to
confirm that this structure is the confirmation of a $N(1880) \frac{1}{2}^{+}$ resonant state. The pole which would explain this
structure in our model is much narrower than the anyways very narrow pole given in refs.~\cite{Anisovich2017,Anisovich2017a}
(1876 MeV - $i$ 31 MeV), so we interpret it as a discontinuity due to data inconsistency rather than a resonance signal.
Preliminary tests have confirmed that if we more strongly constrain the TPWA of the second step of the AA/PWA method to the
smooth amplitudes of the ED BG2017 model, this structure disappears. However, this has to be elaborated in forthcoming
publications.
\\ \\ \indent
As the method shows a strong sensitivity to the self-consistency of the data, we advocate it as a reliable method to perform data
self-consistency testing.
%\begin{comment}

\begin{acknowledgments}
   Y.W. was supported financially by the {\it Transdisciplinary Research Area - Building Blocks of Matter and Fundamental Interactions
   (TRA Matter)} during the completion of this work.
\end{acknowledgments}

\appendix

\section{General photoproduction formalism} \label{sec:PhotoproductionFormalism}

In this appendix, we collect essential parts of the general formalism for pseudoscalar meson photoproduction, in order to keep
the present work self-contained. We consider the following $2 \rightarrow 2$ reaction:
\begin{equation}
\gamma \left( p_{\gamma}; m_{\gamma} \right) + N \left( P_{i}; m_{s_{i}} \right) \longrightarrow K \left( p_{K} \right) + \Lambda \left( P_{f};m_{s_{f}} \right) \mathrm{.} \label{eq:ProcessAppMomentaSpins}
\end{equation}
The $4$-momenta as well as the variables necessary to label the spin-states have been indicated for each particle.

The photoproduction process is conventionally described using the well-known Mandelstam variables $s$, $t$ and $u$. Since
$4$-momentum conservation holds, i.e. $p_{\gamma} + P_{i} = p_{K} + P_{f}$, each of the Mandelstam variables can be written in
two equivalent forms:
\begin{align}
s &= (p_{\gamma} + P_{i})^{2} = (p_{K} + P_{f})^{2} \mathrm{,} \label{eq:sMandelstam} \\
t &= (p_{\gamma} - p_{K})^{2} = (P_{f} - P_{i})^{2} \mathrm{,} \label{eq:tMandelstam} \\
u &= (p_{\gamma} - P_{f})^{2} = (P_{i} - p_{K})^{2} \mathrm{.} \label{eq:uMandelstam}
\end{align}
Since all particles in the initial- and final state of the reaction~\eqref{eq:ProcessAppMomentaSpins} are assumed to be on the
mass-shell, the whole reaction can be described by two independent kinematic invariants. The latter are often chosen to be the
pair $(s,t)$.

In this work, center-of-mass (CMS) coordinates are adopted. The following relations are valid between $(s,t)$ and the
center-of-mass energy $W$ and scattering angle $\theta$ of the reaction:
\begin{align}
 s &= W^{2}    , \label{eq:sEnergyEquation} \\
 t &= m_{K}^{2} - 2 k \sqrt{m_{K}^{2} + q^{2}} + 2 k q \cos \theta  . \label{eq:tAngleEquation}
\end{align}
In these expressions, $k$ and $q$ are the absolute values of the CMS $3$-momenta for the photon and the kaon, respectively. Both
of these variables can be expressed in terms of $W$ and the masses of the particles in the initial- and final states. One can
therefore describe the reaction equivalently in terms of $(W, \theta)$. The phase-space factor for the considered
reaction~\eqref{eq:ProcessAppMomentaSpins} is defined as $\rho = q / k$.

The general decomposition of the reaction amplitude into contributions of individual spin amplitudes has been found by Chew,
Goldberger, Low and Nambu (CGLN)~\cite{CGLN} and it reads as follows
\begin{equation}
\mathcal{F} = \chi_{m_{s_{f}}}^{\dagger} \left( i  \vec{\sigma} \cdot \hat{\epsilon}\; F_{1} + \vec{\sigma} \cdot \hat{q}
  \; \vec{\sigma} \cdot \hat{k} \times \hat{\epsilon}\; F_{2} + i  \vec{\sigma} \cdot \hat{k}\; \hat{q}
  \cdot \hat{\epsilon}\; F_{3} + i \vec{\sigma} \cdot \hat{q} \; \hat{q} \cdot \hat{\epsilon}\; F_{4}  \right) \hspace*{2pt} \chi_{m_{s_{i}}} \mathrm{.} \label{eq:FullAmplitudeCGLN}
\end{equation}
In this expression, $\hat{k}$ and $\hat{q}$ are normalized CMS 3-momenta, $\hat{\epsilon}$ is the normalized photon polarization
vector and $\chi_{m_{s_{i}}}$, $\chi_{m_{s_{f}}}$ are Pauli spinors for the baryons in the initial- and final states. The complex
amplitudes $F_{1}, \ldots, F_{4}$ are called {\it CGLN amplitudes} and they depend on $(W, \theta)$. This set of $4$ amplitudes
contains the full information on the dynamics of the considered process~\eqref{eq:ProcessAppMomentaSpins}.

The so-called {\it transversity amplitudes} $b_{1}, \ldots, b_{4}$ are defined by rotating the spin-quantization axis away from
the $\hat{z}$-axis of the CMS frame, which has been inherent to equation~\eqref{eq:FullAmplitudeCGLN}, to the direction normal to
the so-called reaction plane. The latter plane is spanned by the CMS $3$-momenta $\vec{k}$ and $\vec{q}$. Using the conventions
employed implicitly in the work of Chiang and Tabakin~\cite{Chiang:1996em}, one obtains the following set of linear and
invertible relations between transversity- and CGLN amplitudes~(see also reference~\cite{Wunderlich:2013iga}):
\begin{align}
 b_{1} \left( W, \theta\right) &= - b_{3} \left( W, \theta\right)
  - \frac{1}{\sqrt{2}} \sin \theta \left[ F_{3} \left( W, \theta\right) e^{- i \frac{\theta}{2}} + F_{4} \left( W, \theta\right) e^{ i \frac{\theta}{2}} \right] \mathrm{,} \label{eq:b1BasicForm} \\
 b_{2} \left( W, \theta\right) &= - b_{4} \left( W, \theta\right)
  +  \frac{1}{\sqrt{2}} \sin \theta \left[ F_{3} \left( W, \theta\right) e^{i \frac{\theta}{2}} + F_{4} \left( W, \theta\right) e^{- i \frac{\theta}{2}} \right] \mathrm{,} \label{eq:b2BasicForm} \\
 b_{3} \left( W, \theta\right) &= \frac{i}{\sqrt{2}} \left[ F_{1} \left( W, \theta\right) e^{- i \frac{\theta}{2}} -  F_{2} \left( W, \theta\right) e^{ i \frac{\theta}{2}} \right] \mathrm{,} \label{eq:b3BasicForm} \\
 b_{4} \left( W, \theta\right) &= \frac{i}{\sqrt{2}} \left[ F_{1} \left( W, \theta\right) e^{i \frac{\theta}{2}} -  F_{2} \left( W, \theta\right) e^{- i \frac{\theta}{2}} \right] \mathrm{.} \label{eq:b4BasicForm}
\end{align}
The transversity basis greatly simplifies the mathematical form of the definitions of the polarization observables (see
Table~\ref{tab:PhotoproductionObservables} and the discussion further below). For this reason, this basis is also used in the
analysis performed in the present work (cf. section~\ref{sec:KLambdaDataAnalysis}).

In order to access information on individual resonances, partial waves have to be analyzed. In the present work, we adopt the
well-known expansion of the CGLN amplitudes into electric and magnetic multipoles~$\left\{ E_{\ell \pm}, M_{\ell \pm} \right\}$,
i.e.~\cite{CGLN,Sandorfi2011}:
\begin{align}
F_{1} \left( W, \theta \right) &= \sum \limits_{\ell = 0}^{\infty} \Big\{ \left[ \ell M_{\ell+} \left( W \right) + E_{\ell+} \left( W \right) \right] P_{\ell+1}^{'} \left( \cos \theta \right) \nonumber \\
 & \quad \quad \quad + \left[ \left( \ell+1 \right) M_{\ell-} \left( W \right) + E_{\ell-} \left( W \right) \right] P_{\ell-1}^{'} \left( \cos \theta \right) \Big\} \mathrm{,} \label{eq:MultExpF1} \\
F_{2} \left( W, \theta \right) &= \sum \limits_{\ell = 1}^{\infty} \left[ \left( \ell+1 \right) M_{\ell+} \left( W \right) + \ell M_{\ell-} \left( W \right) \right] P_{\ell}^{'} \left( \cos \theta \right) \mathrm{,} \label{eq:MultExpF2} \\
F_{3} \left( W, \theta \right) &= \sum \limits_{\ell = 1}^{\infty} \Big\{ \left[ E_{\ell+} \left( W \right) - M_{\ell+} \left( W \right) \right] P_{\ell+1}^{''} \left( \cos \theta \right) \nonumber \\
 & \quad \quad \quad + \left[ E_{\ell-} \left( W \right) + M_{\ell-} \left( W \right) \right] P_{\ell-1}^{''} \left( \cos \theta \right) \big\} \mathrm{,} \label{eq:MultExpF3} \\
F_{4} \left( W, \theta \right) &= \sum \limits_{\ell = 2}^{\infty} \left[ M_{\ell+} \left( W \right) - E_{\ell+} \left( W \right) - M_{\ell-} \left( W \right) - E_{\ell-} \left( W \right) \right] P_{\ell}^{''} \left( \cos \theta \right) \mathrm{.} \label{eq:MultExpF4}
\end{align}
The multipoles can be assigned to definite conserved spin-parity quantum numbers $J^{P}$
(resonances with spin $J = \left| \ell \pm \frac{1}{2} \right|$ couple to the multipoles $E_{\ell \pm}$ and $M_{\ell \pm}$).
The multipole expansion of the $F_{i}$ can be formally inverted using a set of known projection integrals~\cite{Ball:1960baa,YannickPhD}. \\
% %
% \begin{align}
% M_{\ell+} &= \frac{1}{2 \left( \ell+1 \right)} \int_{-1}^{1} dx \left[ F_{1} P_{\ell} \left( x \right) - F_{2} P_{\ell+1} \left( x \right) - F_{3} \frac{P_{\ell-1} \left( x \right) - P_{\ell+1} \left( x \right)}{ 2\ell + 1} \right] \mathrm{,} \label{eq:MlplusProjection} \\
% E_{\ell+} &= \frac{1}{2 \left( \ell+1 \right)} \int_{-1}^{1} dx \bigg[ F_{1} P_{\ell} \left( x \right) - F_{2} P_{\ell+1} \left( x \right) + \ell F_{3} \frac{P_{\ell-1} \left( x \right) - P_{\ell+1} \left( x \right)}{ 2\ell + 1} \nonumber \\
%  & \quad \quad \quad \quad \quad \quad \quad \quad + \left( \ell+1 \right) F_{4} \frac{P_{\ell} \left( x \right) - P_{\ell+2} \left( x \right)}{ 2\ell + 3} \bigg] \mathrm{,} \label{eq:ElplusProjection} \\
% M_{\ell-} &= \frac{1}{2 \ell} \int_{-1}^{1} dx \left[ - F_{1} P_{\ell} \left( x \right) + F_{2} P_{\ell-1} \left( x \right) + F_{3} \frac{P_{\ell-1} \left( x \right) - P_{\ell+1} \left( x \right)}{ 2\ell + 1} \right] \mathrm{,} \label{eq:MlminusProjection} \\
% E_{\ell-} &= \frac{1}{2 \ell} \int_{-1}^{1} dx \bigg[ F_{1} P_{\ell} \left( x \right) - F_{2} P_{\ell-1} \left( x \right) - \left( \ell+1 \right) F_{3} \frac{P_{\ell-1} \left( x \right) - P_{\ell+1} \left( x \right)}{ 2\ell + 1} \nonumber \\
%  & \quad \quad \quad \quad \quad \enspace - \ell F_{4} \frac{P_{\ell-2} \left( x \right) - P_{\ell} \left( x \right)}{ 2\ell - 1} \bigg] \mathrm{.} \label{eq:ElminusProjection}
% \end{align}
% %
%In these projection-equations, one has $x = \cos \theta$.

The polarization observables accessible in pseudoscalar meson photoproduction are dimensionless asymmetries among differential
cross sections for different beam-, target- and recoil polarization states:
\begin{equation}
\Ocal = \frac{\beta \left[ \left( \frac{d \sigma}{d \Omega} \right)^{\left( B_{1}, T_{1}, R_{1} \right)} - \left( \frac{d \sigma}{d \Omega} \right)^{\left( B_{2}, T_{2}, R_{2} \right)} \right]}{\sigma_{0}} \mathrm{.} \label{eq:ObservableDefinitionGeneric}
\end{equation}
The factor $\beta$ is a consistency factor introduced in reference~\cite{Sandorfi2011}. It takes the value $\beta = \frac{1}{2}$
for observables which involve only beam- and target polarization and $\beta = 1$ for quantities with recoil polarization. The
unpolarized cross section $\sigma_{0}$ is always the sum of the two polarization configurations present in
equation~\eqref{eq:ObservableDefinitionGeneric}:
$\sigma_{0} = \beta \left[ \left( \frac{d \sigma}{d \Omega} \right)^{\left( B_{1}, T_{1}, R_{1} \right)} + \left( \frac{d \sigma}{d \Omega} \right)^{\left( B_{2}, T_{2}, R_{2} \right)} \right] $. \\
The dimensioned asymmetry $\sigma_{0} \Ocal$ is often called a {\it profile function} \cite{Chiang:1996em,YannickPhD} and it is
distinguished by a hat-mark on the $\Ocal$:
\begin{equation}
\hat{\Ocal} = \beta \left[ \left( \frac{d \sigma}{d \Omega} \right)^{\left( B_{1}, T_{1}, R_{1} \right)} - \left( \frac{d \sigma}{d \Omega} \right)^{\left( B_{2}, T_{2}, R_{2} \right)} \right] \mathrm{.} \label{eq:ProfileFunctionDefinition}
\end{equation}
For single-meson photoproduction, there exist in total $16$ polarization observables~\cite{Chiang:1996em,Sandorfi2011}. They
include the unpolarized cross section $\sigma_{0}$ and $15$ further single- and double-polarization observables. The full set of
$16$ observables can be divided into the four groups of single-spin observables ($\mathcal{S}$), beam-target- ($\mathcal{BT}$),
beam-recoil- ($\mathcal{BR}$) and target-recoil ($\mathcal{TR}$) observables~\cite{Chiang:1996em}. Each group is composed of $4$
quantities. The definitions of the $16$ observables in terms of transversity amplitudes are given in
Table~\ref{tab:PhotoproductionObservables}.

%\vspace*{-10pt}

\begin{table}[hb]
 \begin{center}
 \begin{tabular}{lr}
 \hline
 \hline
  Observable  &  Group  \\
  \hline
  $\sigma_{0} = \frac{1}{2} \left( \left| b_{1} \right|^{2} + \left| b_{2} \right|^{2} + \left| b_{3} \right|^{2} + \left| b_{4} \right|^{2} \right)$  &     \\
  $\hat{\Sigma} = \frac{1}{2} \left( - \left| b_{1} \right|^{2} - \left| b_{2} \right|^{2} + \left| b_{3} \right|^{2} + \left| b_{4} \right|^{2} \right)$  &   $\mathcal{S}$ \\
  $\hat{T} = \frac{1}{2} \left( \left| b_{1} \right|^{2} - \left| b_{2} \right|^{2} - \left| b_{3} \right|^{2} + \left| b_{4} \right|^{2} \right)$  &     \\
  $\hat{P} = \frac{1}{2} \left( - \left| b_{1} \right|^{2} + \left| b_{2} \right|^{2} - \left| b_{3} \right|^{2} + \left| b_{4} \right|^{2} \right)$  &     \\
  \hline
   $\hat{E}  = \mathrm{Re} \left[ - b_{3}^{\ast} b_{1} - b_{4}^{\ast} b_{2} \right]  = - \left| b_{1} \right| \left| b_{3} \right| \cos \phi_{13} - \left| b_{2} \right| \left| b_{4} \right| \cos \phi_{24}$  &  \\
   $\hat{F} = \mathrm{Im} \left[ b_{3}^{\ast} b_{1} - b_{4}^{\ast} b_{2} \right] = \left| b_{1} \right| \left| b_{3} \right| \sin \phi_{13} - \left| b_{2} \right| \left| b_{4} \right| \sin \phi_{24}  $ &  $\mathcal{BT} $ \\
   $\hat{G} = \mathrm{Im} \left[ - b_{3}^{\ast} b_{1} - b_{4}^{\ast} b_{2} \right] = - \left| b_{1} \right| \left| b_{3} \right| \sin \phi_{13} - \left| b_{2} \right| \left| b_{4} \right| \sin \phi_{24} $  &  \\
   $ \hat{H} = \mathrm{Re} \left[ b_{3}^{\ast} b_{1} - b_{4}^{\ast} b_{2} \right] = \left| b_{1} \right| \left| b_{3} \right| \cos \phi_{13} - \left| b_{2} \right| \left| b_{4} \right| \cos \phi_{24} $  &   \\
   \hline
   $\hat{C}_{x'}  = \mathrm{Im} \left[ - b_{4}^{\ast} b_{1} + b_{3}^{\ast} b_{2} \right]  = - \left| b_{1} \right| \left| b_{4} \right| \sin \phi_{14} + \left| b_{2} \right| \left| b_{3} \right| \sin \phi_{23}  $ &  \\
   $\hat{C}_{z'} = \mathrm{Re} \left[ - b_{4}^{\ast} b_{1} - b_{3}^{\ast} b_{2} \right] = - \left| b_{1} \right| \left| b_{4} \right| \cos \phi_{14} - \left| b_{2} \right| \left| b_{3} \right| \cos \phi_{23} $  &  $\mathcal{BR}$   \\
   $\hat{O}_{x'} = \mathrm{Re} \left[ - b_{4}^{\ast} b_{1} + b_{3}^{\ast} b_{2} \right] = - \left| b_{1} \right| \left| b_{4} \right| \cos \phi_{14} + \left| b_{2} \right| \left| b_{3} \right| \cos \phi_{23} $  &  \\
   $\hat{O}_{z'} = \mathrm{Im} \left[ b_{4}^{\ast} b_{1} + b_{3}^{\ast} b_{2} \right] = \left| b_{1} \right| \left| b_{4} \right| \sin \phi_{14} + \left| b_{2} \right| \left| b_{3} \right| \sin \phi_{23} $  &   \\
   \hline
   $\hat{L}_{x'} = \mathrm{Im} \left[ - b_{2}^{\ast} b_{1} - b_{4}^{\ast} b_{3} \right] = - \left| b_{1} \right| \left| b_{2} \right| \sin \phi_{12} - \left| b_{3} \right| \left| b_{4} \right| \sin \phi_{34}$  &   \\
   $\hat{L}_{z'}  = \mathrm{Re} \left[ - b_{2}^{\ast} b_{1} - b_{4}^{\ast} b_{3} \right]  = - \left| b_{1} \right| \left| b_{2} \right| \cos \phi_{12} - \left| b_{3} \right| \left| b_{4} \right| \cos \phi_{34}$  &  $\mathcal{TR}$  \\
   $\hat{T}_{x'} = \mathrm{Re} \left[ b_{2}^{\ast} b_{1} - b_{4}^{\ast} b_{3} \right] = \left| b_{1} \right| \left| b_{2} \right| \cos \phi_{12} - \left| b_{3} \right| \left| b_{4} \right| \cos \phi_{34}$  &    \\
   $\hat{T}_{z'} = \mathrm{Im} \left[ - b_{2}^{\ast} b_{1} + b_{4}^{\ast} b_{3} \right] = - \left| b_{1} \right| \left| b_{2} \right| \sin \phi_{12} + \left| b_{3} \right| \left| b_{4} \right| \sin \phi_{34}$ &   \\
   \hline
   \hline
 \end{tabular}
 \end{center}
 \caption{The definitions of the $16$ polarization ob\-serva\-bles of pseudoscalar meson photoproduction
 are given here in terms of transversity amplitudes $b_{1}, \ldots, b_{4}$ (cf. ref.~\cite{Chiang:1996em};
 sign conventions are consistent with~\cite{YannickPhD}). Expressions are given both in terms of real- and imaginary parts
 of bilinear products of amplitudes and in terms of moduli and relative phases of the amplitudes.
 Furthermore, the phase-space factor $\rho$ has been suppressed in the given expressions (i.e. we have set $\rho = 1$).
 The four different groups of four observables each are indicated as well.}
 \label{tab:PhotoproductionObservables}
\end{table}

\clearpage

\section{Solution-theory for the complete-experiment analysis (CEA) of the investigated database} \label{sec:SolutionTheory}

In the following, we give some more mathematical details on the possible ambiguities of the CEA when it is applied to the
database analyzed in this work (cf. section~\ref{sec:DescriptionDatabase}). We are well aware that the facts given below can be
extracted from the well-known mathematical treatments in the CEA for photoproduction~\cite{Chiang:1996em,Nakayama:2018yzw}.
Still, we hope that the details given in the following can provide some intuition on the mathematical ambiguities one has to be
careful about when analyzing the present database.

\subsection{Lower-energy region ($1625$ - $2179$ $\text{MeV}$): observables
$\left\{ \sigma_{0}, \hat{\Sigma}, \hat{T}, \hat{P}, \hat{O}_{x'} , \hat{O}_{z'},  \hat{C}_{x'} , \hat{C}_{z'} \right\}$} \label{sec:LowerEnergiesSolutionTheory}

We consider the definitions (cf. Table~\ref{tab:PhotoproductionObservables}) of the eight analyzed observables (in the following,
we set $\rho = q/k \equiv 1$):
\begin{align}
 \sigma_{0} &= \frac{1}{2} \left( \left|  b_{1}  \right|^{2} + \left|  b_{2}  \right|^{2} + \left|  b_{3}  \right|^{2} + \left|  b_{4}  \right|^{2} \right)    , \label{eq:DefDCSLowE} \\
 \hat{\Sigma} &= \frac{1}{2} \left( - \left|  b_{1}  \right|^{2} - \left|  b_{2}  \right|^{2} + \left|  b_{3}  \right|^{2} + \left|  b_{4}  \right|^{2} \right)    , \label{eq:DefSigmaLowE} \\
 \hat{T} &= \frac{1}{2} \left(  \left|  b_{1}  \right|^{2} - \left|  b_{2}  \right|^{2} - \left|  b_{3}  \right|^{2} + \left|  b_{4}  \right|^{2} \right)    , \label{eq:DefTLowE} \\
 \hat{P} &= \frac{1}{2} \left( - \left|  b_{1}  \right|^{2} + \left|  b_{2}  \right|^{2} - \left|  b_{3}  \right|^{2} + \left|  b_{4}  \right|^{2} \right)    , \label{eq:DefPLowE} \\
 \hat{O}_{x'} &=   \mathrm{Re} \left[ - b_{1} b_{4}^{\ast} + b_{2} b_{3}^{\ast} \right]   , \label{eq:DefOxLowE}  \\
  \hat{O}_{z'} &=  \mathrm{Im} \left[  b_{1} b_{4}^{\ast} + b_{2} b_{3}^{\ast} \right]    , \label{eq:DefOzLowE}  \\
  \hat{C}_{x'} &=   \mathrm{Im} \left[ - b_{1} b_{4}^{\ast} + b_{2} b_{3}^{\ast} \right]   , \label{eq:DefCxLowE}  \\
  \hat{C}_{z'} &=  \mathrm{Re} \left[ - b_{1} b_{4}^{\ast} - b_{2} b_{3}^{\ast} \right]    . \label{eq:DefCzLowE}
\end{align}
In case the phases of the transversity amplitudes $b_{j} = \left| b_{j} \right| e^{i \phi_{j}}$ are fixed to values coming from
an energy-dependent, unitary PWA model (e.g. BnGa), one can see very quickly that the system of equations composed of the eight
observables~\eqref{eq:DefDCSLowE} to~\eqref{eq:DefCzLowE} is in principle capable of fixing the four moduli $\left| b_{i}
\right|$ uniquely. In fact, the group-$\mathcal{S}$ observables $\left\{ \sigma_{0}, \hat{\Sigma}, \hat{T}, \hat{P} \right\}$
alone are already capable of that feat and the four additional
observables only should make the solution more stable. \\ \\
In case we also wish to determine the phases of the four transversity amplitudes, the situation is as follows: the four moduli
$\left| b_{i} \right|$ are fixed uniquely by the group-$\mathcal{S}$ observables and the four additional observables $\left\{
\hat{O}_{x'}, \hat{O}_{z'},  \hat{C}_{x'}, \hat{C}_{z'} \right\}$ can uniquely pin down the {\it relative phases} $\phi_{14}$ and
$\phi_{23}$. The latter statement is true due to the inverse relations:
\begin{align}
 e^{i \phi_{14}}  &= \frac{\Real \left[ b_{1} b_{4}^{\ast} \right] + i \Imag \left[ b_{1} b_{4}^{\ast} \right]}{\left| b_{1} \right| \left| b_{4} \right|} =  \frac{\left( - \hat{O}_{x'} - \hat{C}_{z'} \right) + i \left( \hat{O}_{z'} - \hat{C}_{x'} \right)}{2 \left| b_{1} \right| \left| b_{4} \right|}    , \label{eq:InverseRelI} \\
 e^{i \phi_{23}}  &= \frac{\Real \left[ b_{2} b_{3}^{\ast} \right] + i \Imag \left[ b_{2} b_{3}^{\ast} \right]}{\left| b_{2} \right| \left| b_{3} \right|} =  \frac{\left( \hat{O}_{x'} - \hat{C}_{z'} \right) + i \left( \hat{O}_{z'} + \hat{C}_{x'} \right)}{2 \left| b_{2} \right| \left| b_{3} \right|}    . \label{eq:InverseRelI}
\end{align}
The complex exponential functions $e^{i \phi_{jk}}$ can be inverted uniquely for phases on the interval $\phi_{jk} \in [0, 2 \pi )$
(via the arctan2 function). \\
The amplitude-arrangement of four transversity amplitudes is however not uniquely fixed\footnote{I.e., uniquely up to one overall
phase for {\it all} $4$~amplitudes.}. One additional 'connecting' relative phase is missing, for instance $\phi_{12}$ or
$\phi_{34}$. In other words, the two subsets of amplitudes $\left\{ b_{1}, b_{4} \right\}$ and $\left\{ b_{2}, b_{3} \right\}$
can be rotated relative to each other in a {\it completely free way} and the observables $\left\{ \hat{O}_{x'} , \hat{O}_{z'},
\hat{C}_{x'}, \hat{C}_{z'} \right\}$ are {\it completely blind to such a rotation}. As an example, consider a rotation of
\textit{only} the two amplitudes $b_{2}$ and $b_{3}$ by the same phase $\tilde{\varphi}$, which can have {\it any} dependence on
energy and angle:
\begin{equation}
  b_{2} (W, \theta) \longrightarrow e^{i \tilde{\varphi} (W, \theta)} b_{2} (W, \theta) \text{ and } b_{3} (W, \theta) \longrightarrow e^{i \tilde{\varphi} (W, \theta)} b_{3} (W, \theta). \label{eq:RotationOfB2AndB3}
\end{equation}
This rotation leaves both relative phases $\phi_{14}$ and $\phi_{23}$ invariant and therefore also all four observables $\left\{
\hat{O}_{x'}, \hat{O}_{z'},  \hat{C}_{x'}, \hat{C}_{z'} \right\}$. However, a rotation like~\eqref{eq:RotationOfB2AndB3}
generally leads to a new set of amplitudes with a very different partial wave decomposition, since it changes the unknown
continuum ambiguity phase (i.e. one overall phase for all~$4$ amplitudes) as well as the 'connecting' relative phases $\phi_{12}$
and $\phi_{34}$. One has to be careful about such effects when analyzing the data although as mentioned above, in our analysis
this ambiguity is removed by fixing the phases of all~$4$ transversity amplitudes to a known ED model (cf.
sections~\ref{sec:SummaryPWAAA} and~\ref{sec:KLambdaDataAnalysis}).

\subsection{Higher-energy region ($2179$ - $2296$ $\text{MeV}$): observables $\left\{ \sigma_{0}, \hat{P}, \hat{C}_{x'},
\hat{C}_{z'} \right\}$} \label{sec:HigherEnergiesSolutionTheory}

We start by considering the definitions (cf. Table~\ref{tab:PhotoproductionObservables}) of the four analyzed observables (again
setting $\rho = q/k \equiv 1$):
\begin{align}
 \sigma_{0} &= \frac{1}{2} \left( \left|  b_{1}  \right|^{2} + \left|  b_{2}  \right|^{2} + \left|  b_{3}  \right|^{2} + \left|  b_{4}  \right|^{2} \right)    , \label{eq:DefDCS} \\
 \hat{P} &= \frac{1}{2} \left( - \left|  b_{1}  \right|^{2} + \left|  b_{2}  \right|^{2} - \left|  b_{3}  \right|^{2} + \left|  b_{4}  \right|^{2} \right)    , \label{eq:DefP} \\
  \hat{C}_{x'} &=   \mathrm{Im} \left[ - b_{1} b_{4}^{\ast} + b_{2} b_{3}^{\ast} \right]   , \label{eq:DefCx}  \\
  \hat{C}_{z'} &=  \mathrm{Re} \left[ - b_{1} b_{4}^{\ast} - b_{2} b_{3}^{\ast} \right]    . \label{eq:DefCz}
\end{align}
We assume that the phases of the four transversity amplitudes are fixed to a model and define:
\begin{equation}
  c_{ij} := \cos \phi_{ij} \text{, and } s_{ij} := \sin \phi_{ij}. \label{eq:CosSinRelPhaseDefinition}
\end{equation}
We re-consider the equations for $\hat{C}_{x'}$ and $\hat{C}_{z'}$:
\begin{align}
 \hat{C}_{x'} &= -  \left| b_{1} \right| \left| b_{4} \right| s_{14} + \left| b_{2} \right| \left| b_{3} \right| s_{23}    , \label{eq:DefCxII}  \\
  \hat{C}_{z'} &= - \left| b_{1} \right| \left| b_{4} \right| c_{14} - \left| b_{2} \right| \left| b_{3} \right| c_{23}      , \label{eq:DefCzII}
\end{align}
and recognize that these equations can be inverted for the following products of moduli:
\begin{align}
  \left| b_{1} \right| \left| b_{4} \right|   &=  \frac{- c_{23} \hat{C}_{x'}  -   s_{23} \hat{C}_{z'}}{s_{14} c_{23} + c_{14} s_{23}} =: \xi_{1} ,  \label{eq:FirstModProduct} \\
  \left| b_{2} \right| \left| b_{3} \right|   &=  \frac{ c_{14} \hat{C}_{x'}  -   s_{14} \hat{C}_{z'}}{s_{14} c_{23} + c_{14} s_{23}} =:\xi_{2} .  \label{eq:SecondModProduct}
\end{align}
The quantities $\xi_{1}$ and $\xi_{2}$ are uniquely fixed from the observables and the employed model phases. \\ \\
We could now choose to eliminate, for instance, the quantities $\left| b_{3} \right|$ and $\left| b_{4} \right|$ in the equations
for the cross section~\eqref{eq:DefDCS} and the observable~$\hat{P}$~\eqref{eq:DefP} and thus obtain:
\begin{align}
 \sigma_{0} &= \frac{1}{2} \left[ \left|  b_{1}  \right|^{2} +  \left|  b_{2}  \right|^{2} + \left( \frac{\xi_{2}}{\left|  b_{2}  \right|} \right)^{2} +  \left( \frac{\xi_{1}}{\left|  b_{1}  \right|} \right)^{2} \right]    , \label{eq:DefDCSModified} \\
 \hat{P} &= \frac{1}{2} \left[ - \left|  b_{1}  \right|^{2} +  \left|  b_{2}  \right|^{2} - \left( \frac{\xi_{2}}{\left|  b_{2}  \right|} \right)^{2} +  \left( \frac{\xi_{1}}{\left|  b_{1}  \right|} \right)^{2} \right]    , \label{eq:DefPModified}
\end{align}
These are two quadratic equations for the two remaining unknowns $\left| b_{1} \right|$ and $\left| b_{2} \right|$. One can make
the following attempt at solving them. We multiply the equation~\eqref{eq:DefPModified} by $\left| b_{2} \right|^{2}$ in order to
obtain:
\begin{equation}
   \left| b_{2} \right|^{4} + \left[  \left( \frac{\xi_{1}}{\left|  b_{1}  \right|} \right)^{2} - \left| b_{1} \right|^{2} - 2 \hat{P} \right] \left| b_{2} \right|^{2} - \left( \xi_{2} \right)^{2} = 0   .  \label{eq:DefPModifiedII}
\end{equation}
This is a quadratic equation for $\left| b_{2} \right|^{2}$ and thus allows for the following two solutions:
\begin{equation}
 \left| b_{2} \right|^{2}_{\text{I,II}} = \frac{1}{2} \left[ 2 \hat{P} + \left| b_{1} \right|^{2} - \left( \frac{\xi_{1}}{\left|  b_{1}  \right|} \right)^{2}  \right] \pm \sqrt{ \frac{1}{4} \left[ 2 \hat{P} + \left| b_{1} \right|^{2} - \left( \frac{\xi_{1}}{\left|  b_{1}  \right|} \right)^{2}  \right]^{2} +  \left( \xi_{2} \right)^{2} } . \label{eq:DefPModifiedTwoSolutions}
\end{equation}
Irrespective of whether solution 'I' or 'II' is the correct one, the '+' branch of the square root has to be taken in order to
arrive at a positive modulus $\left| b_{2} \right|$. In case both solutions 'I' or 'II' are allowed (i.e. larger than zero) in
equation~\eqref{eq:DefPModifiedTwoSolutions}, one obtains two permissible moduli $\left| b_{2} \right|_{\text{I,II}}$. Then, one
has to substitute these solutions into the equation for the cross section~\eqref{eq:DefDCSModified} and solve for $\left| b_{1}
\right|$. In this way, at most a set of four discrete ambiguities can emerge and all continuous ambiguities are
resolved for the four moduli~$\left| b_{1} \right|, \ldots, \left| b_{4} \right|$. \\
Therefore, in case one would attempt to let all four moduli $\left| b_{1} \right|, \ldots, \left| b_{4} \right|$ run freely in
the AA step (i.e. our step '1'), we expect the found solution to lie on a well-defined (approximately) parabolic minimum where
the derivative of the minimized ('$\chi^{2}$-like') function exactly vanishes. Data for the four observables $\left\{ \sigma_{0},
\hat{P}, \hat{C}_{x'}, \hat{C}_{z'} \right\}$ are in principle only capable to distinguish solutions up to the above-mentioned
discrete ambiguity. However, in case the initial conditions are well-chosen for the minimization procedure, we are confident that
the correct minimum can be found, i.e. that the moduli extraction is sufficiently stable.

%\end{comment}


\begin{thebibliography}{99}

\bibitem{Svarc2020} A. \v{S}varc, Y. Wunderlich , and L. Tiator, Phys. Rev. \textbf{C 102}, 064609 (2020).

\bibitem{Anisovich2017} A.V. Anisovich, V. Burkert, M. Had\v{z}imehmedovi\'{c}, D.G. Ireland, E. Klempt, V.A. Nikonov,
R. Omerovi\'{c}, H. Osmanovi\'{c},  A.V. Sarantsev, J. Stahov, A. \v{S}varc, and U. Thoma,  Phys. Rev. Lett. \textbf{119}, 062004
(2017).

\bibitem{Anisovich2017a} A.V. Anisovich, V. Burkert, M. Had\v{z}imehmedovi\'{c}, D.G. Ireland, E. Klempt, V.A. Nikonov,
R. Omerovi\'{c}, H. Osmanovi\'{c},  A.V. Sarantsev, J. Stahov, A. \v{S}varc, and U. Thoma, Eur. Phys. J. A \textbf{53}: 242 (2017).

\bibitem{Martin} A. Martin, J. Richard, Phys. Rev. \textbf{D 101}, 094014 (2020), \emph{and references therein}.

\bibitem{Stefanescu} I. Sabba Stefanescu, Phys. Rev. \textbf{D 21}, 3225 (1980); Fortschr. Phys. \textbf{35}, 8-9, 573-673 (1987).

\bibitem{Continuum-ambiguity} D. Atkinson, P.W. Johnson and R.L. Warnock, Commun. mat. Phys. {\bf 33}, 221 (1973);
J.E. Bowcock and H. Burkhard, Rep. Prog. Phys. {\bf 38}, 1099 (1975); D. Atkinson and I.S. Stefanescu, Commun. Math. Phys. {\bf
101}, 291 (1985).

\bibitem{Svarc2021} A. \v{Svarc}, Phys. Rev. \textbf{ C 104}, 014605 (2021).

\bibitem{Hoehler84} G. H\"{o}hler, Pion Nucleon Scattering, Part 2, Landolt-Bornstein: Elastic and Charge Exchange Scattering of
Elementary Particles, Vol. 9b (Springer-Verlag, Berlin, 1983).


\bibitem{Osmanovic2019}
H. Osmanovi\'{c}, M. Had\v{z}imehmedovi\'{c}, R. Omerovi\'{c}, J. Stahov,  M. Gorchtein,  V. Kashevarov, K. Nikonov, M. Ostrick,
L. Tiator, and A. \v{S}varc, Phys. Rev. \textbf{C 100}, 055203 (2019).


\bibitem{Osmanovic2021}
H. Osmanovi\'{c}, M. Had\v{z}imehmedovi\'{c}, R. Omerovi\'{c}, J. Stahov, V. Kashevarov, M. Ostrick, L. Tiator, and A. \v{S}varc,
Phys. Rev. \textbf{C 104}, 034605 (2021).

\bibitem{Watson} K.M. Watson, Phys. Rev. \textbf{95}, 228 (1954).

\bibitem{Svarc2018}
A. \v{S}varc, Y. Wunderlich, H. Osmanovi\'{c}, M. Had\v{z}imehmedovi\'{c}, R. Omerovi\'{c}, J. Stahov, V. Kashevarov,
 K. Nikonov, M. Ostrick, L. Tiator, and R. Workman, Few-Body Syst. 59:96 (2018).

\bibitem{ETA-MAID15a}
V.L. Kashevarov, L. Tiator, M. Ostrick, Bled Workshops Phys. \textbf{16}, 9 (2015).

\bibitem{BG-web} https://pwa.hiskp.uni-bonn.de/.

\bibitem{Osmanovic2018}  H. Osmanovi\'{c}, M. Had\v{z}imehmedovi\'{c}, R. Omerovi\'{c}, J. Stahov, V. Kashevarov,
K. Nikonov, M. Ostrick, L. Tiator,  and A. \v{S}varc, Phys. Rev.  \textbf{C 97}, 015207 (2018).

\bibitem{Svarc2018a} A.~\v{S}varc, Y. Wunderlich, H. Osmanovi\'{c}, M. Had\v{z}imehmedovi\'{c}, R. Omerovi\'{c}, J. Stahov,
V. Kashevarov, K. Nikonov, M. Ostrich, L. Tiator, R. Workman, Phys.\ Rev.\ C\ {\bf 97}, 054611 (2018).

\bibitem{BoGa} A.~V.~Anisovich {\it et al.},  Phys.\ Rev.\ C\ {\bf 96}, 055202 (2017) and references therein,
https://pwa.hiskp.uni-bonn.de/.

\bibitem{Juelich} D. R\"{o}nchen, M. D\"{o}ring, H. Haberzettl, J. Haidenbauer, U.-G. Mei\"{ss}ner and K. Nakayama,
 Eur. Phys. J.\textbf{ A 51}, 70 (2015).
and references therein; and http://collaborations.fz-juelich.de/ikp/meson-baryon/main.

\bibitem{GWU/SAID} R.L. Workman, R.A. Arndt, W.J. Briscoe, M.W. Paris, and I.I. Strakovsky, Phys. Rev.
\textbf{ C 86}, 035202 (2012); and http://gwdac.phys.gwu.edu/.

\bibitem{MAID} D. Drechsel, S.S. Kamalov and L. Tiator, Eur. Phys. J.\textbf{ A 34 }, 69 (2007); and https://maid.kph.uni-mainz.de/.

\bibitem{GWU-web} https://gwdac.phys.gwu.edu/.

\bibitem{Anisovich-priv-com}  A.V.~Anisovich, private communication (2017).

\bibitem{Bradford} R.~Bradford et al., Phys. Rev.\textbf{C 73}, 035202 (2006).

\bibitem{McCracken} M.E.~McCracken et al., Phys. Rev.\textbf{C 81}, 025201 (2010).

\bibitem{Lleres} A.~Lleres et al., Eur. Phys. J. \textbf{A 31}, 79 (2007).

\bibitem{Paterson} C.A. Paterson et al., Phys. Rev. \textbf{C 93}, 065201 (2016).


\bibitem{Mathematica} Wolfram Research, Inc., Mathematica, Version 11.0, Champaign, IL (2016).

\bibitem{L+P2015} A. \v{S}varc, M. Had\v{z}imehmedovi\'{c}, H. Osmanovi\'{c}, J. Stahov,
and R.L. Workman, Phys. Rev. \textbf{C 91}, {015207} (2015).

\bibitem{Svarc2013} A. \v{S}varc, M. Had\v{z}imehmedovi\'{c}, H. Osmanovi\'{c}, J. Stahov, L. Tiator,
and R.L. Workman, Phys,  Rev.  \textbf{C 88}, 035206 (2013).

\bibitem{Svarc2014} A. \v{S}varc,  M. Had\v{z}imehmedovi\'{c}, R. Omerovi\'{c}, H. Osmanovi\'{c},
and J. Stahov,  Phys, Rev. \textbf{C 89},  0452205  (2014).

\bibitem{Svarc2014a}   A. \v{S}varc, M. Had\v{z}imehmedovi\'{c}, H. Osmanovi\'{c}, J. Stahov, L. Tiator,
and R.L. Workman, Phys,   Rev.  \textbf{C 89}, 65208 (2014).

\bibitem{PDG} P.A. Zyla et al. (Particle Data Group), Prog. Theor. Exp. Phys. 2020, 083C01 (2020) and 2021 update,
mini-review on N and $\Delta$ resonances.

\bibitem{Ciulli}S. Ciulli and J. Fischer in Nucl. Phys. \textbf{24}, 465 (1961).

\bibitem{CiulliFisher}I. Ciulli, S. Ciulli, and J. Fisher, Nuovo Cimento \textbf{23}, 1129 (1962).

\bibitem{Pietarinen} E. Pietarinen, Nuovo Cimento Soc. Ital. Fis. \textbf{12A}, 522 (1972).

\bibitem{Pietarinen1} E. Pietarinen, Nucl. Phys. \textbf{B107}, 21 (1976).

\bibitem{Svarc2016} A. \v{S}varc, M. Had\v{z}imehmedovi\'{c}, H. Osmanovi\'{c}, J. Stahov, L. Tiator, R.L. Workman,
 Phys. Lett. \textbf{B755}, 452 (2016).

\bibitem{Mittag-Leffler}Michiel Hazewinkel: \emph{Encyclopaedia of Mathematics}, Vol.6,  Springer, 31. 8. 1990, p.~251.

\bibitem{CGLN}
G.F. Chew, M. Goldberger, F.E. Low, and Y. Nambu, Phys. Rev. {\bf 106}, 1345 (1957).
%\cite{Chiang:1996em}

\bibitem{Chiang:1996em}
  W.~T.~Chiang and F.~Tabakin,
  %``Completeness rules for spin observables in pseudoscalar meson photoproduction,''
  Phys.\ Rev.\ C {\bf 55}, 2054 (1997).
%  doi:10.1103/PhysRevC.55.2054
%  [nucl-th/9611053].
  %%CITATION = doi:10.1103/PhysRevC.55.2054;%%
  %187 citations counted in INSPIRE as of 19 Dec 2019

\bibitem{Wunderlich:2013iga}
Y.~Wunderlich, R.~Beck and L.~Tiator,
%``The complete-experiment problem of photoproduction of pseudoscalar mesons in a truncated partial-wave analysis,''
Phys. Rev. C \textbf{89}, no.5, 055203 (2014).
%doi:10.1103/PhysRevC.89.055203
%[arXiv:1312.0245 [nucl-th]].
%40 citations counted in INSPIRE as of 11 May 2021

\bibitem{Sandorfi2011} A.M. Sandorfi, S. Hoblit, H. Kamano,  and T.-S.H. Lee, J. Phys. G: Nucl. Part. Phys. {\bf 38},  053001 (2011).


\bibitem{Ball:1960baa}
J.S.~Ball,
%``The application of the Mandelstam representation to photoproduction of Pions from Nucleons,''
Phys. Rev. \textbf{124}, 2014 (1961).
%1 citations counted in INSPIRE as of 18 May 2020

\bibitem{YannickPhD}
Y.~Wunderlich, "The complete experiment problem of pseudoscalar meson photoproduction in a truncated partial wave analysis", PhD
thesis, University of Bonn (2019) [arXiv:2008.00514 [nucl-th]].

%\cite{Nakayama:2018yzw}
\bibitem{Nakayama:2018yzw}
  K.~Nakayama,
  %``Explicit derivation of the completeness condition in pseudoscalar meson photoproduction,''
  Phys.\ Rev.\ C {\bf 100}, no. 3, 035208 (2019).
  %doi:10.1103/PhysRevC.100.035208
  %[arXiv:1809.00335 [nucl-th]].
  %%CITATION = doi:10.1103/PhysRevC.100.035208;%%


\end{thebibliography}
\end{document}